\documentclass[letterpaper,12pt,unknownkeysallowed]{article} % unknownkeysallowed fixes compilation outside ShareLatex
\usepackage[utf8]{inputenc}
\usepackage{verbatim}
\usepackage[english]{babel}
\usepackage{amsfonts}
\usepackage{amssymb}
\usepackage{amsmath}
\usepackage{enumitem}
\usepackage{bbm}
\usepackage{mathtools}
\usepackage{subcaption}
\usepackage{booktabs}
\usepackage{setspace}
\usepackage{graphicx}
\usepackage{lscape}
\usepackage{rotating}
\usepackage{fullpage}
\usepackage[table]{xcolor}
\usepackage{multirow}

\usepackage[capposition=top]{floatrow} % for notes under charts
	\usepackage[utf8]{inputenc}
	\usepackage[english]{babel}
	\usepackage{amsfonts}
	\usepackage{amssymb}
	\usepackage{enumitem}
	\usepackage{bbm}
	\usepackage{mathtools}
    \usepackage{subcaption}
	\usepackage{booktabs}
	\usepackage{setspace}
	\usepackage{graphicx}
	\usepackage{lscape}
	\usepackage{rotating}
	\usepackage{fullpage}
    \usepackage[capposition=top]{floatrow} % for notes under charts
    \usepackage{array}
    \newcolumntype{H}{>{\setbox0=\hbox\bgroup}c<{\egroup}@{}}

	\usepackage{natbib}
	\usepackage[framemethod=tikz]{mdframed}
	\usepackage{todonotes}
	\presetkeys{todonotes}{color=purple!10,inline}{} % noline

	\newcommand{\rf}[1]{\comment{Reference: \url{#1}}}

\newcommand{\bs}{\boldsymbol}
\renewcommand{\appendix}{\small\parindent 0cm\parskip 5pt\setcounter{equation}{0}
\setcounter{section}{0}
\renewcommand{\thesection}{A.\arabic{section}}
\renewcommand{\theequation}{A.\arabic{equation}}
\setcounter{lemma}{0}\renewcommand{\thelemma}{A.\arabic{lemma}}
\setcounter{theorem}{0}\renewcommand{\thetheorem}{A.\arabic{theorem}}
}

\newtheorem{assump}{Assumption}

  \makeatletter
\newcommand{\oset}[3][0ex]{%
  \mathrel{\mathop{#3}\limits^{
    \vbox to#1{\kern-2\ex@
    \hbox{$\scriptstyle#2$}\vss}}}}
\makeatother

\makeatletter
\def\blfootnote{\xdef\@thefnmark{}\@footnotetext}
\makeatother

\makeatletter
\def\@sect#1#2#3#4#5#6[#7]#8{\ifnum #2>\c@secnumdepth
     \let\@svsec\@empty\else
     \refstepcounter{#1}\edef\@svsec{\csname the#1\endcsname. \hskip 0.4em}\fi
     \@tempskipa #5\relax
      \ifdim \@tempskipa>\z@
        \begingroup #6\relax
          \@hangfrom{\hskip #3\relax\@svsec}{\interlinepenalty \@M #8\par}%
        \endgroup
       \csname #1mark\endcsname{#7}\addcontentsline
         {toc}{#1}{\ifnum #2>\c@secnumdepth \else
                      \protect\numberline{\csname the#1\endcsname}\fi
                    #7}\else
        \def\@svsechd{#6\hskip #3\relax  %% \relax added 2 May 90
                   \@svsec #8\csname #1mark\endcsname
                      {#7}\addcontentsline
                           {toc}{#1}{\ifnum #2>\c@secnumdepth \else
                             \protect\numberline{\csname the#1\endcsname}\fi
                       #7}}\fi
     \@xsect{#5}}
\makeatother

\makeatletter
\renewcommand{\section}{\@startsection{section}{1}{0mm}{-\baselineskip}{0.25\baselineskip}{\centering\normalfont\normalsize\bf}}
\renewcommand{\subsection}{\@startsection{subsection}{2}{0mm}{-\baselineskip}{0.25\baselineskip}{\raggedright\normalfont\normalsize\bf}}
\renewcommand{\subsubsection}{\@startsection{subsubsection}{3}{0mm}{-\baselineskip}{0.25\baselineskip}{\raggedright\normalfont\small}}
\def\@begintheorem#1#2{\trivlist \item[\hskip \labelsep{\bf #1\ #2}]\it}
\makeatother

\renewcommand{\thesection}{\arabic{section}}

\renewenvironment{abstract}
 {\begin{center}\normalsize\bf\text{Abstract}
 \end{center}\begin{quote}\normalsize}
 {\end{quote}}

\allowdisplaybreaks

\begin{document}
\vspace*{0.2cm}
%\vskip 10pt
%\hfill {\tt Revision in progress}
\centerline{\Large\bf Synthetic Controls in Action}
\begin{center}%
    \vskip 10pt
    {\large
     \lineskip .5em%
      \begin{tabular}[t]{ccc}%
       Alberto Abadie&\hspace*{.5cm}&Jaume Vives-i-Bastida\\[.2ex]
       MIT&&MIT\\
      \end{tabular}
      \par}%
      \vskip 1em%
    {\large \today } \par%\monthname \ \number\year
   \vskip 1em%
\end{center}\par

\bigskip
 \begin{abstract}
 \noindent In this article we propose a set of simple principles to guide empirical practice in synthetic control studies. The proposed principles follow from formal properties of synthetic control estimators, and pertain to the nature, implications, and prevention of over-fitting biases within a synthetic control framework, to the interpretability of the results, and to the availability of validation exercises. We discuss and visually demonstrate the relevance of the proposed principles under a variety of data configurations.
 \end{abstract}

\blfootnote{\hspace*{-0.25in}Alberto Abadie, Department of Economics, MIT, abadie@mit.edu. Jaume Vives-i-Bastida, Department of Economics, MIT, vives@mit.edu. We greatly benefited form  comments by Kaspar Wuthrich and Yinchu Zhu. NSF support through grant SES-1756692 is gratefully acknowledged.
}

\vskip 10pt

\section{Introduction}

Synthetic control methods have become widely applied in empirical research in economics and other disciplines. Recent empirical applications of the synthetic control framework include studies of the effects of political connections \citep{acemoglu2016turbulent}, legalized prostitution \citep{cunningham2017prostitution}, right-to-carry laws \citep{donohue2017crime}, and many other \citep[see][for a more elaborate review of the applications of synthetic control estimators]{abadie2021using}. With the increasing popularity of synthetic control estimators, it has become particularly important to understand the settings where synthetic controls produce reliable estimates and those where they do not.

In this article, we explore in detail how the properties of synthetic controls translate to actual performance. From the properties of synthetic control estimators we distill a set of simple principles to guide empirical practice, and demonstrate the practical relevance of our proposed principles via simulation exercises.

First, we use a variety of data configurations to demonstrate the crucial role of pre-treatment fit in the performance of synthetic control estimators. Close pre-treatment fit, however, does not guarantee good performance of synthetic control estimators because of the possibility of over-fitting. We describe the nature of over-fitting within the context of synthetic control estimation, explain how over-fitting induces biases, demonstrate the practical relevance of these biases, and discuss how a careful study design can minimize or avoid them. Finally, we discuss validation exercises that can be applied to assess the validity of synthetic control estimators in empirical applications. 

The rest of the article is organized as follows. Section \ref{section:primer} introduces synthetic control estimators and discusses their formal properties under a linear factor structure in the data generating process. Section \ref{section:performance} employs a grouped version of the linear factor model of Section \ref{section:performance} to study how the performance of synthetic controls is related to pre-treatment fit and over-fitting. Section \ref{section:validation} discusses how to validate synthetic control estimates. Section \ref{section:trimming} discusses trimming as a way to alleviate over-fitting and interpolation biases. Section \ref{section:observed} discusses the role of covariates. The applicability of synthetic control estimators is not confined, however, to the linear factor model. In Section \ref{section:autoregressive}, we use a simple auto-regressive model to illustrate this point. Section \ref{section:conclusions} concludes. 

\section{A Primer on Synthetic Controls}
\label{section:primer}

Suppose we observe $j=1,\ldots, J+1$ aggregate units, such as states or countries for $t=1,\ldots, T$ periods. The first unit (that is, $j=1$) is exposed to a policy intervention, or some other event or treatment of interest, at time $t=T_0+1$, with $T_0+1\leq T$. The remaining $J$ units are not exposed to the event or intervention of interest. We aim to estimate the effect of the treatment on some outcome of interest during the post-treatment periods, $T_0+1, \ldots, T$. We will use the terms ``intervention'', ``event'', and ``treatment'' interchangeably. 

To define treatment effects, we formally adopt a model of potential outcomes \citep{rubin1974estimating}. Let $Y^N_{jt}$ be the potential outcome observed for unit $j\in \{1,\ldots, J+1\}$ and time $t=\{1,\ldots, T\}$ in the absence of the intervention. Let $Y^I_{1t}$ be the potential outcome observed for the treated unit at time $t=T_0+1,\ldots, T$ under the intervention. For each unit and time period, $Y_{jt}$ is the observed outcome. Therefore, observed outcomes for untreated units, $j=2,\ldots,, J+1$, are equal to $Y^N_{jt}$. For the treated unit, the observed outcome is equal to $Y^N_{1t}$ for $t=1, \ldots, T_0$, and equal to $Y^I_{1t}$ for $t=T_0+1, \ldots, T$. The object of interest is the treatment effect on the treated unit,
\[
\tau_t = Y^I_{1t} - Y_{1t}^N, 
\]
for $t=T_0+1,\ldots, T$. Because $Y^I_{1t}=Y_{1t}$ in the post-treatment periods, we obtain
\[
\tau_t = Y_{1t} - Y_{1t}^N, 
\]
for $t=T_0+1,\ldots, T$. That is, because $Y^I_{1t}$ is observed in the post-treatment periods, estimating $\tau_{1t}$ for $t=T_0+1,\ldots, T$ boils down to estimating $Y_{1t}^N$. A synthetic control estimator of $Y_{1t}^N$ is a weighted average of the outcomes for the ``donor pool'' of $J$ untreated units,
\[
\widehat Y_{1t}^N = \sum_{j=2}^{J+1} W_j Y_{jt},
\]
where $W_2, \ldots, W_{J+1}$ are non-negative and sum to one. A synthetic control estimator of $\tau_{1t}$ is equal to the difference between the outcome values for the treated units and the outcomes values for the synthetic control,
\[
\widehat\tau_t = Y_{1t} - \sum_{j=2}^{J+1} W_j Y_{jt}.
\]
The weights, $W_2, \ldots, W_{J+1}$, represent the contribution of each untreated observation to the estimate of the counterfactual of interest, $\widehat Y^N_{1t}$. While these weights could, in principle, be directly chosen by the analyst, the synthetic control literature provides a host of data-driven weight selectors. For $j=1, \ldots, J+1$, let $\bs X_j= (X_{1j}, \ldots, X_{kj})'$ be a $(k\times 1)$-vector of pre-intervention values of predictors of $Y^N_{jt}$, with $t=T_0+1, \ldots, T$. Let $\bs X_0$ be the $(k\times J)$-matrix that concatenates $\bs X_2, \ldots, \bs X_{J+1}$. A simple data-driven selector $\bs W^*=(W^*_2, \ldots, W^*_{J+1})'$ of the synthetic control weights minimizes
\begin{equation}
\label{equation:objective}
\|\boldsymbol{X}_1 - \boldsymbol{X}_0\boldsymbol{W}\| = \left( \sum_{h=1}^k v_h(X_{h1} - W_2X_{h2} - \dots - W_{J+1}X_{hJ+1})^2\right)^{1/2},
\end{equation}
subject to the constraints that the weights are non-negative and sum up to one. The non-negative constants, $v_1, \ldots, v_k$, can be used to standardize the predictors (for example, by making $v_h$ equal to the inverse of the variance of the $h$-th predictor) or chosen to reflect their predictive power on the outcome of interest.

\begin{figure}[ht!]
\centering
  \includegraphics[width=0.6\linewidth]{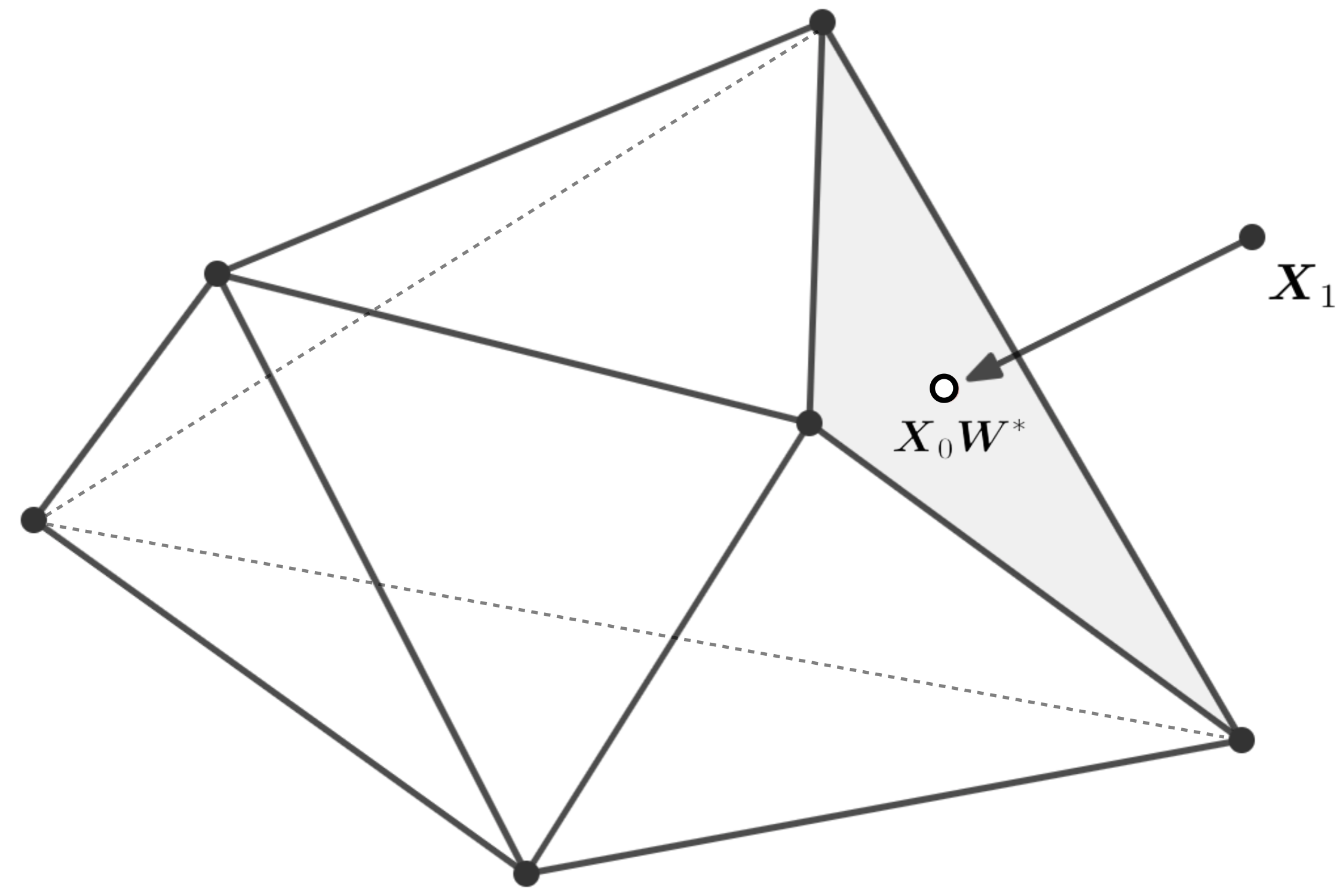}
  \caption{Sparsity of Synthetic Controls: A Geometric Interpretation.}
  \label{figure:convex}
\floatfoot{{\em Note:} Projection of $\boldsymbol X_1$ on the convex hull of the columns of $\bs X_0$.}
\end{figure}

Synthetic control estimates, as defined above, are typically sparse, meaning that only a few units in the donor pool obtain weigths, $W^*_j$, different from zero. Figure \ref{figure:convex} provides a geometric interpretation of this property. The synthetic control $\boldsymbol X_0\boldsymbol W^*$ is given by the projection of $\boldsymbol X_1$ on the convex hull of the columns of $\boldsymbol X_0$. The curse of dimensionality implies that, even in moderately large dimension, $\boldsymbol X_1$ may often be outside the convex hull of the columns of $\boldsymbol X_0$. This is the case depicted in Figure \ref{figure:convex}. As indicated in the figure, when $\boldsymbol X_1$ is outside the convex hull of the columns of $\boldsymbol X_0$, the synthetic control $\boldsymbol X_0\boldsymbol W^*$ is unique and sparse provided that the columns of $\boldsymbol X_0$ are in general position (see \citeauthor{abadie2021penalized}, \citeyear{abadie2021penalized}, for the meaning of ``general position''). Then, the number of non-zero weights, $W^*_j$, is not larger than $k$, the dimension of $\boldsymbol{X}_j$. In Figure \ref{figure:convex}, the only untreated units that contribute to the synthetic control are represented by the three vertices of the shaded facet, the one that contains $\boldsymbol X_0\boldsymbol W^*$. 
%More generally, if the columns of $\boldsymbol X_0$ are in general position, and $\boldsymbol X_1$ is outside the convex hull of the columns of $\boldsymbol X_0$, then the number of non-zero synthetic control weights is not greater than $k$, the dimension of $\boldsymbol X_j$. 
If $\boldsymbol X_1$ is in the convex hull of the columns of $\boldsymbol X_0$, then the synthetic control does not need to be unique or sparse. However, sparse solutions with no more than $k+1$ non-zero weights exist by Carath\'{e}odory's theorem. \cite{abadie2021penalized} provide a penalized synthetic control estimator that is unique and sparse even in the case that $\boldsymbol X_1$ is in the convex hull of the columns of $\boldsymbol X_0$. The weight selector in \cite{abadie2021penalized} favors synthetic controls composed by untreated units with values of the predictors in $\boldsymbol X_j$ close to the values of the predictors for the treated unit, $\boldsymbol X_1$. The result is a procedure that not only produces unique and sparse estimates, but also ameliorates potential interpolation biases that could emerge from averaging outcomes between untreated units that are far from the treated unit in the space of the predictors, $\boldsymbol X_j$. Sparsity plays an important role in facilitating the interpretability of synthetic control estimates. The exact nature of the synthetic control estimate, the contribution of each unit in the donor pool to this estimate, and the size and direction of potential biases that may arise if the untreated units contributing to a synthetic control are indirectly affected by the treatment or by other known idiosyncratic shocks to their outcomes, are greatly facilitated by the non-negativity and sparsity of the weights and by the fact that they sum up to one, so they can be interpreted as proper weights.

To study the properties of synthetic control estimators, we will posit a generative model for the outcomes, $Y_{jt}^N$. \cite{AbaDiaHai2010} consider the linear factor model,
\begin{equation}
\label{equation:factor}
Y^N_{jt} = \delta_t + \boldsymbol{\theta}_t \boldsymbol{Z}_j + \boldsymbol{\lambda}_t \boldsymbol{\mu}_j + \epsilon_{jt},
\end{equation}
where $\delta_t$ is a time trend, and $\bs Z_j$ and $\bs\mu_j$ are vectors of observed and unobserved predictors, respectively, with time varying coefficients, $\bs\theta_t$ and $\bs\lambda_t$. We will also use the term ``covariates'' to refer to $\bs Z_j$ and $\bs\mu_j$. The variable $\epsilon_{jt}$ is a transitory shock that we will model as white noise. We will take all the components on the right-hand side of equation \eqref{equation:factor} as given (conditioned on) except for $\epsilon_{jt}$. 

% While equation \eqref{equation:factor} does not directly restrict the process that determines selection for treatment, it imposes a zero mean restriction on $\epsilon_{jt}$ regardless of treatment status, which can be regarded as a form of unconfoundedness conditional on $\bs Z_j$ and $\bs\mu_j$. However, notice that in the contexts where synthetic controls are typically applied in practice---with only one or a small number of treated units---a precise knowledge of the selection mechanism may not buy much in the absence of a model for the outcomes. Consider the usual setting with only one treated units. Even if the treatment is randomized--- which is extremely unlikely in setting with aggregate units, like states or countries--a simple difference in means is ex-ante unbiased (i.e., unbiased before randomization), but be heavily biased ex-post because of large difference between the characteristics of the treated unit and the units that are not exposed to the treatment. 

Equation \eqref{equation:factor} imposes a zero mean restriction on $\epsilon_{jt}$ regardless of treatment status, which can be interpreted as a form of unconfoundedness conditional on $\bs Z_j$ and $\bs\mu_j$. However, equation \eqref{equation:factor} does not directly restrict the process that determines selection for treatment. 
%Instead, it models the effect of observables, $\bs Z_j$, and unobservables, $\bs\mu_j$, on the potential outcome, $Y^N_{jt}$. 
In the contexts where synthetic controls are typically applied in practice---with only one or a small number of treated units---a precise knowledge of the selection mechanism may not be particularly useful in the absence of a model for the outcomes. Consider the usual setting with only one treated unit. Randomization of the treatment--something that is extremely unlikely in settings with aggregate units, like states or countries---implies that a simple difference in means between outcomes of treated and untreated units is ex-ante unbiased for the average treatment effect (that is, unbiased before randomization). However, a simple difference in means could be heavily biased ex-post if randomization produces large differences between the characteristics of the treated unit and the units that are not exposed to the treatment. Equation \eqref{equation:factor} accounts for these differences by modeling the effect of observables, $\bs Z_j$, and unobservables, $\bs\mu_j$, on the potential outcomes, $Y^N_{jt}$. 

Suppose that, with probability one, the solution to the minimization of \eqref{equation:objective} subject to the weight constraints yields
\begin{equation}
\label{equation:fit}
\sum_{j=2}^J W_j^* \bs Z_j = \bs Z_1\quad\mbox{ and }\quad\sum_{j=2}^J W_j^* Y_{jt} = Y_{1t},
\end{equation}
for $t = 1, \ldots, T_0$. Under these conditions, a bound can be established on the magnitude of the bias, $|E[\widehat\tau_t-\tau_t]|$, of the synthetic control estimator. \cite{AbaDiaHai2010} provide a precise expression for this bound. For the purpose of the present article, however, it is enough to know that the bound increases with {\em (i)} the ratio between the scale of the transitory shocks, $\epsilon_{jt}$, and the number of pre-intervention periods, $T_0$, {\em (ii)} the number of units in the donor pool, $J$, and {\em (iii)} the dimension of $\bs\mu_j$ (i.e., the number of unobserved factors). The bias bound and equation \eqref{equation:fit} motivate including $\bs Z_j$ and pre-treatment values of the outcomes in the vector of matching variables, $\bs X_j$. This equation may only hold approximately in practice if $\bs X_1$ is outside the convex hull of the columns of $\bs X_0$.

We next provide intuition for why {\em (i)} to {\em (iii)} may affect the size of the bias. Notice that, under the model in \eqref{equation:factor}, synthetic control weights that reproduce the values of observed and unobserved covariates,
\begin{equation}
\label{equation:fitfactors}
\sum_{j=2}^J W_j^* \bs Z_j = \bs Z_1\quad\mbox{ and }\quad\sum_{j=2}^J W_j^* \bs\mu_j = \bs\mu_1,
\end{equation}
would yield an unbiased estimator of the treatment effect. Although the unobserved covariates, $\bs\mu_j$, cannot be fitted directly, a synthetic control such that \eqref{equation:fit} holds employs pre-treatment outcomes as proxies for the unobserved factors. This would clearly be justified if the scale of the transitory shocks is small, so most of the heterogeneity in pre-treatment outcomes that does not come from $\bs Z_j$ is generated by heterogeneity in $\bs\mu_j$. In that case, fitting pre-treatment outcomes, as in \eqref{equation:fit}, comes very close to fitting the unobserved factors, as in \eqref{equation:fitfactors}. However, a substantial amount of variability in $\epsilon_{jt}$ opens the door to over-fitting: the possibility that pre-treatment outcomes are fitted in \eqref{equation:fit} out of variation in $\epsilon_{jt}$. In that case, the fitted value for the unobserved factors given by the synthetic control,
\begin{equation*}
\sum_{j=2}^{J+1}W^*_j\bs\mu_j,
%\label{equation:sync_mu}
\end{equation*}
may substantially differ from $\bs\mu_1$, inducing bias. The probability that \eqref{equation:fit}, or an approximate version of it, is mainly produced by over-fitting is small when the scale of $\epsilon_{jt}$ is small and the number of pre-treatment periods, $T_0$, to be fitted is large. A large donor pool, however, increases the chances that the pre-treatment outcomes are fitted out of variation in $\epsilon_{jt}$. Moreover, given that the linear model in \eqref{equation:factor} is nothing but a local approximation to a data generating process that could be non-linear, fitting the values of the treated units with untreated units that are far from each other in the space of the predictors may result in sizable interpolation biases. Finally, the dimension of $\bs\mu_j$ reflects the variability induced by unobserved covariates. In the absence of unobserved covariates, a synthetic control that reproduces the value of $\bs Z_1$ only  (as in the first part of \eqref{equation:fitfactors}) would be exactly unbiased.    

From our discussion of the synthetic control design and the bias bound, we can now distill seven guiding principles for empirical practice with synthetic control estimators:
\begin{enumerate}
    \item Low volatility. Closely fitting a highly volatile series is likely the result of over-fitting at work, especially if it happens over a short pre-intervention period. Synthetic controls were designed for settings with aggregate series, where aggregation attenuates the magnitude of the noise.\label{item:noise} 
    \item Extended pre-intervention period. A good control unit must closely reproduce the trajectory of the outcome variable for the treated unit over an extended pre-intervention period. A good fit, if it is the result of a secular agreement between the treated and the synthetic control units in their responses to unobserved factors, should persist in time.\label{item:time} 
    \item Small donor pool. A larger donor pool is not necessarily better than a smaller one. Adopting a small donor pool of untreated units that are close to the treated unit in the space of the predictors helps reduce over-fitting and interpolation biases. \label{item:small}
    \item Sparsity makes synthetic controls interpretable.\label{item:wisdom}
    \item Covariates matter. A component of $\bs Z_i$ that is not controlled for (that is, not included in $\bs X_j$) is effectively thrown into $\bs\mu_j$. Fitting $\bs Z_j$ is easier than fitting $\bs\mu_j$.\label{item:covariates} 
    \item Fit matters. The bound on the bias is predicated on close fit. A deficient fit raises concerns about the validity of a synthetic control \citep[see, however,][for exceptions and qualifications on this rule]{ferman2021properties, ferman2021synthetic}.\label{item:fit} 
    \item Out-of-sample validation is key. The goal of synthetic controls is to predict the trajectory of the outcome variable for the treated unit in the absence of the intervention of interest. The quality of a synthetic control can be assessed by measuring predictive power in pre-intervention periods that are left out of the sample used to calculate the synthetic control weights.\label{item:validation}  
\end{enumerate}

We will study and illustrate the empirical relevance of the seven principles described above. Principles \ref{item:noise}-\ref{item:fit} derive directly from the bias bound in \cite{AbaDiaHai2010}. Because the formal validity of the bias bound rests on strong assumptions, it is useful to investigate the extent to which these principles effectively translate to empirical practice. We will also illustrate the effectiveness of validation techniques derived from the last principle on the list. 

We will employ simulations under a variety of data generating processes to investigate how features of the data affect the performance of synthetic control estimators in ways that can be predicted from the seven guiding principles above. An important take-away of our analysis is that, as previously argued in \cite{abadie2021using} and others, contextual and data requirements are key for the performance of synthetic control estimators, and should be carefully checked in empirical applications. Mechanistic applications of the method, without regard to the guiding principles described above, leave the results of the empirical exercise vulnerable to biases created by over-fitting and by discrepancies between the values of the predictors for the treated units and for the units that contribute to the synthetic control. When conscientiously applied, however, synthetic controls become powerful and transparent design tools for estimating the effects of aggregate interventions.

\section{Performance of the Synthetic Control Estimator}
\label{section:performance}

To begin our investigation of the performance of the synthetic control estimator under a variety of features in the data, we adopt the grouped factor model of \cite{ferman2021properties}, which is a special case of \eqref{equation:factor}. We will use a generative model with $F$ equally-sized groups of units denoted $f(i) \in \{1, \dots, F\}$ and $F$ common factors, with each unit $i$ loading exclusively on factor $f(i)$,
\begin{align}
    Y_{it}^N = \delta_t + \lambda_{f(i)t} + \epsilon_{it}.
    \label{equation:group_factor}
\end{align}
The common factors, $\lambda_{ft}$, follow an $AR(1)$ processes with autoregressive coefficient $\rho$ and standard Gaussian innovations. $\epsilon_{it}$ follows a Gaussian distribution, with mean zero and variance $\sigma^2$, independent of any other component. 

We assume that only the first unit is treated and that only the second unit loads on the same factor as the treated unit, so $f(1)=f(2)$. The other untreated units---that is, units $j=3, \ldots, J+1$---are also grouped into pairs that load on a pair-specific factor, $f(3)=f(4)$, $f(5)=f(6)$, and so on. An unbiased synthetic control in this setting is one with $W^*_2 = 1$ and $W_3^* = W^*_4 = \cdots = W^*_{J+1}=0$. This paired design represents the least favorable case for synthetic control estimators under the grouped factor model of \cite{ferman2021properties}: there is only one unit that reproduces the trajectory of $\lambda_{f(i)t}$ for the treated unit in the absence of the intervention. 

Unless we state otherwise, we will impose $Y^I_{1t}=Y^N_{1t}$ for $t=T_0+1, \ldots, T$, so the treatment effect on the treated unit is equal to zero. This allows us to interpret deviations from zero in treatment effect estimates as reflective of lack of accuracy or precision.  
We fix $T=30$ and study the performance of the synthetic control estimator (and the extent to which the estimated weights approximate $W^*_2=1$) under a variety of values for $\sigma$, $T_0$, $\rho$ and $F$. We will also investigate the extent to which variations on the basic estimator improve performance or diminish it. 

\begin{figure}[ht!]
\centering
  \begin{subfigure}[b]{0.5\textwidth}
  \centering
  	\includegraphics[width=1\linewidth]{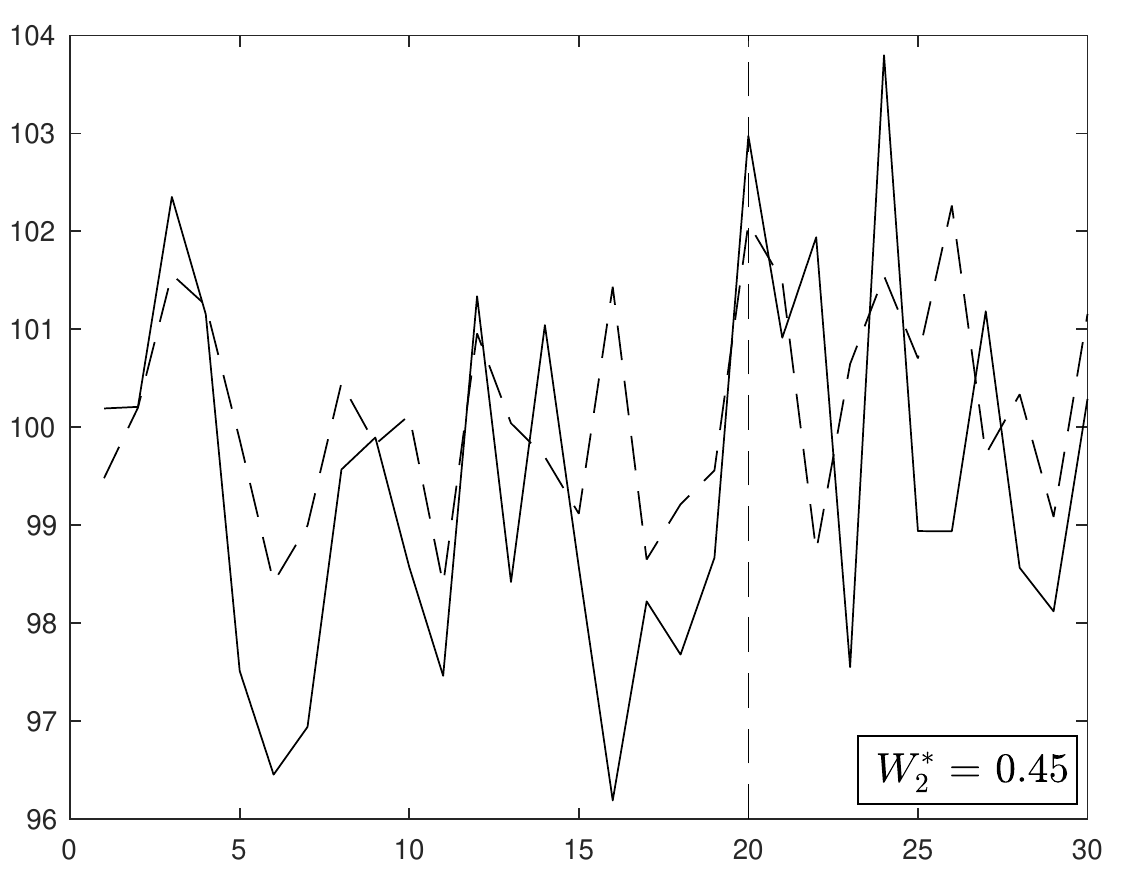}
  	\caption{$\sigma = 2$}
  \end{subfigure}%
  \begin{subfigure}[b]{0.5\textwidth}
  \centering
  	\includegraphics[width=1\linewidth]{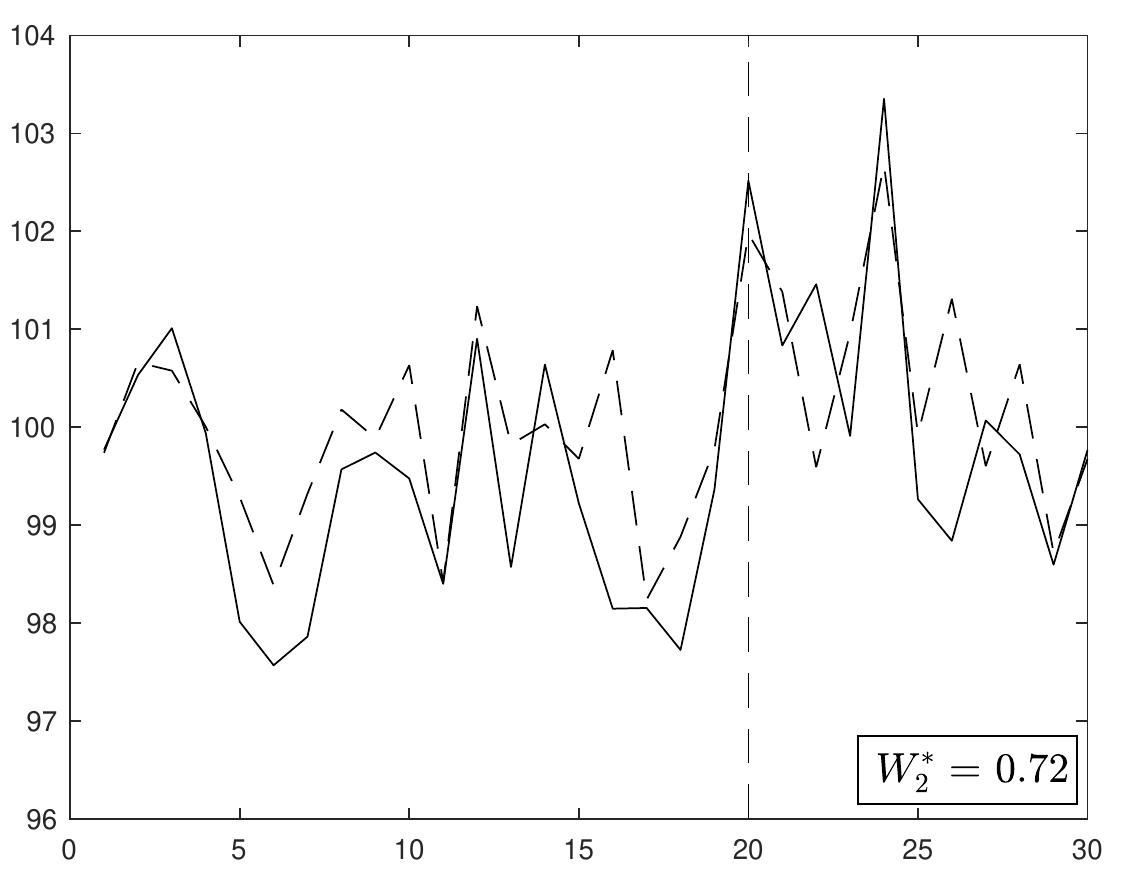}
  	\caption{$\sigma = 1$}
  \end{subfigure}
  
   \begin{subfigure}[b]{0.5\textwidth}
  \centering
  	\includegraphics[width=1\linewidth]{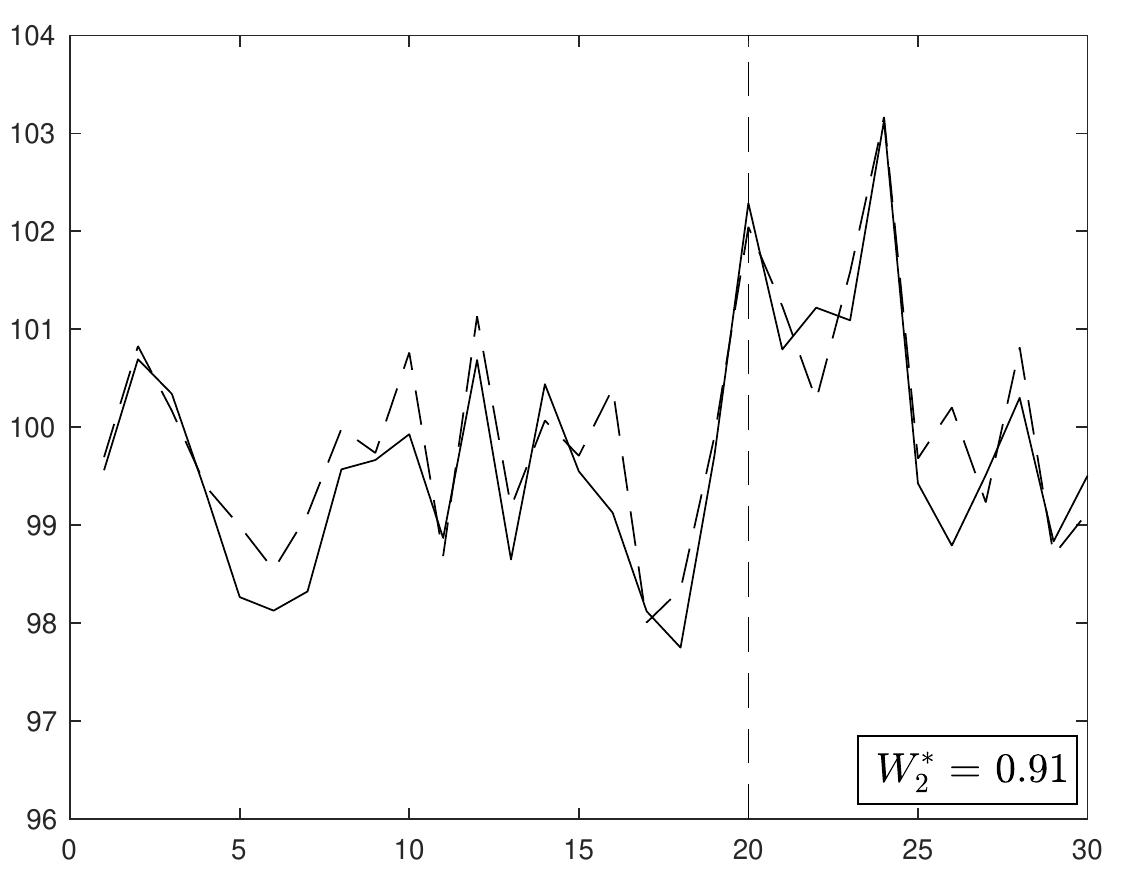}
  	\caption{$\sigma = 0.5$}
  \end{subfigure}%
  \begin{subfigure}[b]{0.5\textwidth}
  \centering
  	\includegraphics[width=1\linewidth]{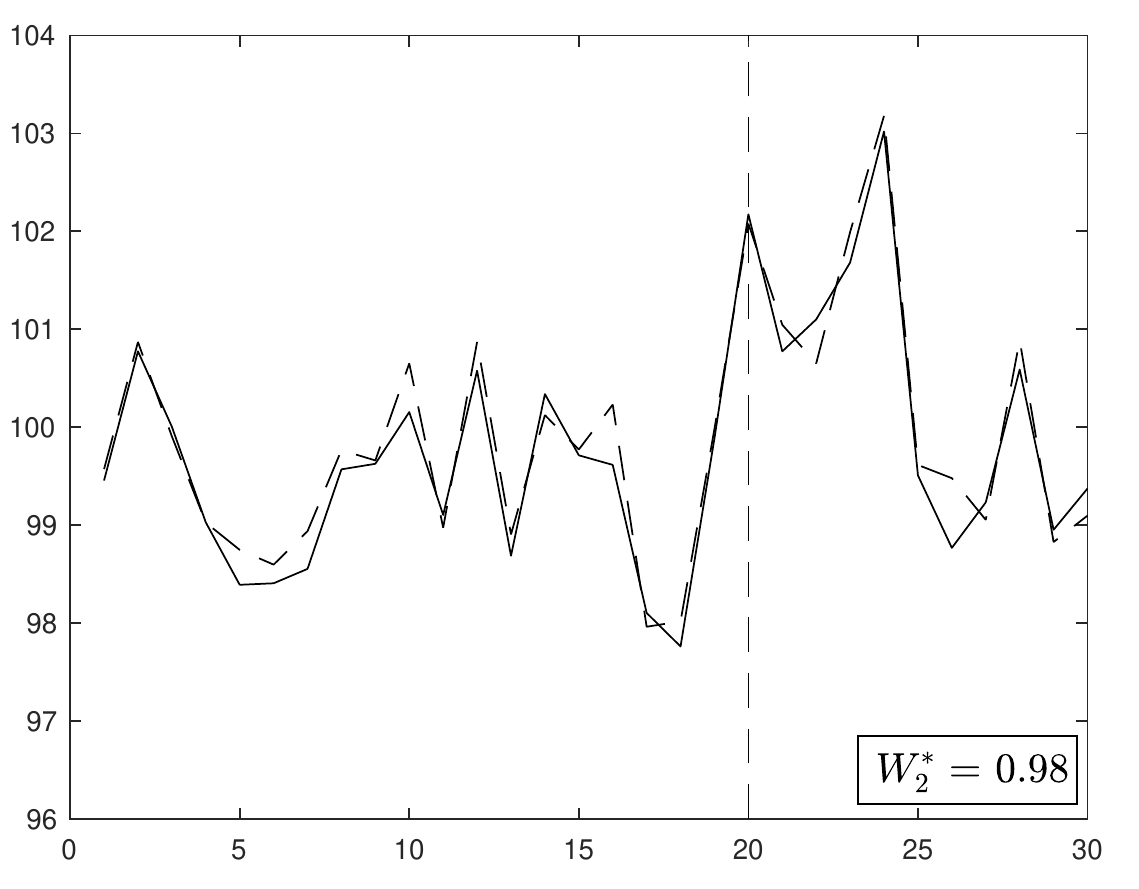}
  	\caption{$\sigma = 0.25$}
  \end{subfigure}
  \caption{Pre-treatment fit and estimation error.}
\vspace{-8mm}
\center
\includegraphics[width=0.2\linewidth]{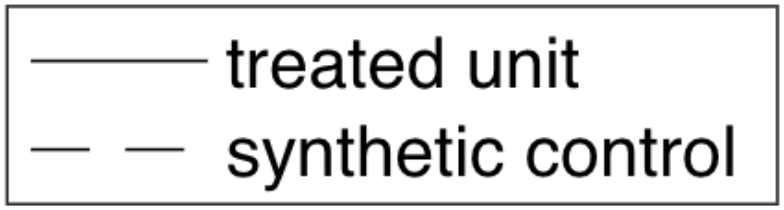}  
\label{figure:noise}
\floatfoot{{\em Note:} Synthetic control estimates for the grouped factor model in equation \eqref{equation:factor}, with   $\rho=0.5$, $T_0=20$, and $J+1=20$ ($F=10$). Each panel reports results for a different value of $\sigma$.}

\end{figure}

Figure \ref{figure:noise} reports simulations of the synthetic control estimator for the case $T_0=20$, $\rho=0.5$, and $J=19$ (so $F=10$). Each panel contains results for a different value for $\sigma$. With a large value of $\sigma$ in panel (a), the pre-treatment fit is poor, which translates in large estimation errors during the post-treatment periods. As $\sigma$ decreases in the subsequent panels, the pre-treatment fit improves noticeably and it translates into similar precision gains during the post-intervention periods. These gains in estimation accuracy are reflected in substantial increases in the value of $W^*_2$ as $\sigma$ decreases. With $\sigma=2$ in panel (a), the synthetic control assigns less than half of the weight the the ``correct'' unit, $j=2$. When $\sigma=0.25$ in panel (d), the weight of unit $j=2$ on the synthetic control estimator is almost one. The pattern of results in Figure \ref{figure:noise} is reflective of the fact that the bias of the synthetic control estimator and the quality of the pre-treatment fit both depend on the scale of the individual transitory shocks, $\epsilon_{jt}$. Under the generative process in \eqref{equation:factor}, lack of pre-treatment fit can arise because of noise in the series (large $\sigma$) or because the values of $\bs Z_1$ and $\bs\mu_1$ cannot be closely reproduced with a convex combination of the values of $\bs Z_j$ and $\bs\mu_j$ for the units in the donor pool. In both cases, post-treatment synthetic control estimates could incorporate sizable biases. The results in Figure \ref{figure:noise} underscore the importance of a good pre-treatment fit. 

\begin{figure}[ht!]
\centering
  \begin{subfigure}[b]{0.5\textwidth}
  \centering
  	\includegraphics[width=1\linewidth]{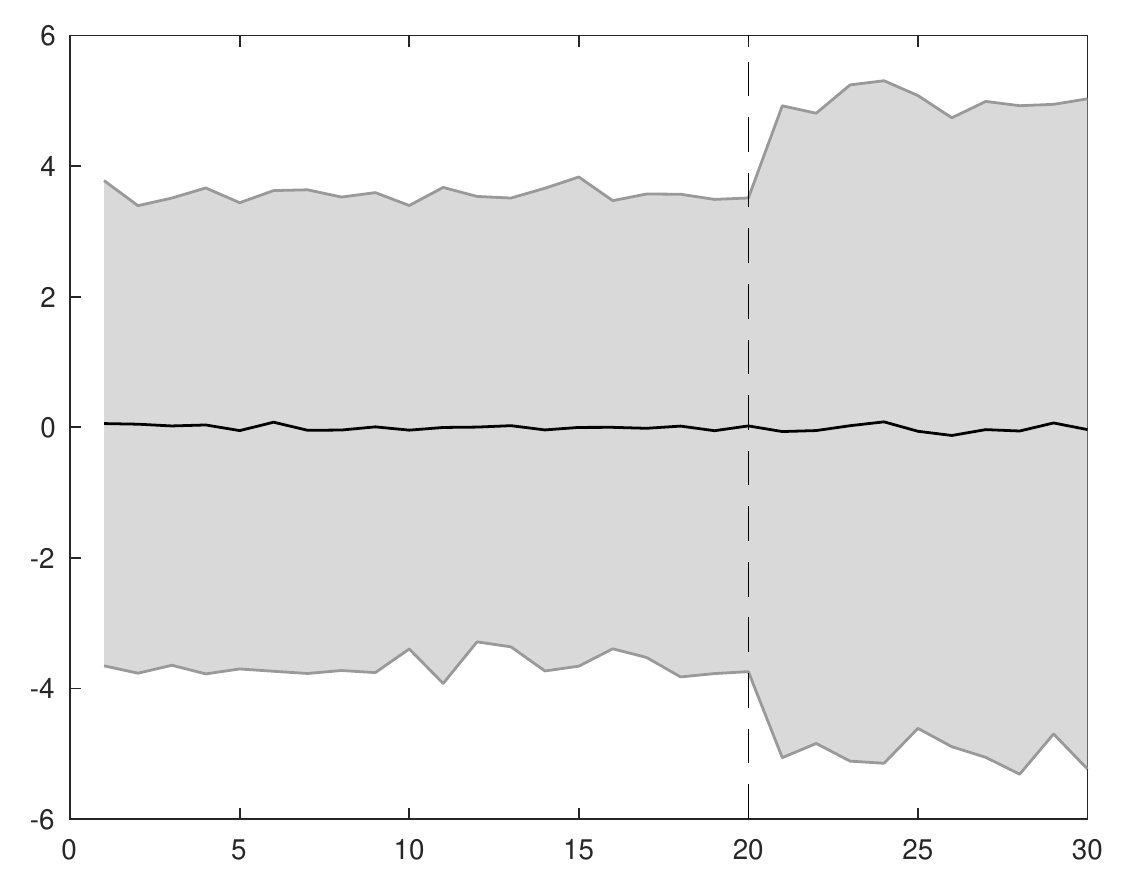}
  	\caption{$\sigma = 2$}
  \end{subfigure}%
  \begin{subfigure}[b]{0.5\textwidth}
  \centering
  	\includegraphics[width=1\linewidth]{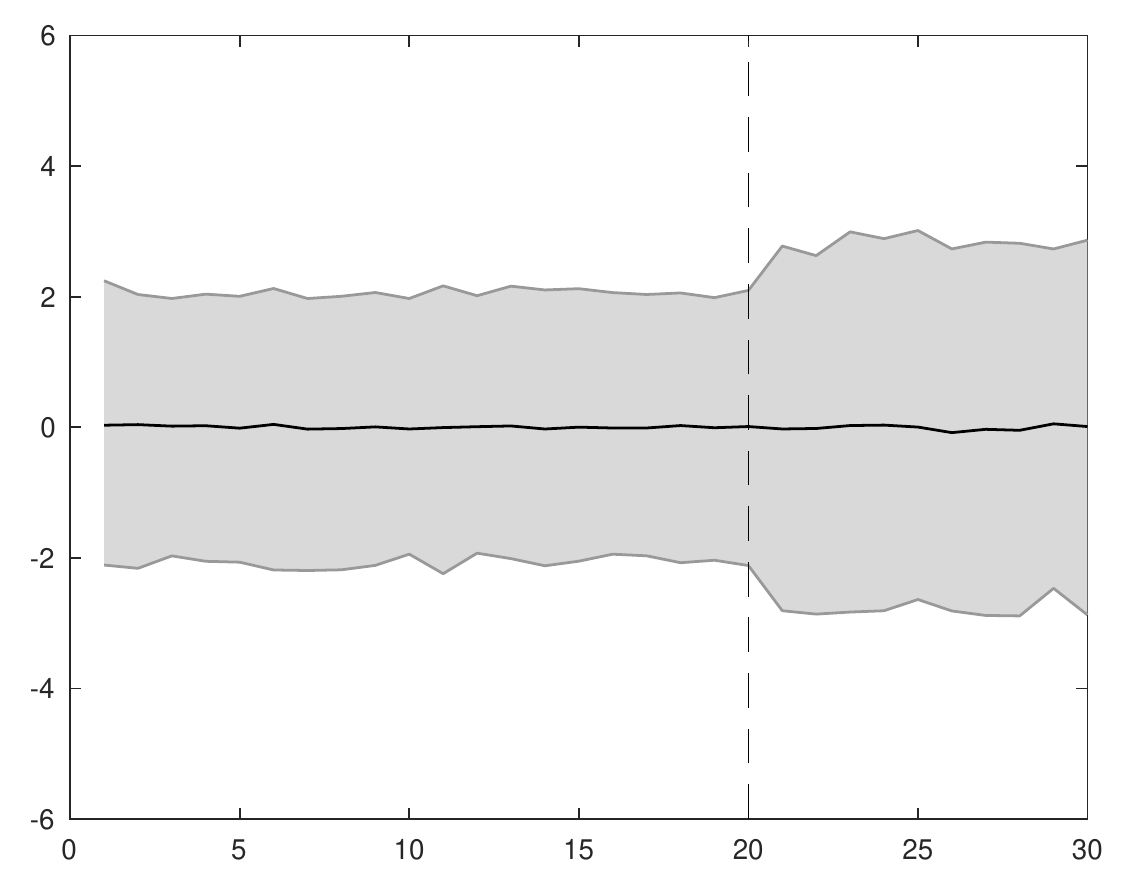}
  	\caption{$\sigma = 1$}
  \end{subfigure}%
  
   \begin{subfigure}[b]{0.5\textwidth}
  \centering
  	\includegraphics[width=1\linewidth]{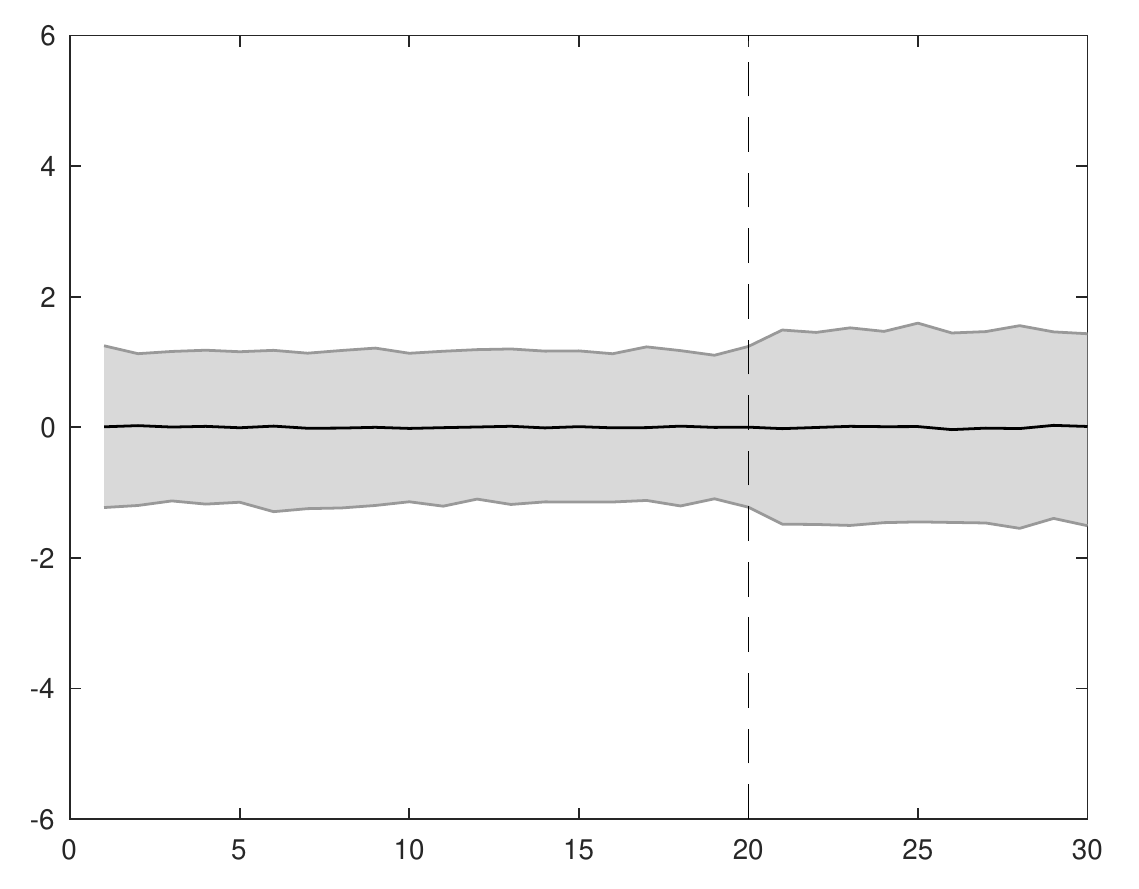}
  	\caption{$\sigma = 0.5$}
  \end{subfigure}%
  \begin{subfigure}[b]{0.5\textwidth}
  \centering
  	\includegraphics[width=1\linewidth]{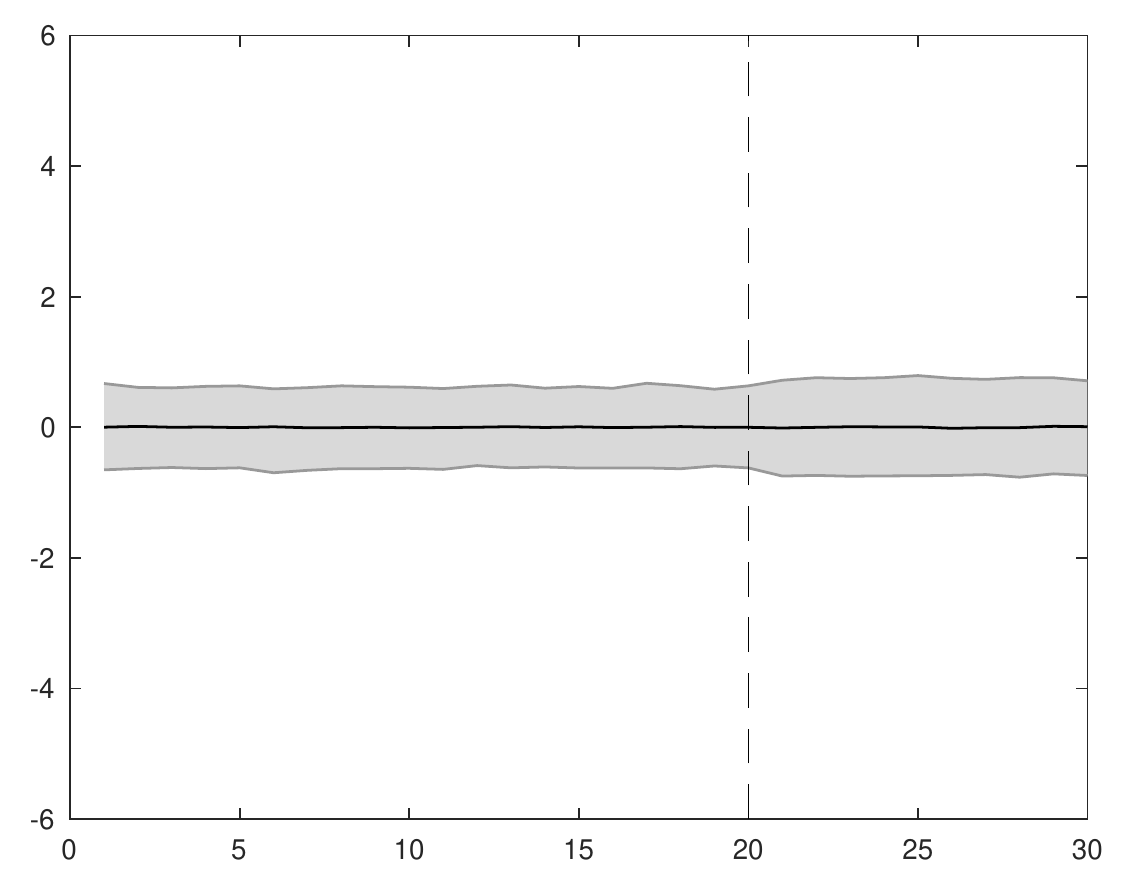}
  	\caption{$\sigma = 0.25$}
  \end{subfigure}
  \caption{Prediction error over 10000 simulations}
  \label{figure:noise_sim}
  \floatfoot{{\em Note:} 95\% bands for the simulation design of Figure \ref{figure:noise}.}
\end{figure}

In order to visually demonstrate the performance of synthetic controls ``in action'', Figure \ref{figure:noise} reports the results for one random realization in our simulation design only. One could, instead, run a large number of simulations and plot or tabulate results on pre-treatment fit and post-treatment bias. Figure \ref{figure:noise_sim} plots results for 10000 simulations, with the same simulation design as in Figure \ref{figure:noise}. Like Figure \ref{figure:noise}, the new Figure \ref{figure:noise_sim} indicates substantial increases in estimation accuracy that are associated with pre-intervention fit. For the rest of this article, we report results of single random simulations in the main text of the article. In the appendix we report the analogous results calculated over a large number of simulations.    

\begin{figure}[ht!]
\centering
  \begin{subfigure}[b]{0.5\textwidth}
  \centering
  	\includegraphics[width=1\linewidth]{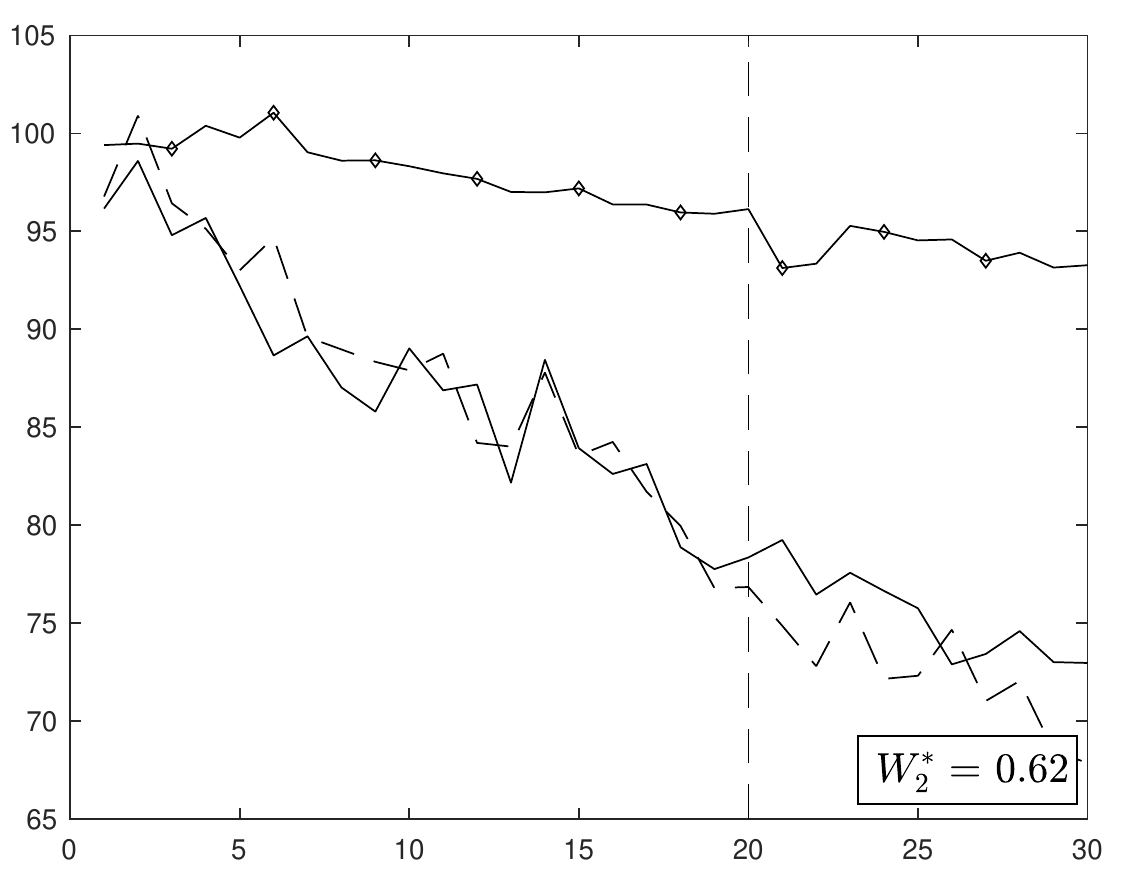}
  	\caption{$\sigma = 2$}
  \end{subfigure}%
  \begin{subfigure}[b]{0.5\textwidth}
  \centering
  	\includegraphics[width=1\linewidth]{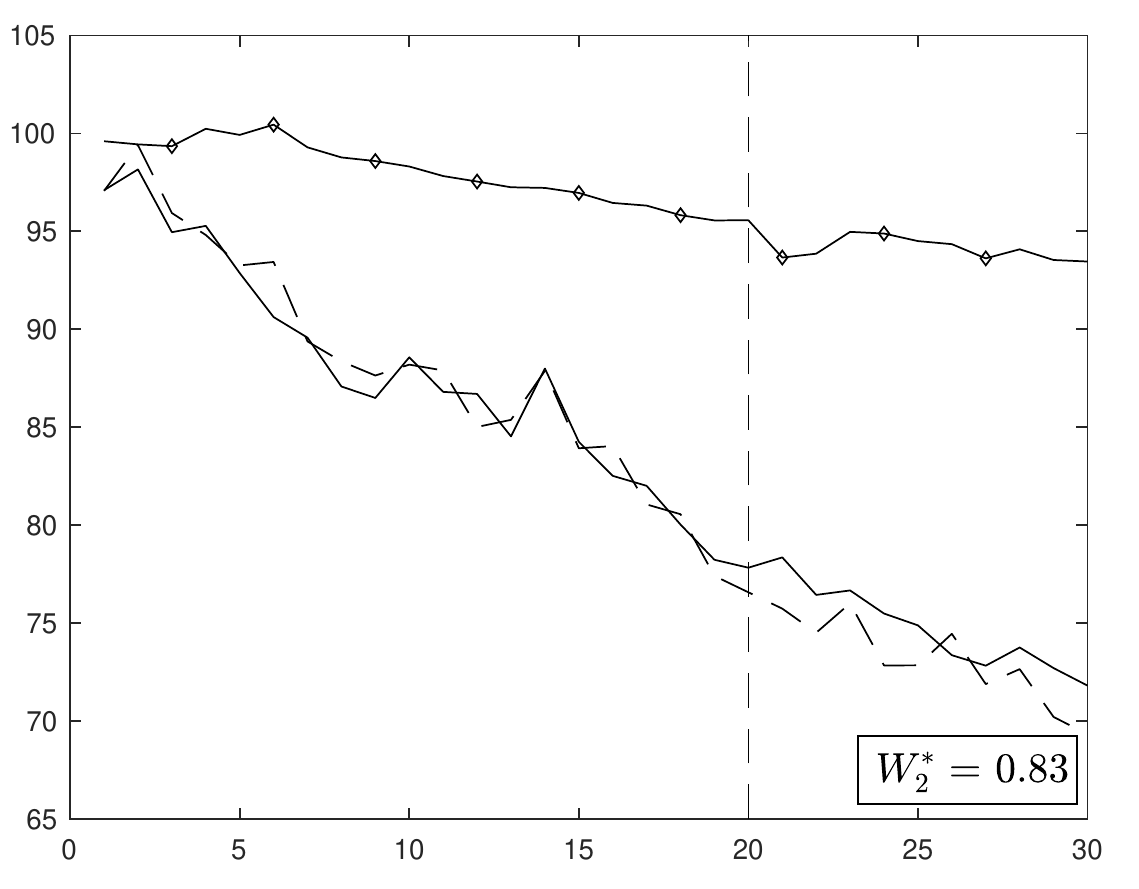}
  	\caption{$\sigma = 1$}
  \end{subfigure}%
  
   \begin{subfigure}[b]{0.5\textwidth}
  \centering
  	\includegraphics[width=1\linewidth]{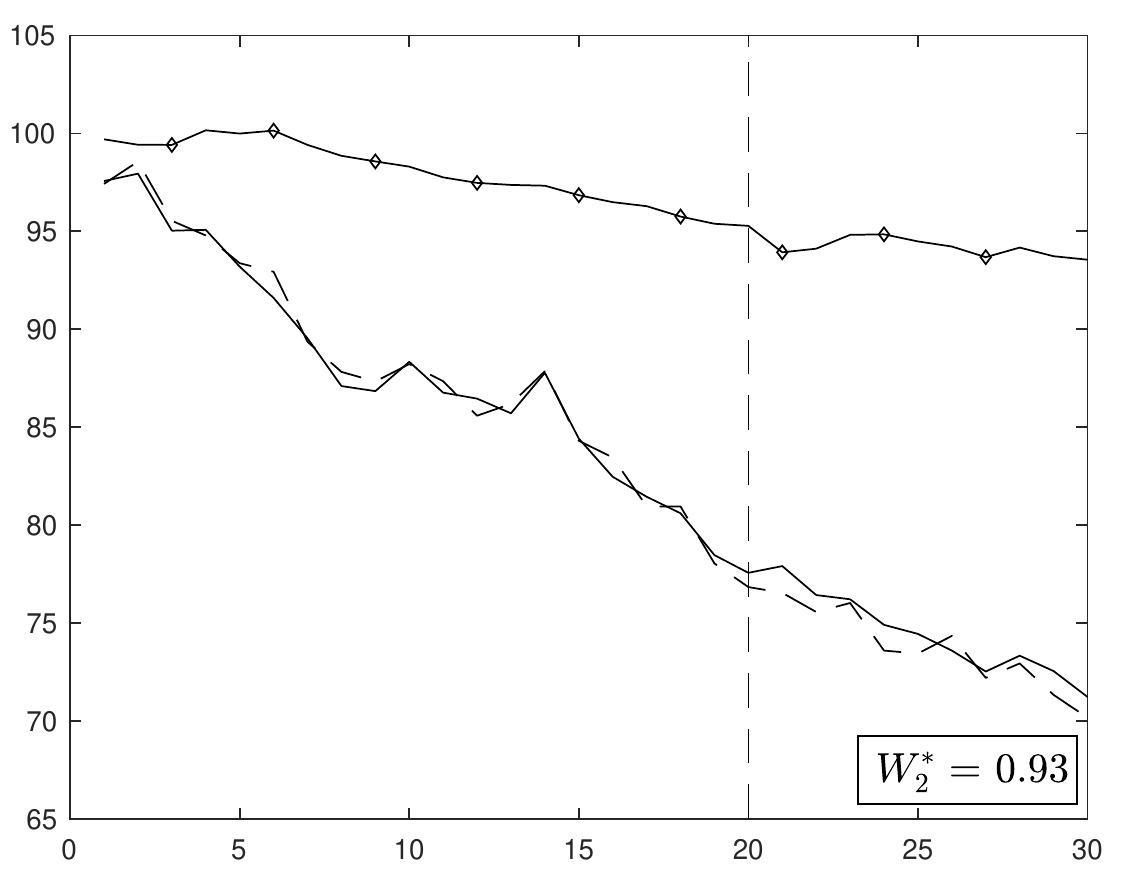}
  	\caption{$\sigma = 0.5$}
  \end{subfigure}%
  \begin{subfigure}[b]{0.5\textwidth}
  \centering
  	\includegraphics[width=1\linewidth]{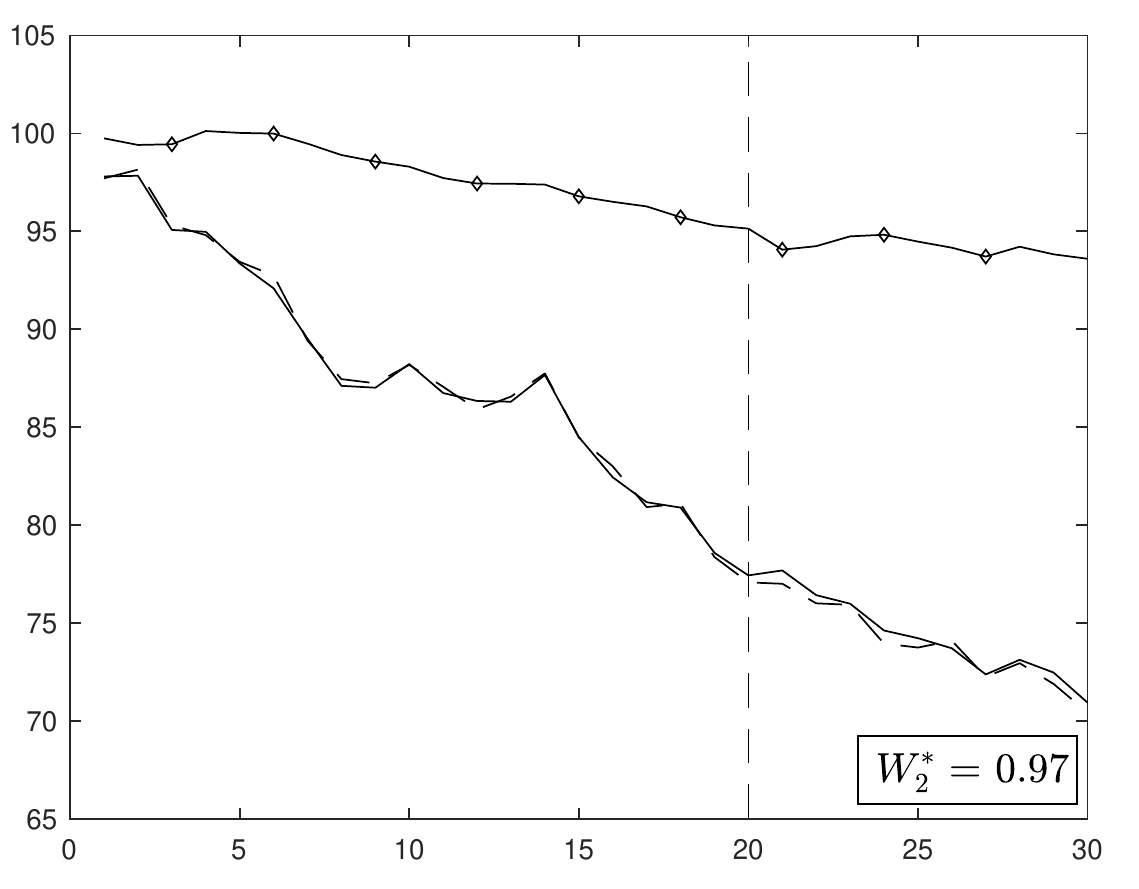}
  	\caption{$\sigma = 0.25$}
  \end{subfigure}%
  \caption{Pre-treatment fit and estimation error with a stochastic trend}
    \center
    \vspace{-8mm}
    \includegraphics[width=0.2\linewidth]{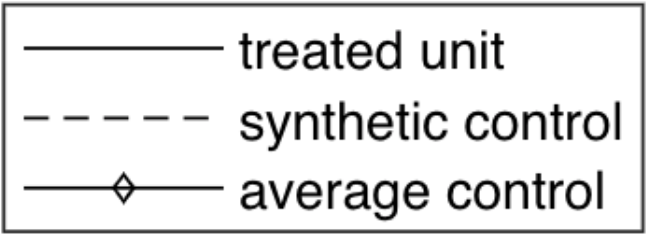}  
  \label{figure:noise_st}
  \floatfoot{{\em Note:} Synthetic control estimates for the grouped factor model in equation \eqref{equation:factor}, with   $\rho=0.5$, $T_0=20$, and $J+1=20$ ($F=10$). Relative to Figure \ref{figure:noise}, the series $\lambda_{ft}$ incorporate stochastic trends. The average control assigns equal weights to all the untreated units.}
\end{figure}

Figure \ref{figure:noise_st} shows the results of a simulation that employs the same data generating process as in Figure \ref{figure:noise} with the exception that now the series $\lambda_{ft}$ incorporate stochastic trends. As for Figure \ref{figure:noise}, the magnitude of the noise in $\epsilon_{jt}$ is key for pre-treatment fit and post-treatment prediction error. Notice, however, that heterogeneity in the trending behavior of the series helps with the selection of a synthetic control heavily based on the outcome series of unit 2, which is affected by the same stochastic trend as the treated unit. In addition to the outcome series for the treated unit and the synthetic controls, Figure \ref{figure:noise_st} also reports the average outcome path for all untreated units.
Figure \ref{figure:noise_st} shows large gains from synthetic control estimation, relative to a simple average of the outcome for the untreated units. These gains are created by heterogeneity across units in the outcome trends, something to which the synthetic control method is able to adapt.

\begin{figure}[ht!]
\centering
  \begin{subfigure}[b]{0.5\textwidth}
  \centering
  	\includegraphics[width=1\linewidth]{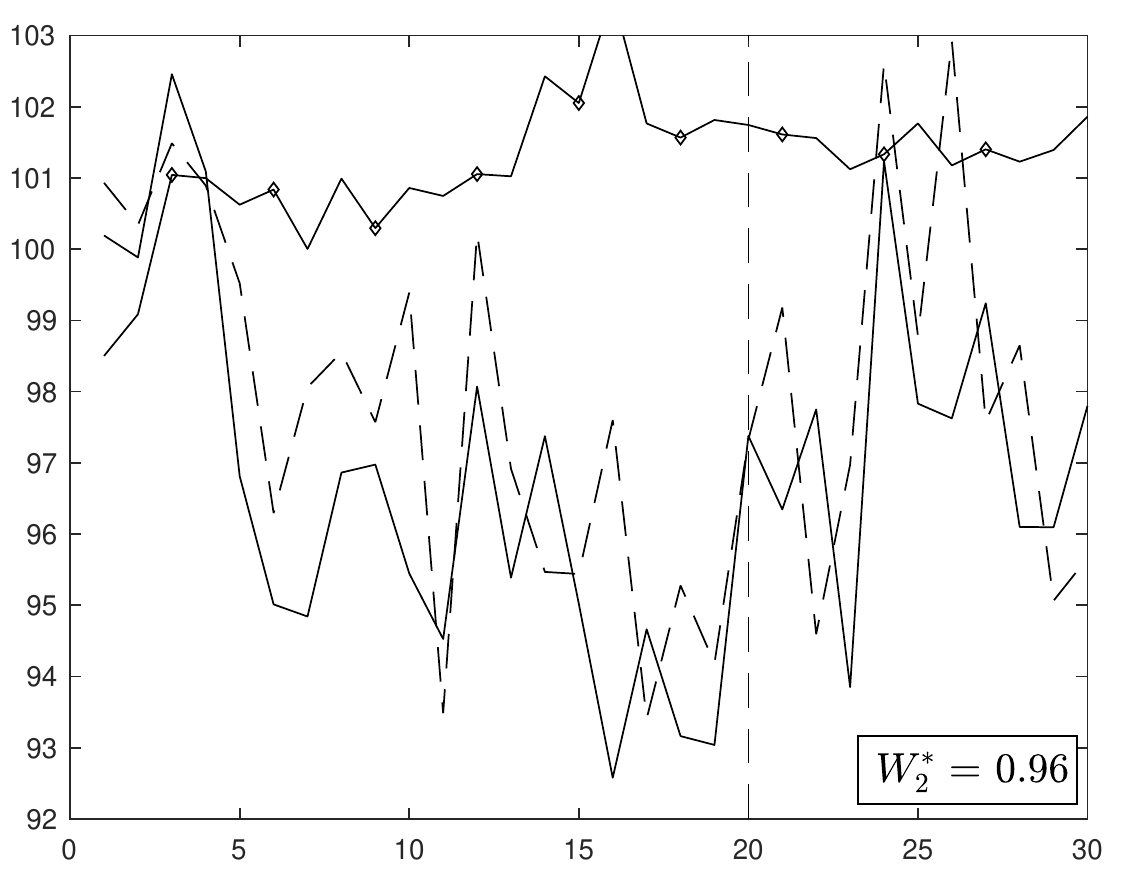}
  	\caption{$\sigma = 2$}
  \end{subfigure}%
  \begin{subfigure}[b]{0.5\textwidth}
  \centering
  	\includegraphics[width=1\linewidth]{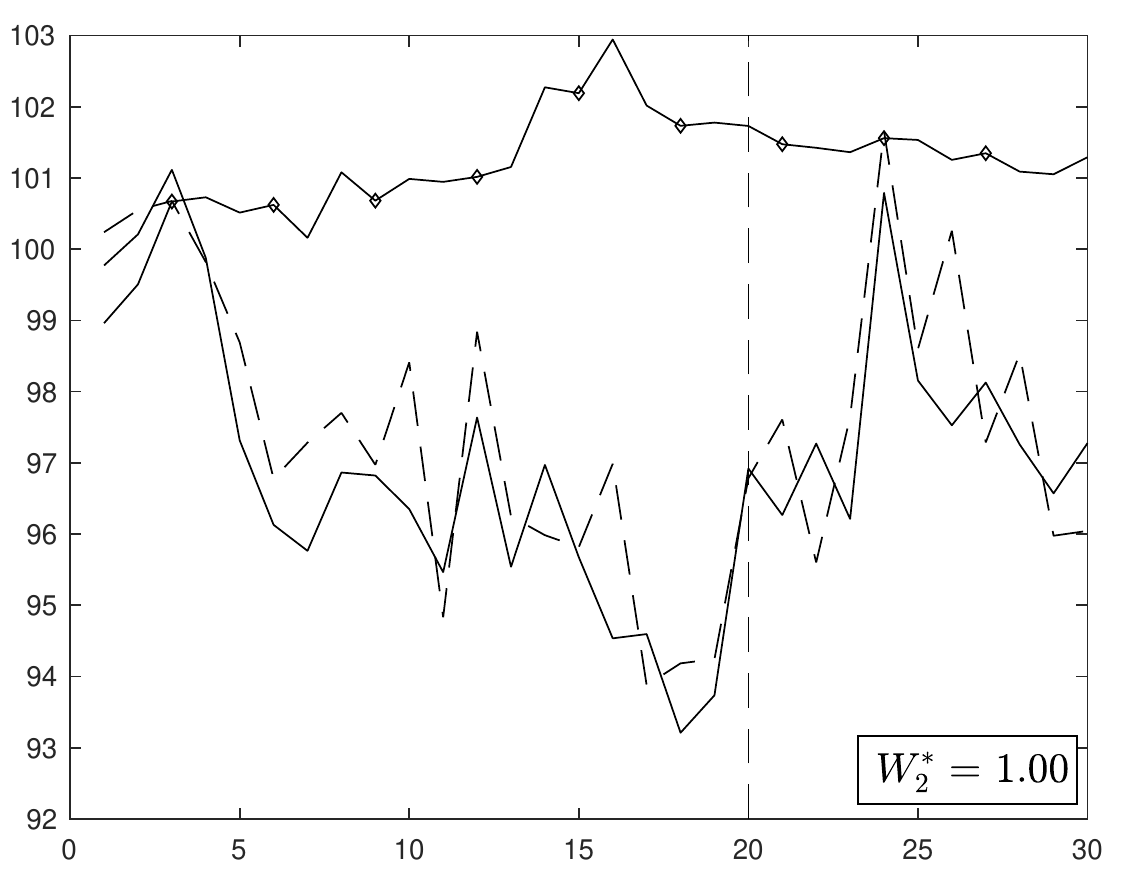}
  	\caption{$\sigma = 1$}
  \end{subfigure}%
  
   \begin{subfigure}[b]{0.5\textwidth}
  \centering
  	\includegraphics[width=1\linewidth]{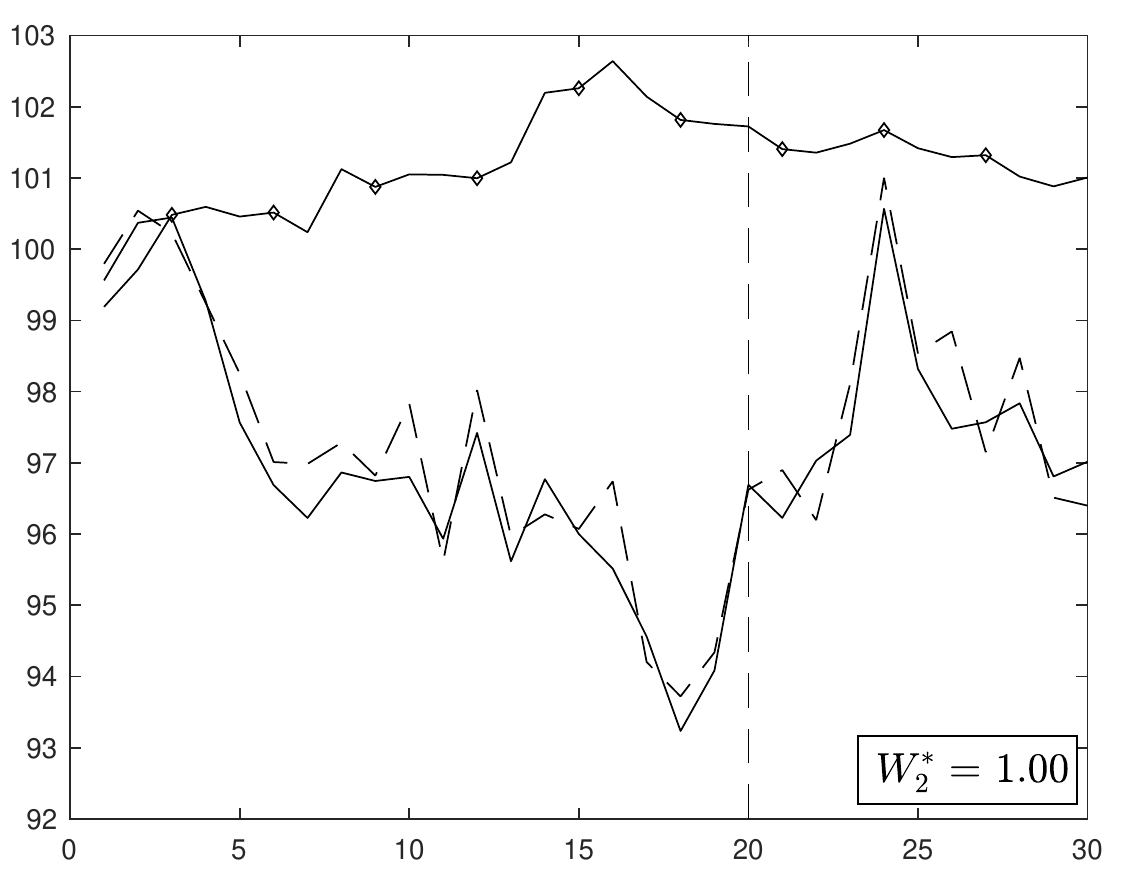}
  	\caption{$\sigma = 0.5$}
  \end{subfigure}%
  \begin{subfigure}[b]{0.5\textwidth}
  \centering
  	\includegraphics[width=1\linewidth]{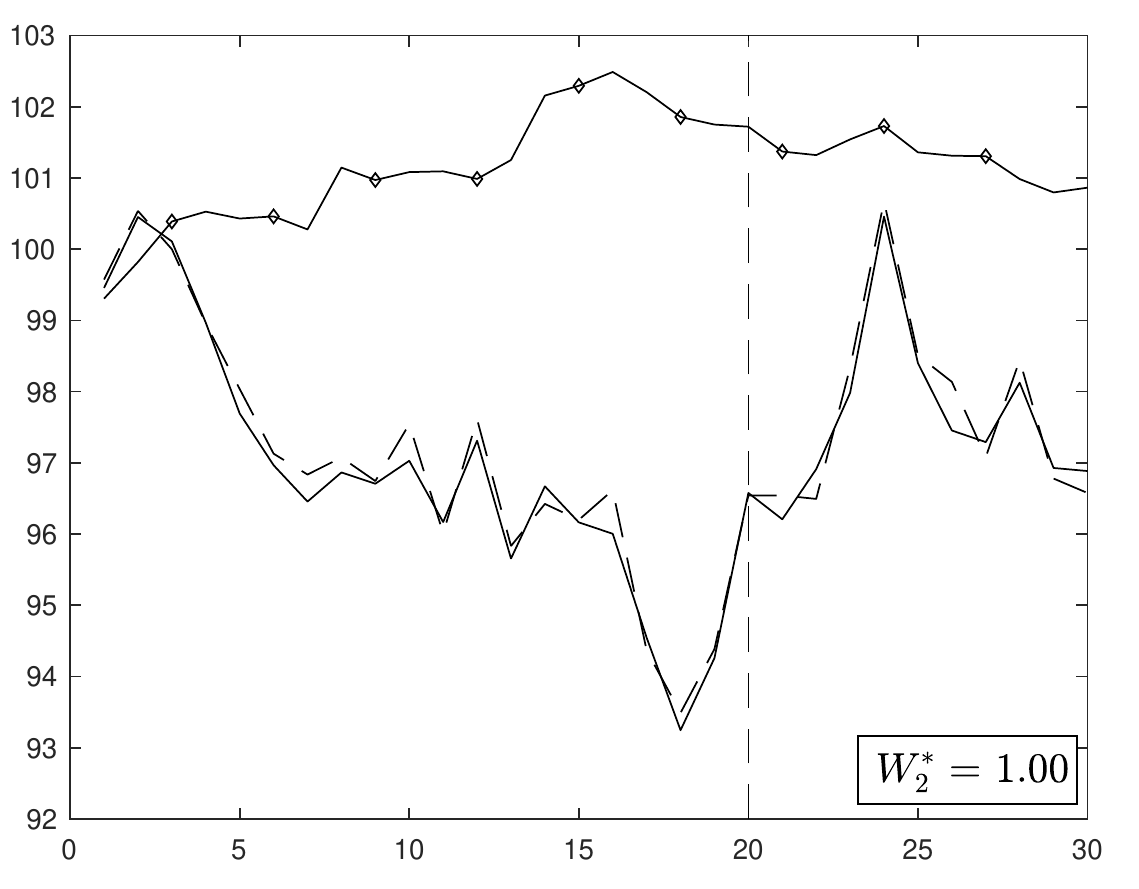}
  	\caption{$\sigma = 0.25$}
  \end{subfigure}%
  \caption{Pre-treatment fit and estimation error with $\rho = 1$}
    \center
    \vspace{-8mm}
    \includegraphics[width=0.2\linewidth]{Tables_and_Figures/legend2.png}  
  \label{figure:noise_rho1}
  \floatfoot{{\em Note:} Synthetic control estimates for the grouped factor model in equation \eqref{equation:factor}, with   $\rho=1$, $T_0=20$, and $J+1=20$ ($F=10$). Each panel reports results for a different value of $\sigma$. The average control assigns equal weights to all the untreated units.}
\end{figure}

The result of Figure \ref{figure:noise_st} is rather general. In a model like \eqref{equation:factor}, non-stationary factor components help identify synthetic controls that reproduce the values of the $\bs\mu_j$ for the treated. This is again the case in Figure \ref{figure:noise_rho1}, which repeats the analysis of Figure \ref{figure:noise} but with $\rho=1$, so the series $\lambda_{ft}$ are non-stationary. As in Figure \ref{figure:noise_st}, the large variation in factor components cannot be compensated by variation in $\epsilon_{jt}$. This results in minimal over-fitting in Figure \ref{figure:noise_rho1}, with $W^*_2$ close to one for all values of $\sigma$ in the figure. 

Obtaining a good pre-treatment fit in the outcome variable (but also in other predictors of the outcome, as we will discuss later) is an important requisite for the performance of synthetic control estimators. However, good pre-treatment fit is not in itself a guarantee of good performance. The reason is that, in some scenarios, a good pre-treatment fit may be obtained from variation in the individual transitory shocks, $\epsilon_{jt}$, even when the selected synthetic control does not come close to reproducing the values of $\bs\mu_j$ in equation \eqref{equation:factor} for the treated. This is what we have previously referred to as over-fitting. 

A researcher employing synthetic controls should be able to identify settings conductive of over-fitting, and to modify the design of a study in order to avoid or attenuate over-fitting biases. We devote much of the rest of the article to discussing prevention, detection, and correction of over-fitted estimators. We discuss the scenarios where over-fitting is likely to bias synthetic control estimators and how to avoid those biases. 

\begin{figure}[ht!]
\centering
  \begin{subfigure}[b]{0.5\textwidth}
  \centering
  	\includegraphics[width=1\linewidth]{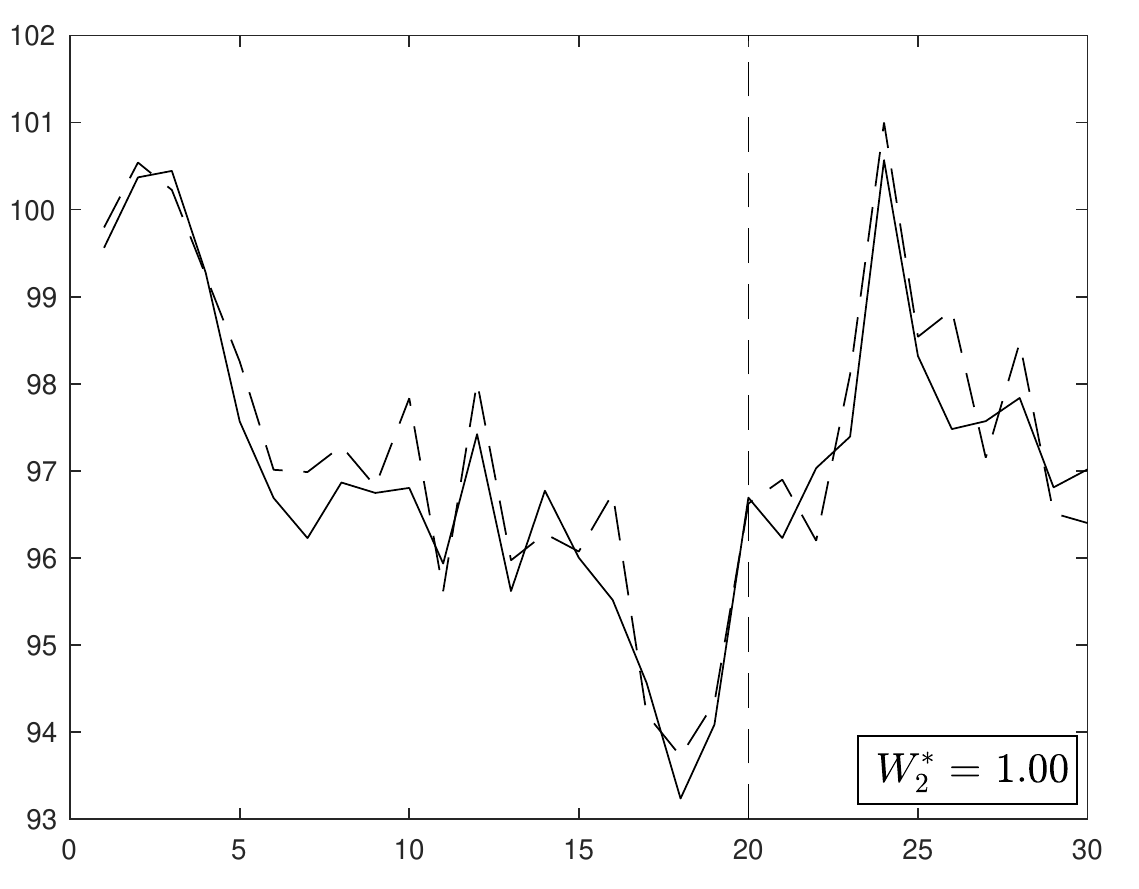}
  	\caption{$T_0 = 20$, $F = 10$ $(J = 19)$}
  \end{subfigure}%
  \begin{subfigure}[b]{0.5\textwidth}
  \centering
  	\includegraphics[width=1\linewidth]{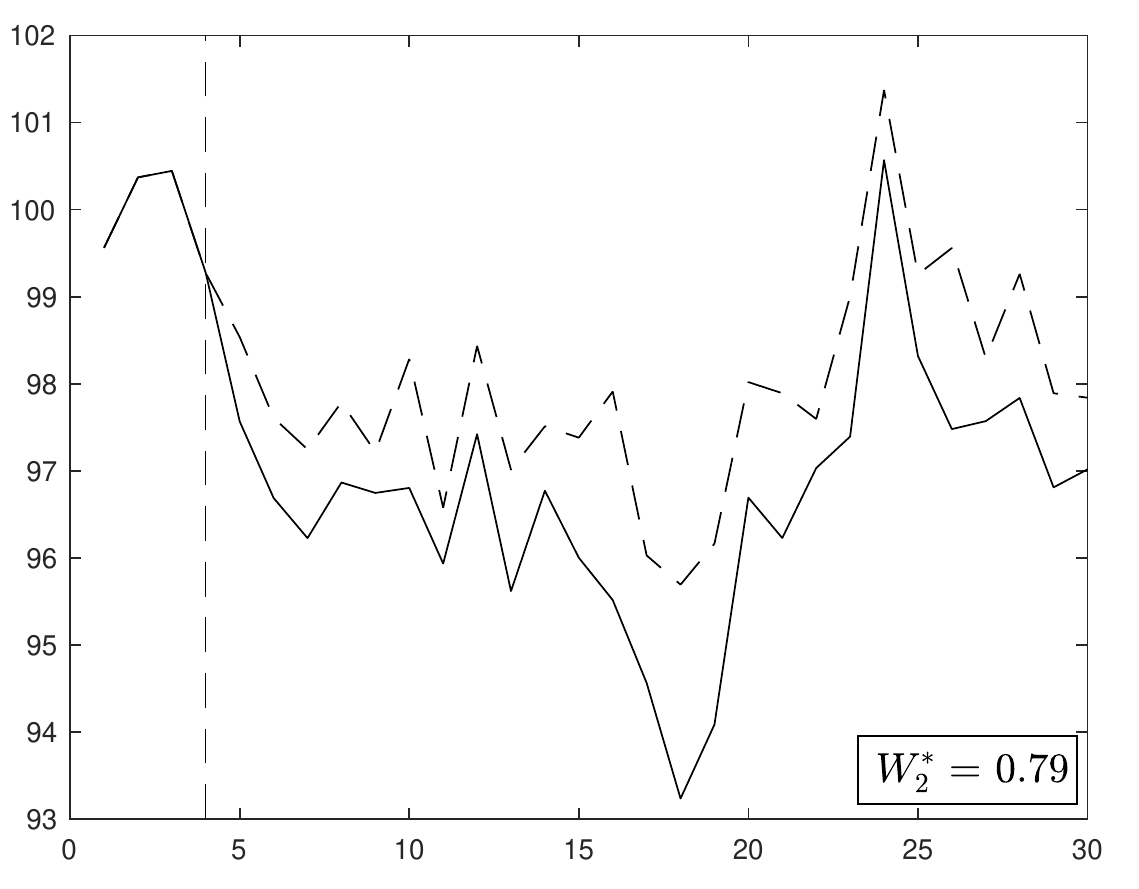}
  	\caption{$T_0 = 4$, $F = 10$ $(J = 19)$}
  \end{subfigure}%
  
   \begin{subfigure}[b]{0.5\textwidth}
  \centering
  	\includegraphics[width=1\linewidth]{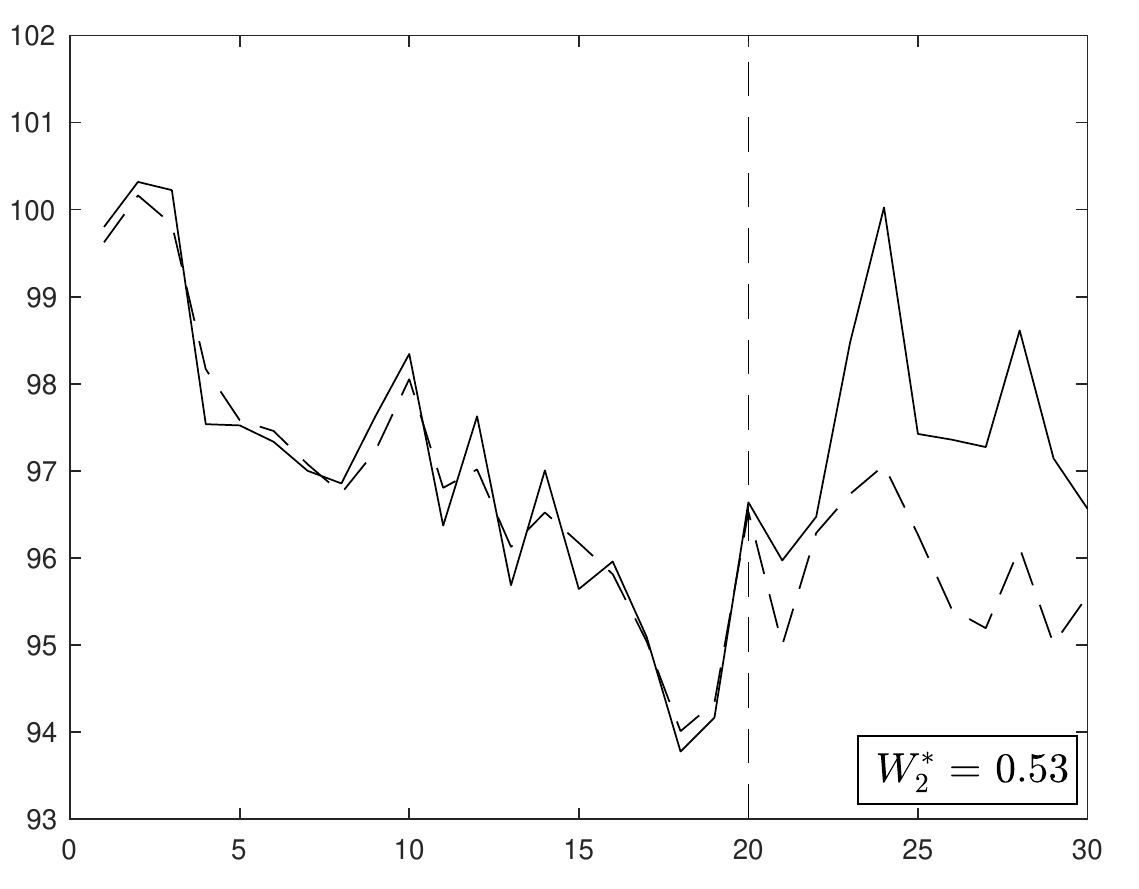}
  	\caption{$T_0 = 20$, $F = 100$ $(J = 199)$}
  \end{subfigure}%
  \begin{subfigure}[b]{0.5\textwidth}
  \centering
  	\includegraphics[width=1\linewidth]{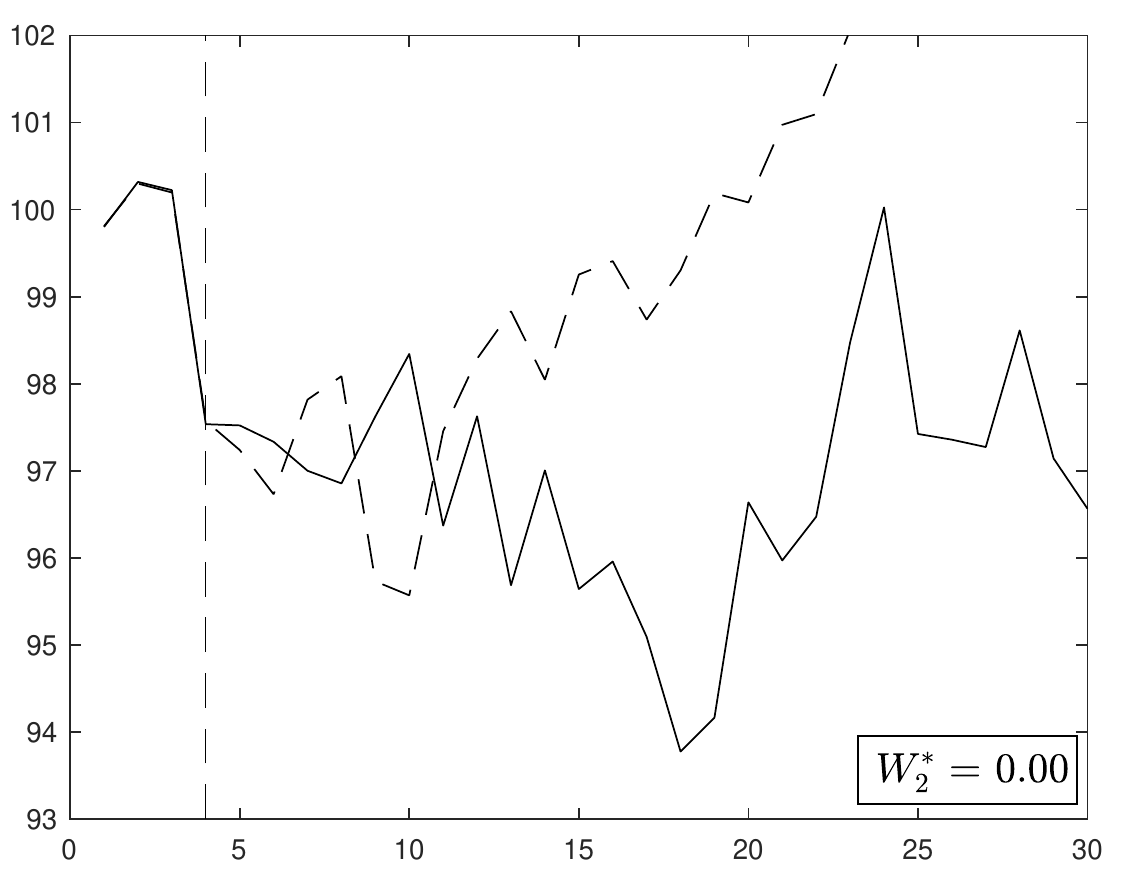}
  	\caption{$T_0 = 4$, $F = 100$ $(J = 199)$}
  \end{subfigure}%
  \caption{Over-fitting with a short pre-treatment period and many donor units}
\vspace{-4mm}
\center
\includegraphics[width=0.2\linewidth]{Tables_and_Figures/legend1.png}  
\label{figure:overfitting}
\floatfoot{{\em Note:} Synthetic control estimates for different values of $T_0$ and $J$, with $\rho = 1$ and $\sigma = 0.5$.}
\end{figure}

Figure \ref{figure:overfitting} illustrates the effect of small $T_0$ and large $J$ on over-fitting. Panel (a) of  Figure \ref{figure:overfitting} reproduces panel (c) in Figure \ref{figure:noise_rho1}, so $\rho=1$ and $\sigma=0.5$. In panel (a), the number of pre-treatment periods is $T_0=20$. Heterogeneity in the processes that govern the trajectories of the factors, $\lambda_{ft}$, identifies the ideal synthetic control with $W^*_2=1$. That is, panel (a) depicts a setting with no over-fitting or bias. Variation in the transitory shocks, $\epsilon_{jt}$, is equally well represented in the pre-treatment and post-treatment differences between the outcomes for the treated unit and the outcomes for the synthetic control. As we will see next, however, a small number of pre-intervention periods, $T_0$, or a larger number of donor pool units, $J$, may create over-fitting, breaking the close correspondence between pre-treatment fit and post-treatment error. Consider now panel (b), where the design of the simulation is the same as in panel (a) but with only $T_0=4$ pre-intervention periods. The fit in pre-intervention outcomes is now perfect. However, the small number of pre-intervention periods induces over-fitting, which creates large post-treatment estimation errors and a reduced role of the second sample unit in the synthetic control, with $W^*=0.79$. Panel (c) goes back to $T_0=20$, as in panel (a), but increases the number of units in the donor pool to $J=199$ (that is, $F=100$). Similar to the effect of a small number of pre-treatment periods in panel (b), the increased number of units in the donor pool opens the door to over-fitting in panel (c). The post-treatment estimation error in panel (c) is much larger than in panel (a), even when the pre-treatment fit is better in panel (c). Moreover, the role of the second unit in the synthetic control is reduced to $W^*_2=0.53$. Finally, panel (d) depicts the setting with $T_0=4$ and $J=199$. As in panel (b), the pre-treatment fit is perfect. However, post-treatment estimation error is very large and the second unit does not contribute to the synthetic control, $W_2^*=0$. 
On the whole, Figure \ref{figure:overfitting} illustrates how the number of pre-treatment periods and the number of untreated units crucially affect the bias of the synthetic control estimator. 

Apart from the possibility of over-fitting bias created by an excessively large $J$, preventing interpolation bias is another good reason to place restrictions on the untreated units that are allowed to enter the donor pool. Consider the factor model in equation \eqref{equation:factor}. Even if $\epsilon_{jt}$ lacks variation and, therefore, over-fitting bias is not a concern, the linearity assumption in equation \eqref{equation:factor} may only be a local approximation. If that is the case, reproducing the predictor values for the treated unit by averaging untreated units that are far from the treated unit in the space of the predictors may result in interpolation biases. To attenuate interpolation biases, it is useful to restrict the units allowed in the donor pool to untreated units with similar values in the predictors, $\bs X_j$, as the treated unit. 

As we have discussed above, researchers applying synthetic control estimators should recognize settings conducive to biases---because of small values of $T_0$, large values of $J$, or because the units in the donor pool have values of the predictor that are substantially different from those of the treated unit---and adapt their designs to ameliorate bias concerns. In Section \ref{section:validation}, we will discuss validation techniques to assess the potential for bias in concrete empirical settings. 

\begin{figure}[ht!]
\centering
  \begin{subfigure}[b]{0.5\textwidth}
  \centering
  	\includegraphics[width=1\linewidth]{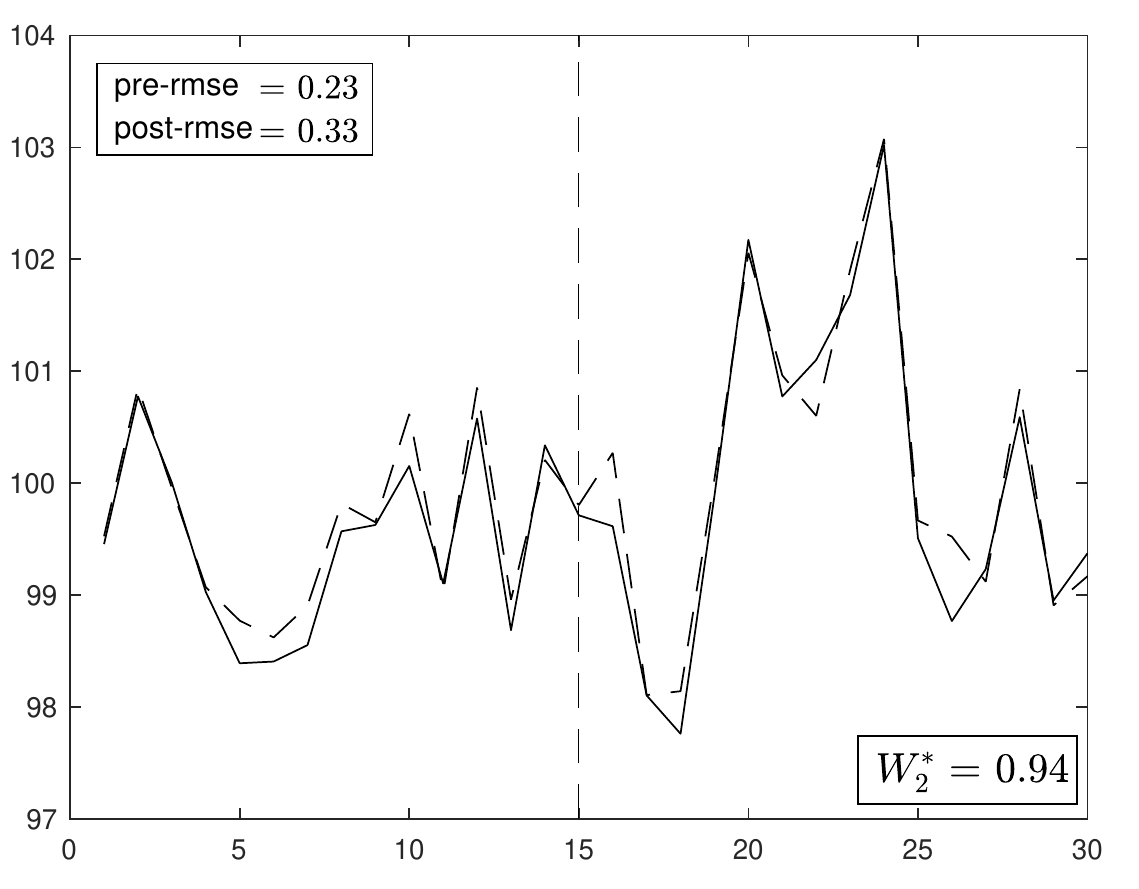}
  	\caption{Synthetic control}
  \end{subfigure}%
  \begin{subfigure}[b]{0.5\textwidth}
  \centering
  	\includegraphics[width=1\linewidth]{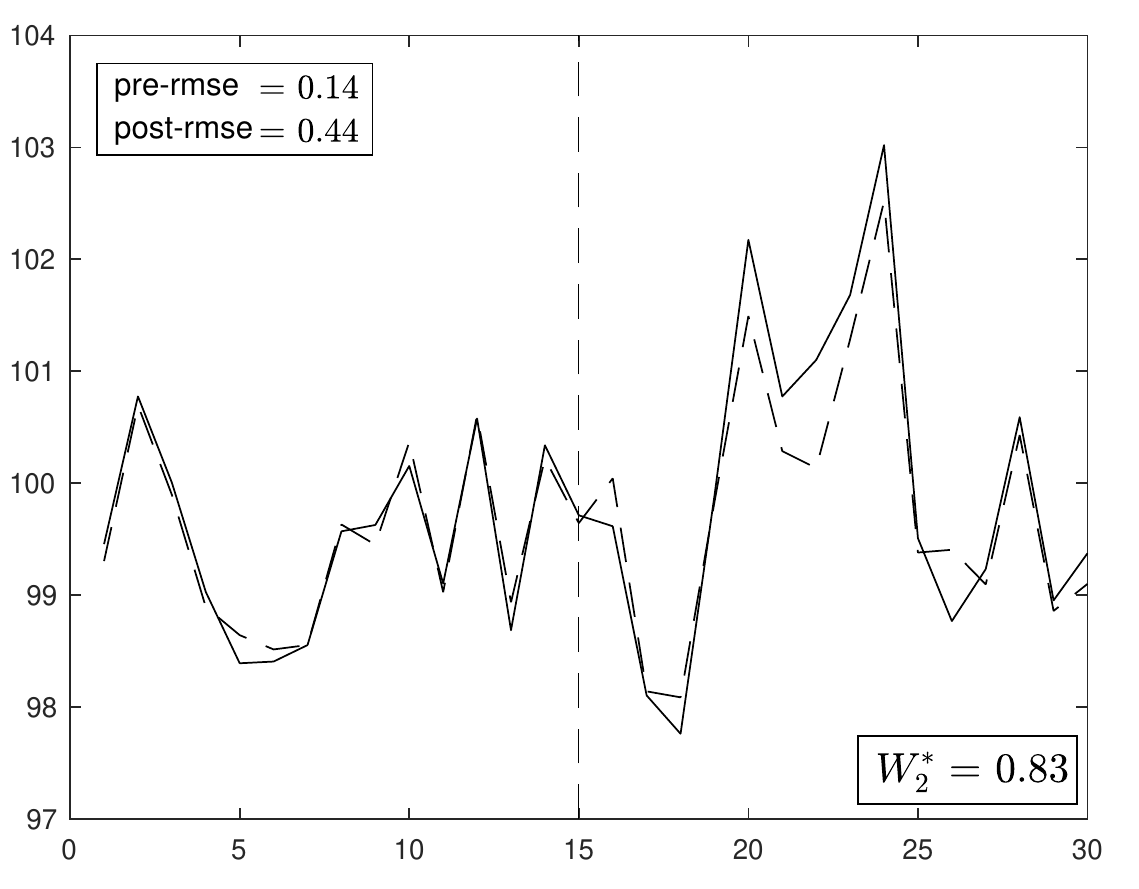}
  	\caption{Synthetic control with constant shift}
  \end{subfigure}%
  
  \begin{subfigure}[b]{0.5\textwidth}
  \centering
  	\includegraphics[width=1\linewidth]{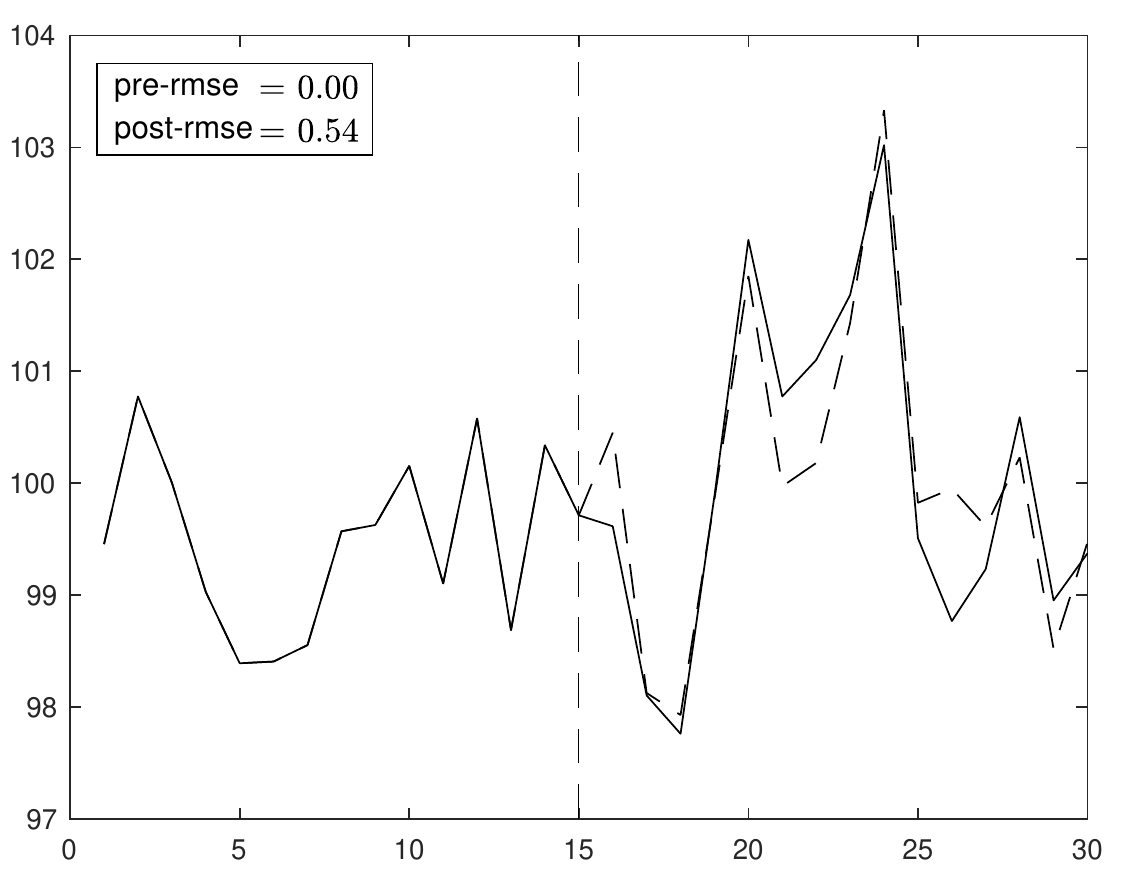}
  	\caption{Synthetic control via unrestricted regression}
  \end{subfigure}%
  \begin{subfigure}[b]{0.5\textwidth}
    %\raggedleft
    \hspace{4.8mm}\raisebox{6.19cm}{\includegraphics[width=0.4\linewidth]{Tables_and_Figures/legend1.png}}
  \end{subfigure}%
  \caption{Over-fitting with added flexibility}
  \label{figure:flex}
  \floatfoot{{\em Note:} Synthetic control estimates with added flexibility. Panel (a) reports results for the standard synthetic control estimator with weight constraints. Panel (b) reports results for a synthetic control model that includes a constant shift. Panel (c) reports results for a synthetic control estimated via unrestricted regression. $\sigma = 0.25$, $T_0 = 15$, $\rho = 0.5$ and $J = 19$.}
\end{figure}

We finish this section with a discussion of how modifications of the synthetic control design may affect the performance of the estimates. Panel (a) of Figure \ref{figure:flex} plots the outcome values for a treated unit and its synthetic control, for a simulation design that is the same as for panel (d) or Figure \ref{figure:noise} except that $T_0=15$. Because this is a simple setting without covariates, the synthetic control estimate of $Y^N_{1t}$ in panel (a) is 
\[
\sum_{j=2}^{J+1}W^*_j Y_{jt},
\]
with $W^*_2, \ldots, W^*_{J+1}$  chosen to minimize
\[
\sum_{t=1}^{T_0} \Big(Y_{1t}- \sum_{j=2}^{J+1}W_jY_{jt}\Big)^2
\]
with respect to $W_2, \ldots, W_{J+1}$, subject to the constrain that the weights, $W_2, \ldots, W_{J+1}$, are non-negative and sum up to one. 
The pre-treatment root-mean-square error (pre-RMSE),
\begin{equation}
\left(\frac{1}{T_0}\sum_{t=1}^{T_0}\Big(Y_{1t}-\sum_{j=2}^{J+1} W^*_j Y_{jt}\Big)^2\right)^{1/2},
\label{equation:pre-rmse}
\end{equation}
is equal to 0.23. The post-treatment root-mean-square error (post-RMSE)--which is analogous to \eqref{equation:pre-rmse}, but calculated for the post-treatment periods---is somewhat higher, 0.33. Still, in this example, pre-intervention fit comes reasonably close to estimation accuracy in the post-intervention periods. As we will see next, adding extra flexibility to the synthetic control estimator may break down the correspondence between pre-RMSE and post-RMSE. Consider now panel (b) of Figure \ref{figure:flex}. It reports the results of a synthetic control estimator with an added constant shift \citep{doudchenko2016balancing}. That is, panel (b) plots the result of estimating of $Y^N_{1t}$ as
\begin{equation}
\widehat Y^N_{1t} = \widehat\alpha+\sum_{j=2}^{J+1}\widehat W_j Y_{jt},
\label{equation:least_squares_estimate}
\end{equation}
where $\widehat\alpha, \widehat W_2, \ldots, \widehat W_{J+1}$  minimize
\begin{equation}
\sum_{t=1}^{T_0} \Big(Y_{1t}-\alpha- \sum_{j=2}^{J+1}W_jY_{jt}\Big)^2
\label{equation:least_squares}
\end{equation}
with respect to $\alpha, W_2, \ldots, W_{J+1}$, subject to the constrain that the weights, $W_2, \ldots, W_{J+1}$, are non-negative and sum up to one. Introducing the constant shift $\alpha$, as in \eqref{equation:least_squares} is equivalent to computing a synthetic control with the outcome variables measured in deviations with respect to their pre-treatment means
\[
Y_{jt}-\frac{1}{T_0}\sum_{k=1}^{T_0} Y_{jk}.
\]
The availability of the constant $\alpha$ increases the degrees of freedom of the synthetic control estimator. This extra flexibility could be useful in certain instances, especially when the trajectory of the outcome for the treated unit is extreme relative to the trajectory of the outcomes in the donor pool, so a good fit is not attainable on the levels of $Y_{jt}$, but may be attainable on the demeaned outcomes. 
It is good to remember, however, that additional flexibility increases the potential for over-fitting. In the example of Figure \ref{figure:flex}, introducing a constant shift in the estimate may seem to be warranted at first sight. After all, the pre-treatment trajectory of the synthetic control seems to over-estimate the values of the treated outcomes. Indeed, panel (b) shows that the inclusion of the constant shift noticeably improves pre-treatment fit, with pre-RMSE dropping to 0.14. However, post-RMSE increases to 0.44.  
Panel (c) allows for additional flexibility relaxing the constraints on the weights \citep{hsiao2012panel,doudchenko2016balancing}. A simple regression estimator of this model takes the same form as $\widehat Y^N_{1t}$ in equation \eqref{equation:least_squares_estimate}, but with parameters $\widehat\alpha, \widehat W_2, \ldots, \widehat W_{J+1}$ that minimize \eqref{equation:least_squares} for unrestricted values of $\alpha, W_2, \ldots, W_{J+1}$. Panel (c) in Figure \ref{figure:flex} plots the result. Now, pre-RMSE is equal to zero, the unrestricted regression perfectly fits the trajectory of the outcome variable for the treated in the pre-treatment periods. However, the post-RMSE is 0.54, indicating a larger out-of-sample estimation error than in panels (a) and (b). 

The increased risk of over-fitting that comes from increased flexibility in synthetic control estimation is well-recognized in the literature. At least since \cite{doudchenko2016balancing}, researchers extending synthetic controls beyond the convex hull of the predictors' values for donor pool units have often combined the extra flexibility in the parameters with some regularization device---usually, but not exclusively, in the form of $L_1$ and/or $L_2$ penalizations on the synthetic control parameters \citep[see, in particular,][]{amjad2018robust, agarwal2021synthetic, arkhangelsky2019synthetic, athey2021matrix, ben2021augmented, chernozhukov2021conformal,doudchenko2016balancing}. 

Figure \ref{figure:flex} illustrates the effectiveness of restricting synthetic control weights to be non-negative and sum up to one as a regularization device. Relaxing the restrictions on the synthetic control weights, or adopting bias-correction techniques, as in \cite{abadie2021penalized} and \cite{ben2021augmented}, extends the applicability of the synthetic control method to settings where the values of the predictors for the treated unit cannot be closely fitted by a weighted average of the units in the donor pool. However, the increased flexibility in the synthetic control weights is obtained at the expense of the straightforward interpretability of the estimates. In addition, relaxing the restrictions on the synthetic control weights allows for unchecked extrapolation outside the support of the data, increasing the risk that the estimates reflect extreme counterfactuals, for which the data contain little information \citep{king2006danger}.   

\section{Validating the Synthetic Control Estimator}
\label{section:validation}

In Section \ref{section:performance} we have discussed how the scale of the transitory shocks, the number of pre-treatment periods, and the size of the donor pool influence the bias of the synthetic control estimator. In any particular empirical setting, however, a researcher could be uncertain about potential exposure to over-fitting and interpolation biases. In this section, we discuss validation techniques to help assess the magnitude of potential biases in concrete empirical applications. 

\begin{figure}[ht!]
\centering
  \begin{subfigure}[b]{0.5\textwidth}
  \centering
  	\includegraphics[width=1\linewidth]{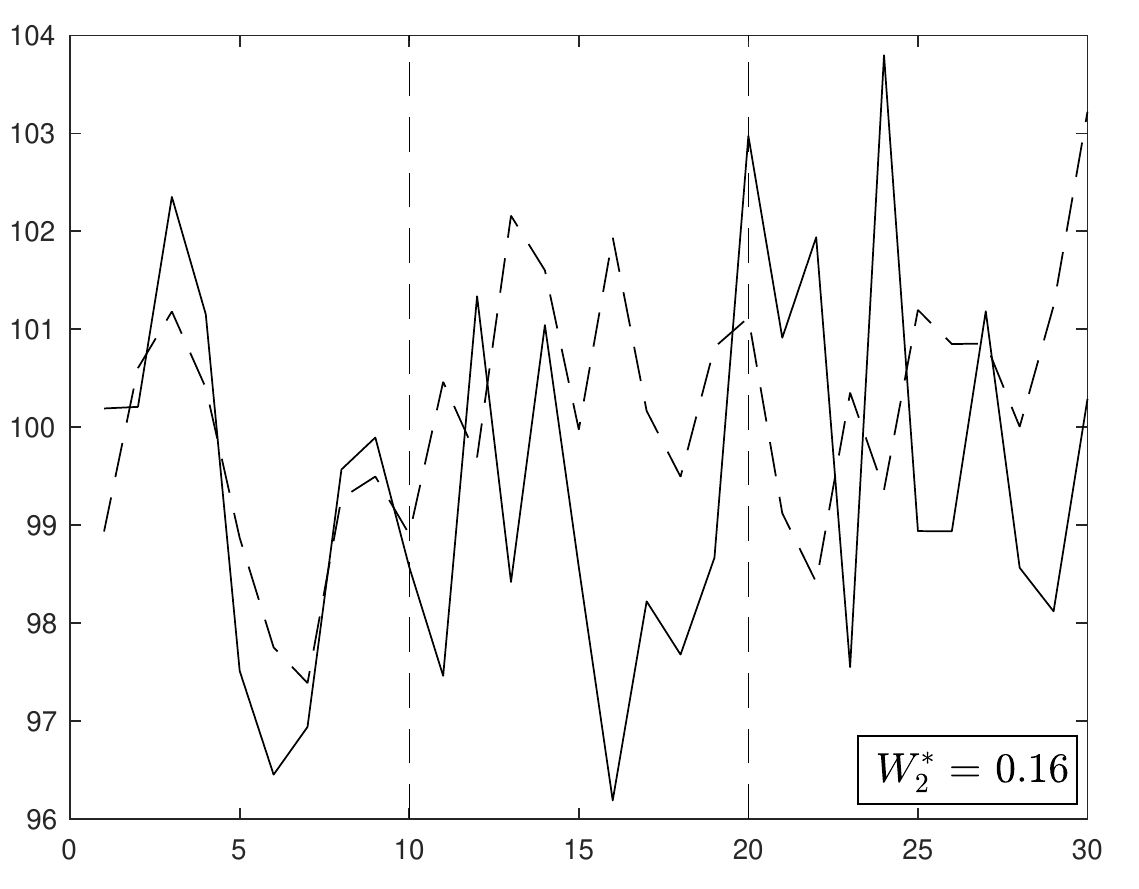}
  	\caption{$\sigma = 2$}
  \end{subfigure}%
  \begin{subfigure}[b]{0.5\textwidth}
  \centering
  	\includegraphics[width=1\linewidth]{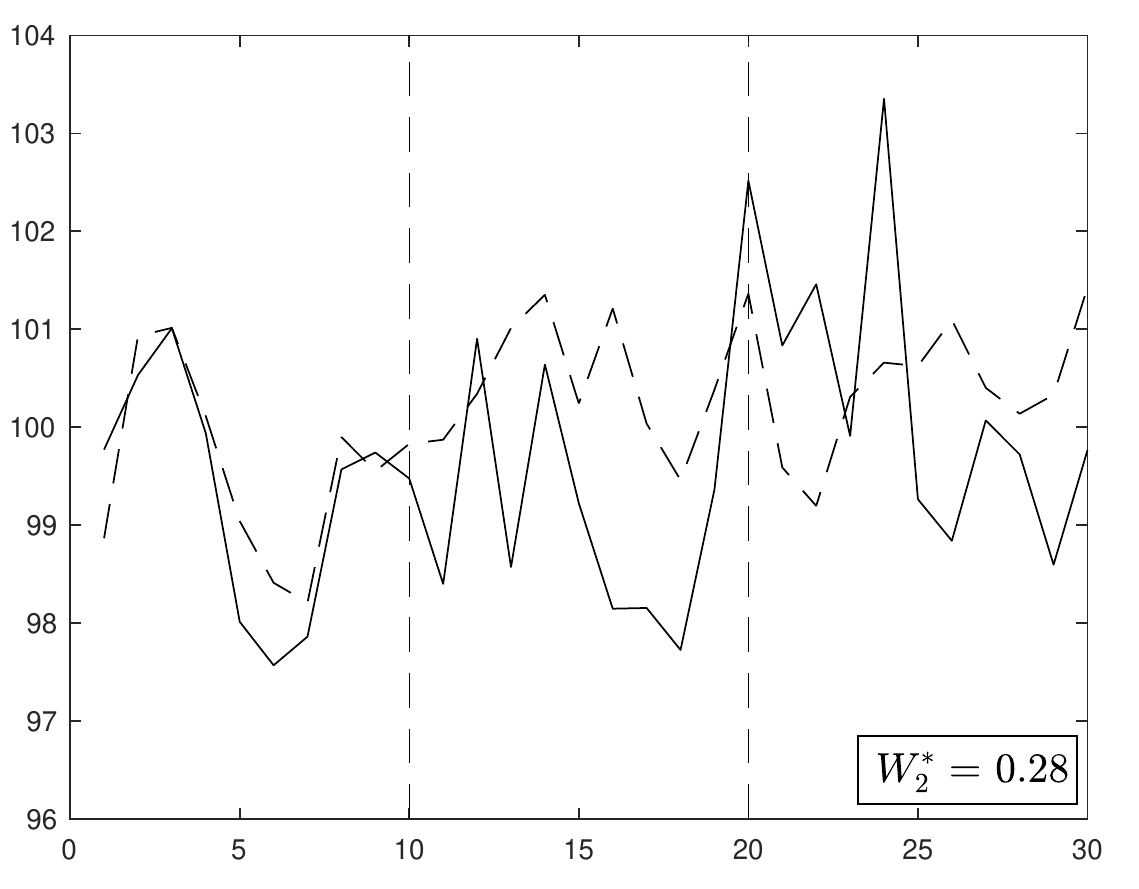}
  	\caption{$\sigma = 1$}
  \end{subfigure}%
  
   \begin{subfigure}[b]{0.5\textwidth}
  \centering
  	\includegraphics[width=1\linewidth]{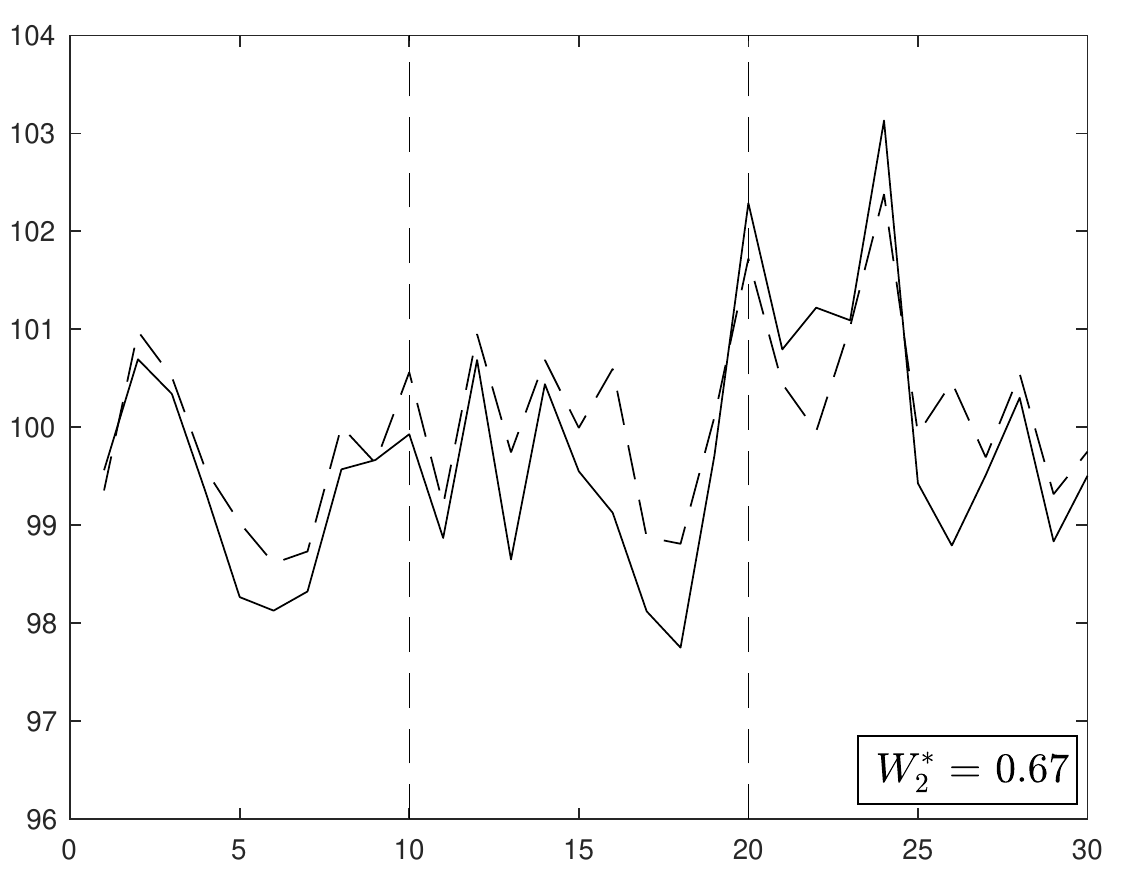}
  	\caption{$\sigma = 0.5$}
  \end{subfigure}%
  \begin{subfigure}[b]{0.5\textwidth}
  \centering
  	\includegraphics[width=1\linewidth]{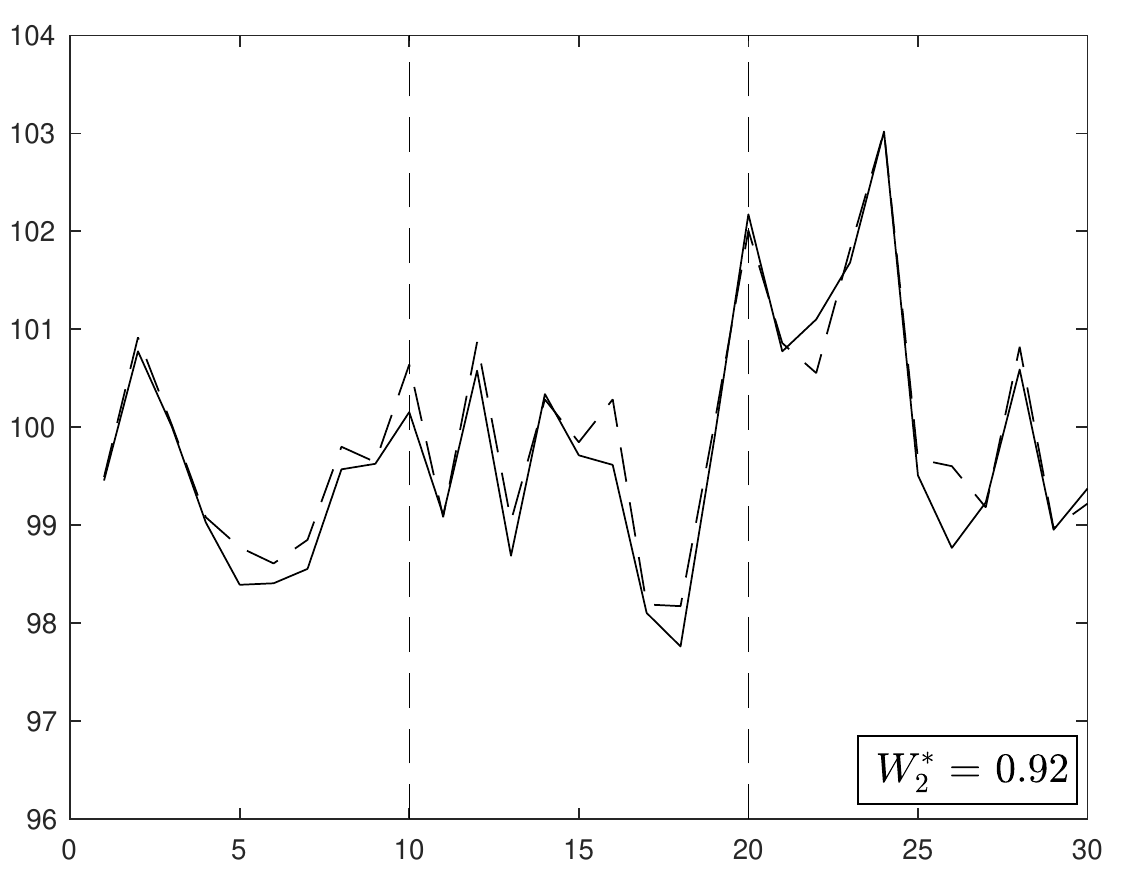}
  	\caption{$\sigma = 0.25$}
  \end{subfigure}%
  \caption{Synthetic control validation}
\vspace{-8mm}
\center
\includegraphics[width=0.2\linewidth]{Tables_and_Figures/legend1.png} 
  \label{figure:validation}
  \floatfoot{{\em Note:} Synthetic control estimates with treatment backdated to $T_0^b=10$ for different values of $\sigma$. $T_0 = 20$, $\rho = 1$ and $J = 19$.}
\end{figure}

An effective and visually interpretable validation exercise for synthetic controls can be obtained by artificially backdating the treatment of interest in the data. Figure \ref{figure:validation} reports the results of backdating the treatment in the simulation setting of Figure \ref{figure:noise}. Recall that in this simulated scenario, the last pre-treatment period is $T_0=20$. In Figure \ref{figure:validation}, we have backdated the treatment so that the last pre-treatment period is now $T_0^{\,b}=10$. This implies that we calculate the synthetic control weights using data for $t=1, \ldots, T_0^{\, b}$ only. Backdating the treatment to $T_0^{\,b}=10$ creates ten hold-out periods, $t=11, \ldots, 20$, available to validate the predictions of the synthetic control estimator. Notice that this validation exercise could be performed at $t=T_0$, before post-treatment data can possibly be observed, to inform study design decisions. An analyst carrying out this validation exercise at $t=T_0$ would observe large out-of-sample prediction errors for $\sigma=2$ and $\sigma =1$ in panels (a) and (b), but much more reliable predictions for $\sigma=0.5$ and especially for $\sigma=0.25$, in panels (c) and (d) respectively. In Figure \ref{figure:validation}, the quality of the prediction in the post-treatment periods, $t=21, \ldots, 30$ closely mirrors the prediction error for the hold-out periods, $t=11, \ldots, 20$. 

\begin{figure}[ht!]
\centering
  \begin{subfigure}[b]{0.5\textwidth}
  \centering
  	\includegraphics[width=1\linewidth]{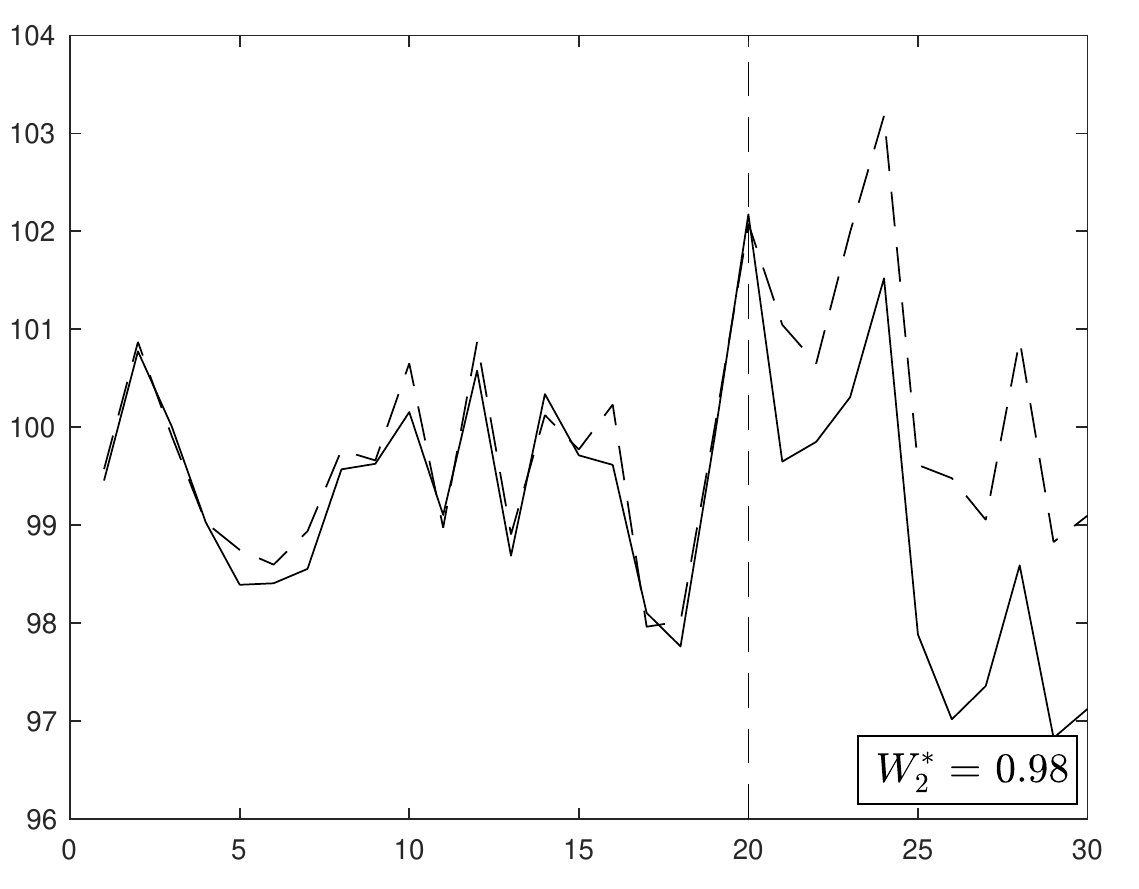}
  	\caption{Intervention at $T_0 =20$}
  \end{subfigure}%
  \begin{subfigure}[b]{0.5\textwidth}
  \centering
  	\includegraphics[width=1\linewidth]{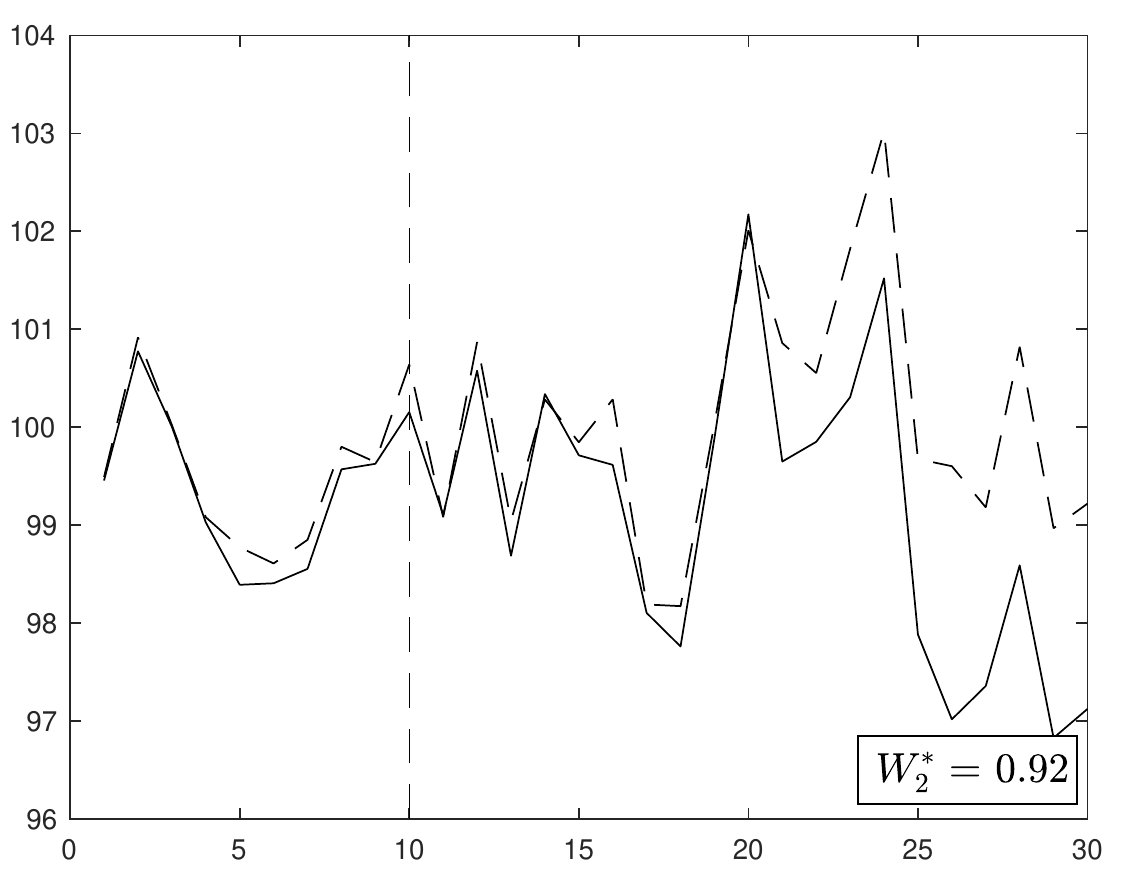}
  	\caption{Backdating to $T_0^{\, b}=10$}
  \end{subfigure}%
  \caption{Backdating with a treatment effect.}
\vspace{-4mm}
\center
\includegraphics[width=0.2\linewidth]{Tables_and_Figures/legend1.png} 
  \label{figure:validation_teffect}
    \floatfoot{{\em Note:} Synthetic control estimates with a non-zero treatment effect and treatment backdated to $T_0^b=10$. $\sigma=0.25$, $T_0 = 20$, $\rho = 0.5$ and $J = 19$.}
\end{figure}

Figure \ref{figure:validation_teffect} employs the same simulation design as in Figure \ref{figure:validation}, but with a positive treatment effect in the post-treatment periods. The effect of the treatment is detected in panel (a), where the synthetic control is trained using data up to period $T_0$. Panel (b) reports the result of the same backdating exercise as in Figure \ref{figure:validation}. Not only is the out-of-sample fit good during the hold-out periods, $t=11,\ldots, 20$, but the fact that an effect emerges at $T_0$, exactly the time when the true (in this case, simulated) treatment takes place and that the magnitudes of the estimates are similar in panels (a) and (b) provides additional credibility to the synthetic control estimates of panel (a).

\section{Trimming the Donor Pool}
\label{section:trimming}

In Sections \ref{section:primer} and \ref{section:performance}, we have discussed the importance controlling the size of the donor pool. In this section, we will show how actively trimming the donor pool to units with values of the predictors that are close to the values of the predictors for the treated unit can substantially improve the performance of synthetic control estimators. 

\begin{figure}[ht!]
\centering

  \begin{subfigure}[b]{0.5\textwidth}
  \centering
  	\includegraphics[width=1\linewidth]{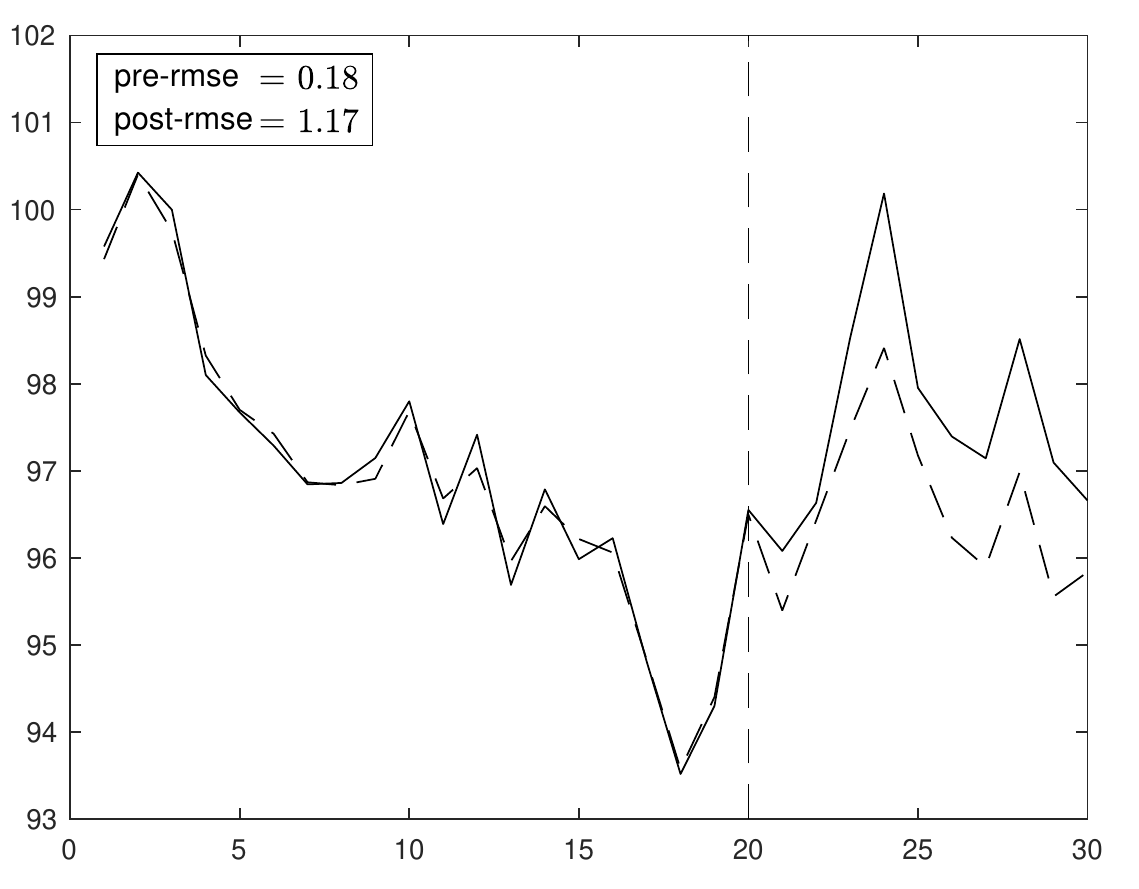}
  	\caption{Without trimming.}
  \end{subfigure}%
  \begin{subfigure}[b]{0.5\textwidth}
  \centering
  	\includegraphics[width=1\linewidth]{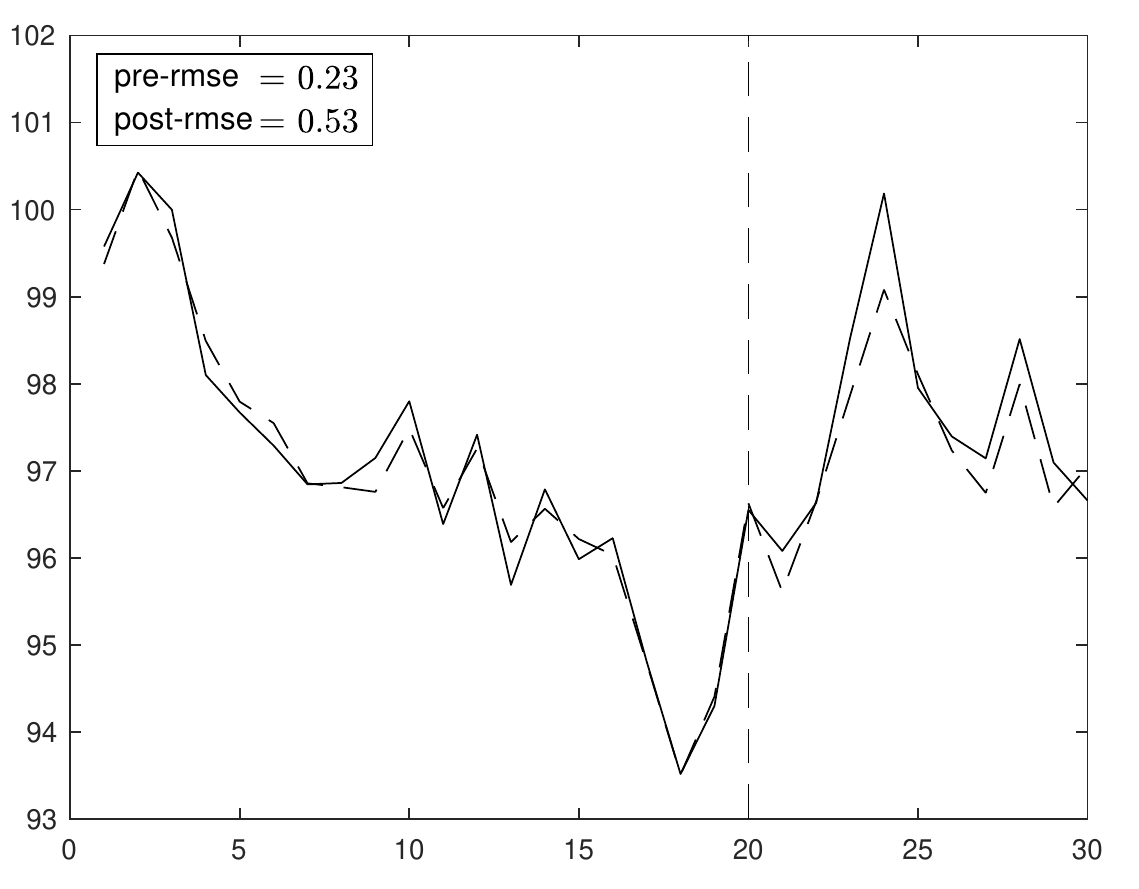}
  	\caption{75\% trimming.}
  \end{subfigure}
    \begin{subfigure}[b]{0.5\textwidth}
  \centering
  	\includegraphics[width=1\linewidth]{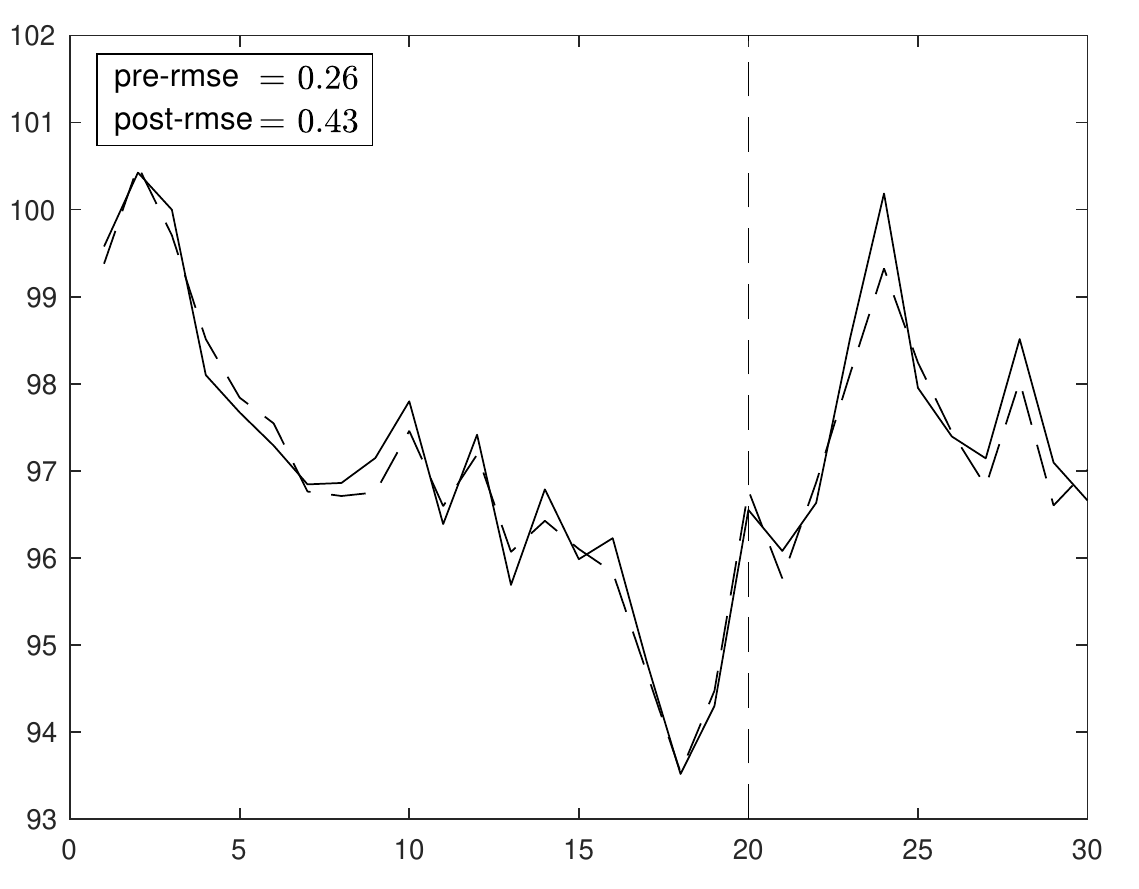}
  	\caption{90\% trimming.}
  \end{subfigure}%
  \begin{subfigure}[b]{0.5\textwidth}
  \centering
  	\includegraphics[width=1\linewidth]{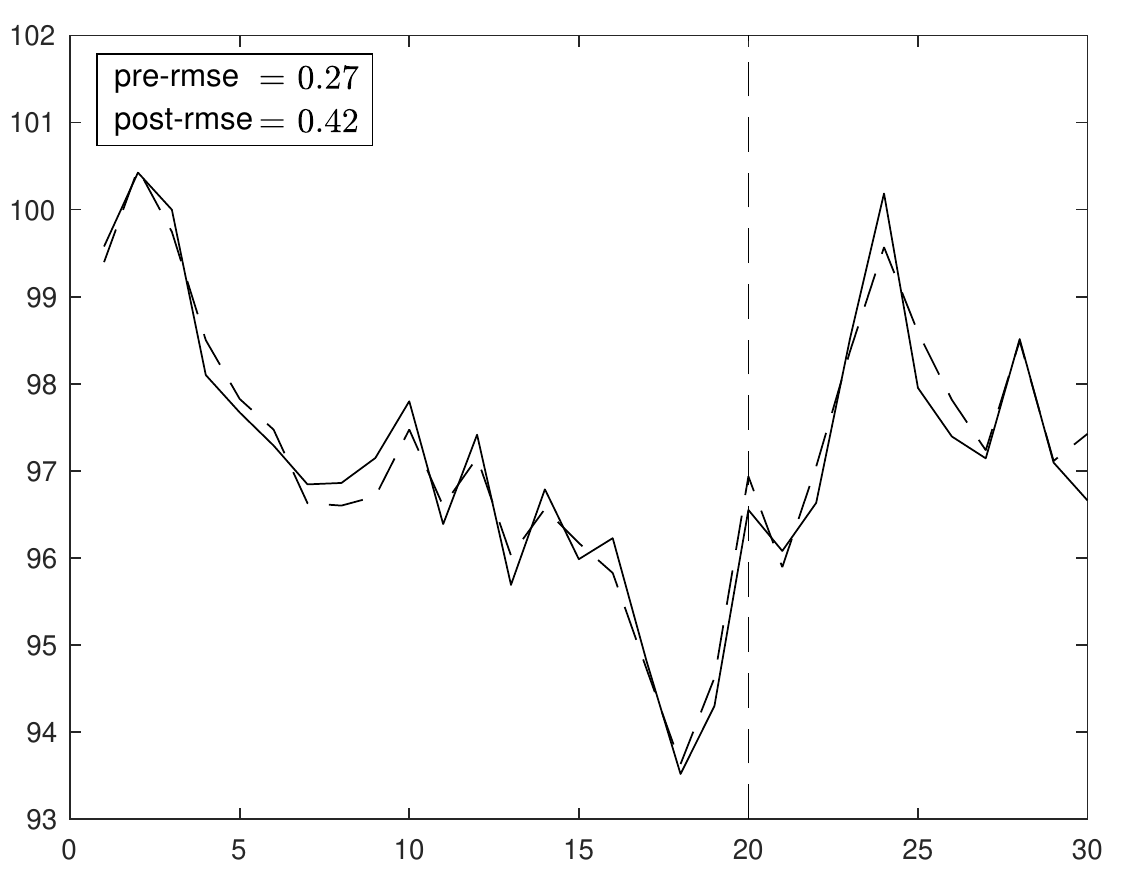}
  	\caption{95\% trimming.}
  \end{subfigure}
  \caption{Synthetic control with trimming.}
\vspace{-8mm}
\center
\includegraphics[width=0.2\linewidth]{Tables_and_Figures/legend1.png} 
  \label{figure:trimming}
  \floatfoot{{\em Note:} Synthetic control with and without trimming for $\sigma=0.25$, $T_0 = 20$, $\rho = 1$ and $J = 199$. Panel (a) shows the standard synthetic control. Panel (b) shows the synthetic control with the donor pool trimmed to the $25\%$ closest units to the treated unit according to the Euclidean distance. Panel (c) and (d) are like panel (b) with trimming to $10\%$ and $5\%$ of the units respectively.}
\end{figure}

The first panel of Figure \ref{figure:trimming} reports the result of a simulation based on the grouped factor model in equation \eqref{equation:group_factor}, with $\rho=1$, $\sigma=0.25$, $T_0=20$, and $F=100$ ($J=199$). In this simulation, the large size of the donor pool promotes over-fitting resulting in large post-treatment estimation errors. In panels (b)-(d) of Figure \ref{figure:trimming} we have only retained the 50, 20, and 10 donor pool units, respectively, that best fit the values of the predictors for the treated (in this simple case, pre-treatment outcomes only). Trimming mechanically increases pre-treatment differences between the treated unit and the synthetic control, but noticeably improves the performance of the estimator in the post-treatment periods. 

\section{The Role of Observed Covariates}
\label{section:observed}

Thus far, we have only considered settings without covariates, $\boldsymbol Z_j$, in our simulations. In this section we discuss how the inclusion of observed covariates influences the performance of synthetic control estimators. 

\begin{figure}
\centering
  \begin{subfigure}[b]{0.45\textwidth}
  \centering
  	\includegraphics[width=1\linewidth]{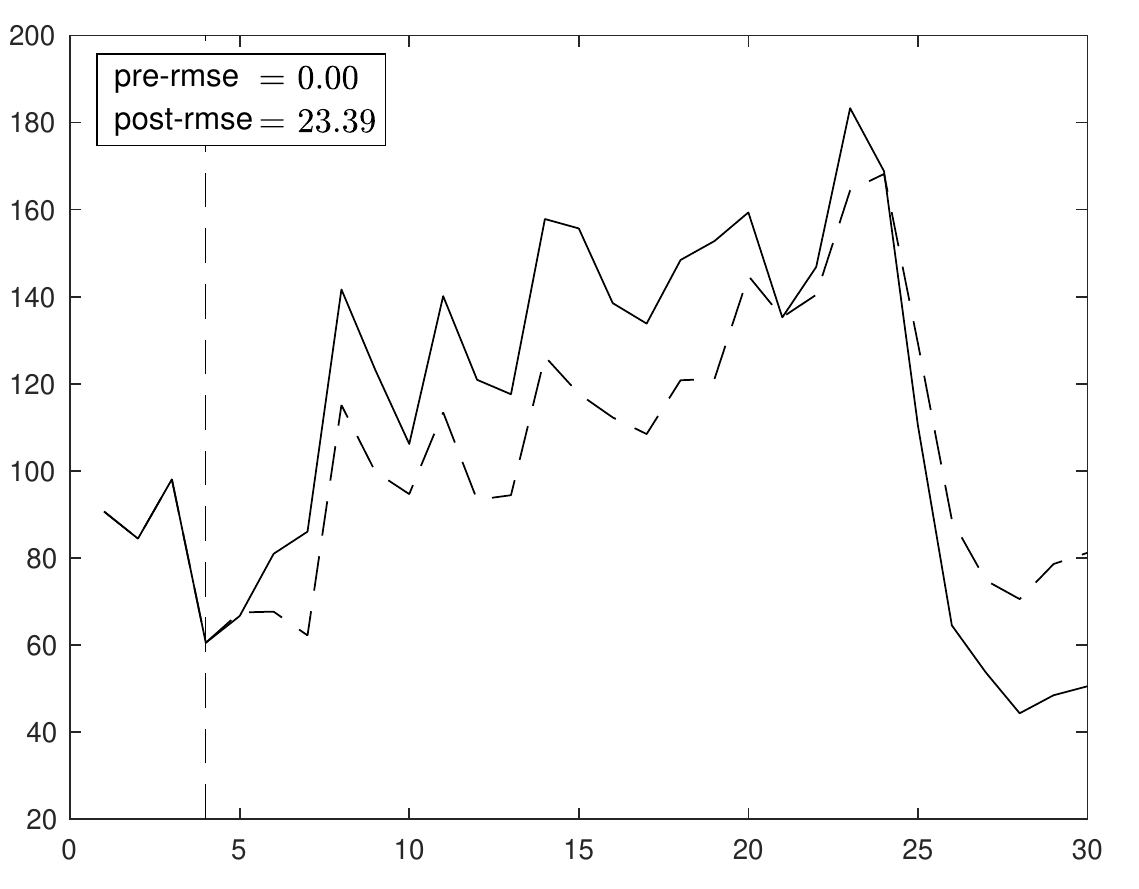}
  	\caption{$F = 5$}
  \end{subfigure}%
  \begin{subfigure}[b]{0.45\textwidth}
  \centering
  	\includegraphics[width=1\linewidth]{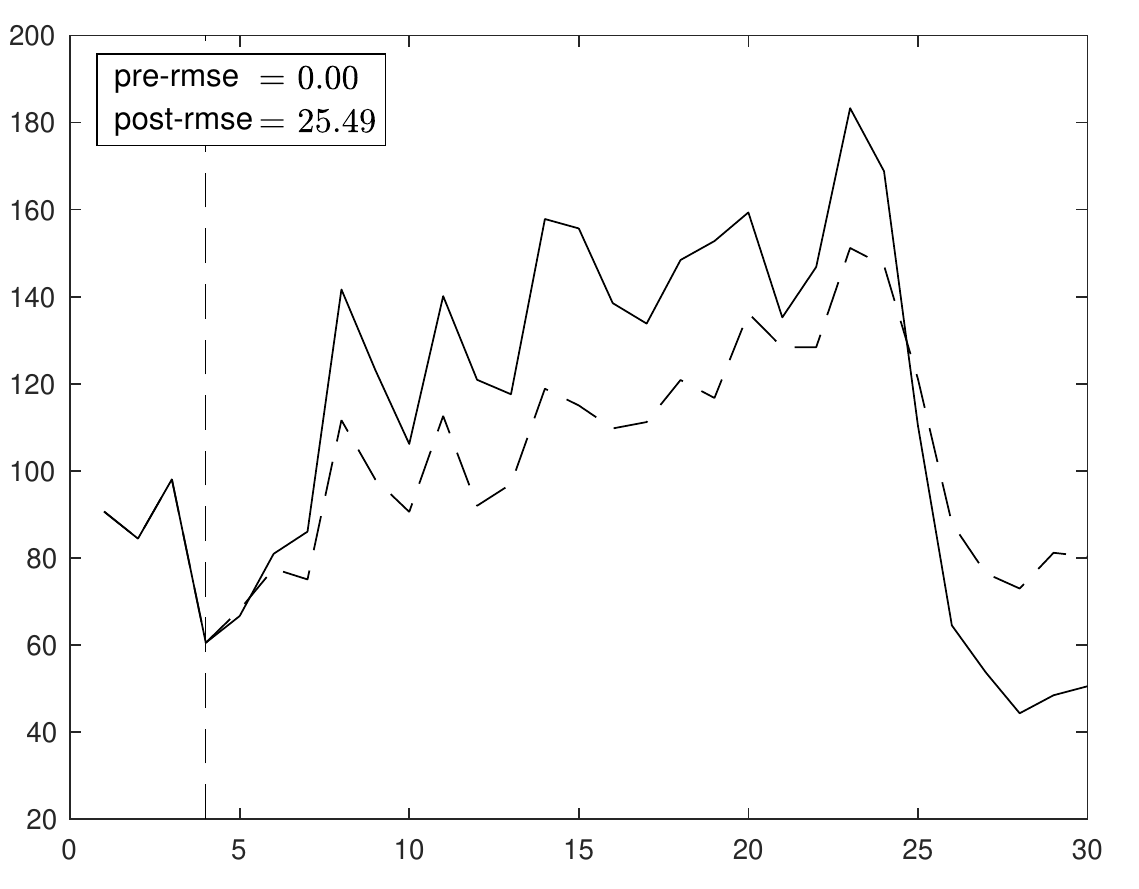}
  	\caption{$F = 4$}
  \end{subfigure}%
  
   \begin{subfigure}[b]{0.45\textwidth}
  \centering
  	\includegraphics[width=1\linewidth]{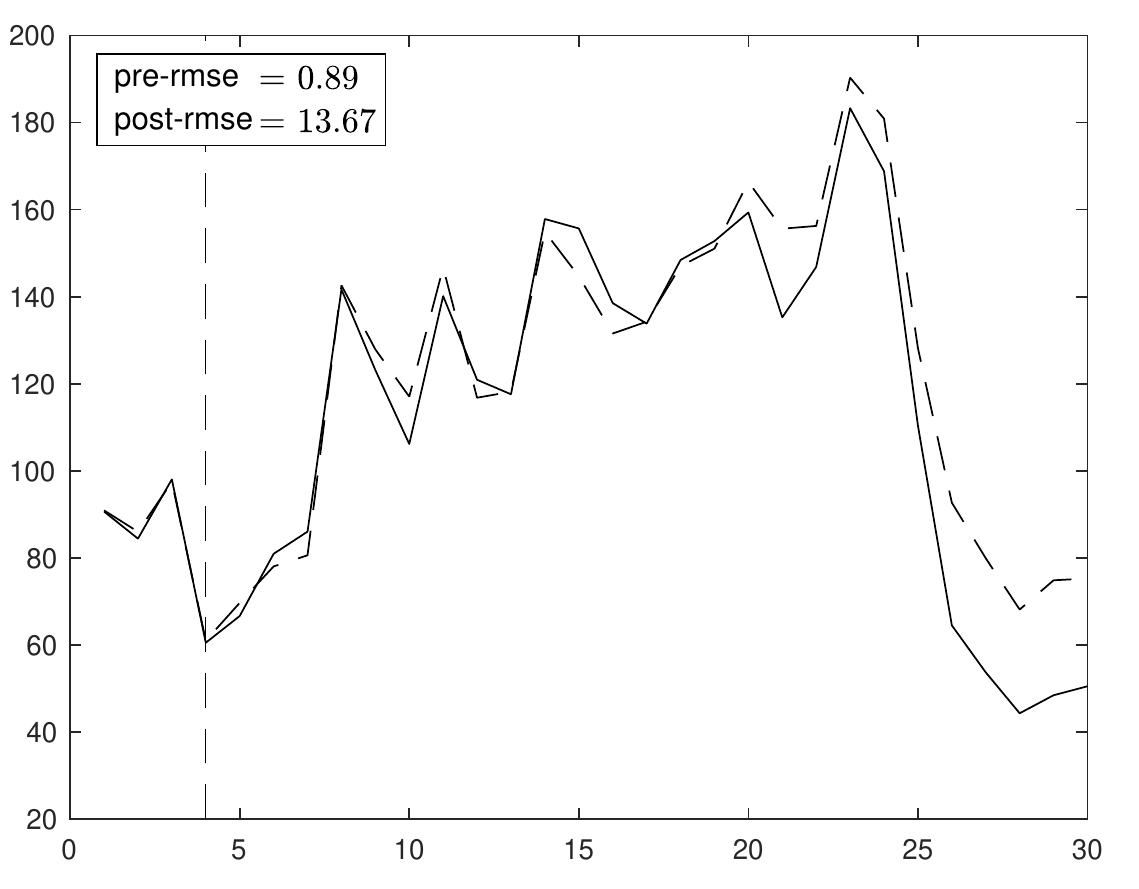}
  	\caption{$F = 3$}
  \end{subfigure}%
  \begin{subfigure}[b]{0.45\textwidth}
  \centering
  	\includegraphics[width=1\linewidth]{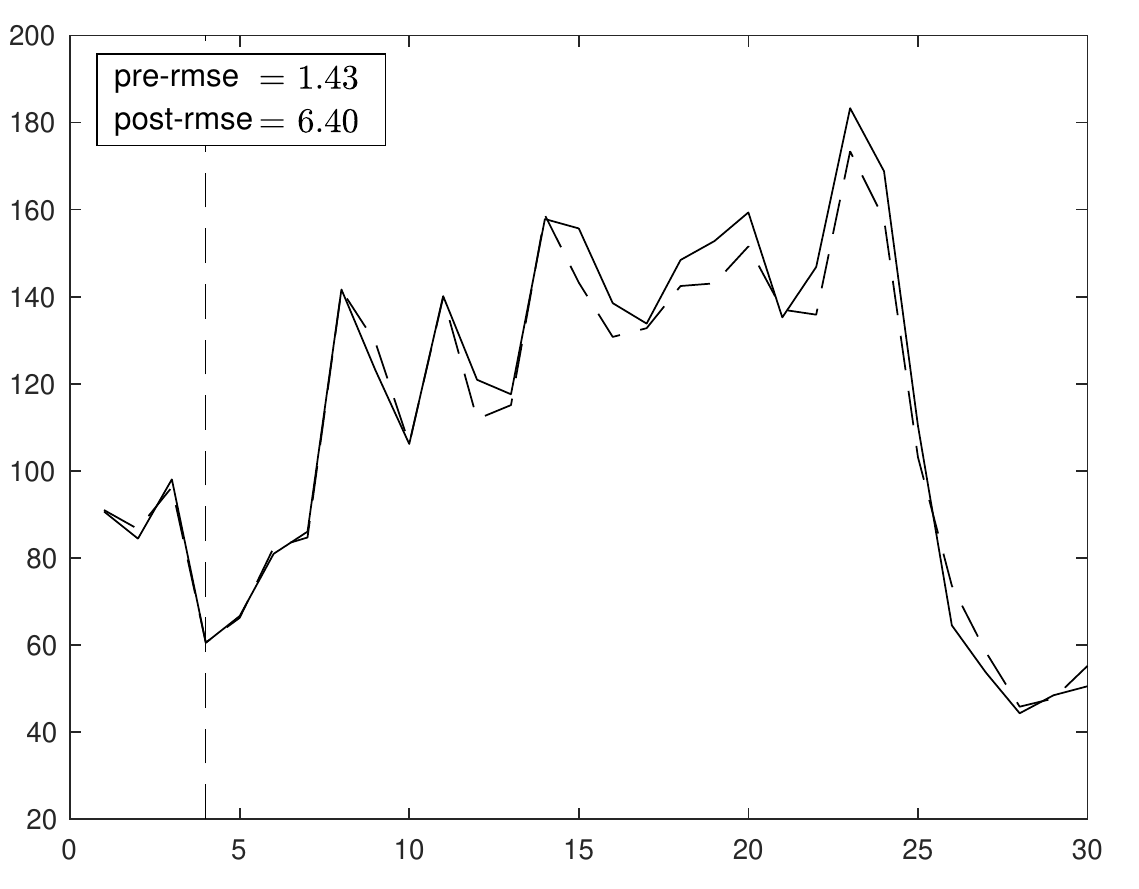}
  	\caption{$F = 2$}
  \end{subfigure}%
  
     \begin{subfigure}[b]{0.45\textwidth}
  \centering
  	\includegraphics[width=1\linewidth]{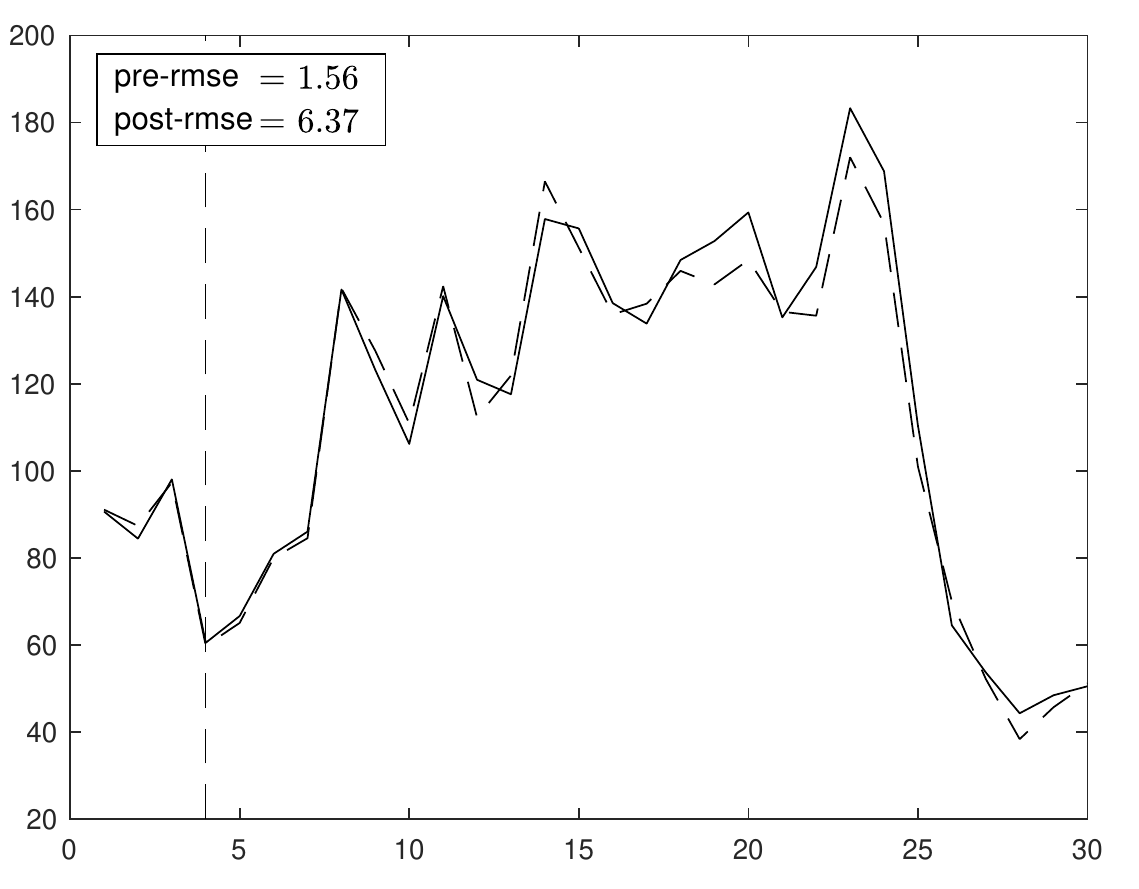}
  	\caption{$F = 1$}
  \end{subfigure}%
  \begin{subfigure}[b]{0.45\textwidth}
  \centering
  	\includegraphics[width=1\linewidth]{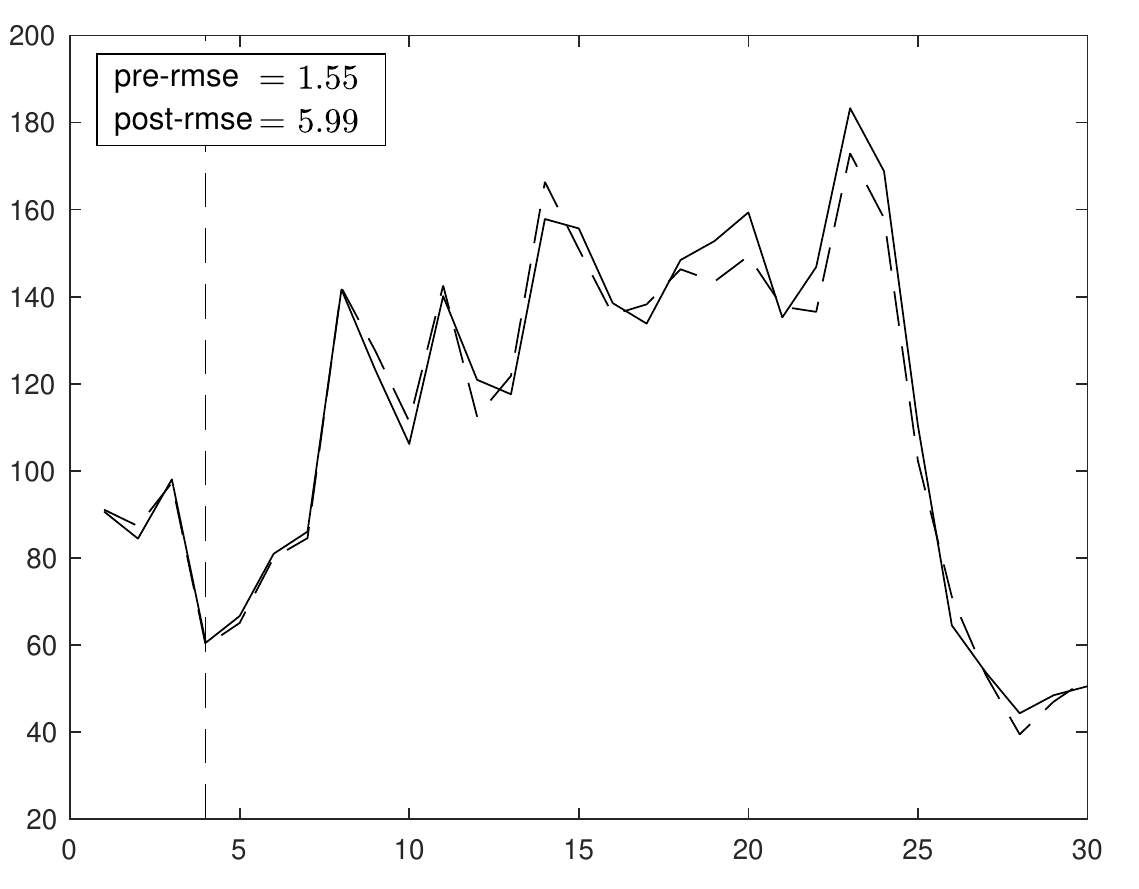}
  	\caption{$F = 0$}
  \end{subfigure}%
  \caption{Including observed covariates}
\vspace{-8mm}
\center
\includegraphics[width=0.2\linewidth]{Tables_and_Figures/legend1.png} 
  \label{figure:factors}
  \floatfoot{{\em Note:} Synthetic control under different values of $F$, the number of covariates not included in $\bs X_j$. See main text for a detailed description of the data generating process.}
\end{figure}

To illustrate the role of observed covariates, we consider the linear factor model in  \eqref{equation:factor}, with $\delta_t=100$, $\boldsymbol Z_j$ and $\boldsymbol \mu_j$ uniformly distributed on $[0,20]$, and where the components of $\boldsymbol \theta_t$ and $\boldsymbol \lambda_t$ follow independent random walks with standard Gaussian innovations. The idiosynchratic errors, $\epsilon_{jt}$, are modelled as Gaussian noise with standard deviation equal to five. In Figure \ref{figure:factors} we adopt a setting with $J=1000$ and $T_0=4$ and five covariates. Panel (a) considers the case where the five covariates are unobserved (so $F=5$) and there are no observed covariates, so a synthetic control estimator is based on pre-treatment outcomes only. The large value of $J$, the small value of $T_0$, and the presence of unobserved factors creates substantial over-fitting and estimation bias. In panels (b)-(f), we incrementally shift covariates from $\boldsymbol\mu_j$ to $\boldsymbol Z_j$, by including these covariates in the vectors $\bs X_j$ of predictive variables. That is, in contrast to panel (a), panels (b)-(f) use incrementally larger numbers of observed covariates, in addition to pre-treatment outcomes, to fit a synthetic control. In panel (f) there are no unobserved factors ($F=0$): all five covariates are included in $\boldsymbol Z_j$ (and, therefore, in $\boldsymbol X_j$). The patterns of fit and estimation error in Figure \ref{figure:factors} illustrate the importance of including predictive covariates in $\boldsymbol X_j$. Adding covariates to $\boldsymbol X_j$ induces a slight deterioration in pre-treatment fit in the outcomes, given that other covariates (apart from pre-treatment outcomes) enter $\boldsymbol X_j$. However, reductions in $F$ are associated with substantial decreases in estimation error. As we increase the number of observed covariates, we rely less on pre-treatment outcomes as imperfect proxies for the values of the covariates, obtaining more reliable estimates. Figure \ref{figure:factors} underscores the potential importance of including observed predictors of the outcomes, beyond pre-treatment outcomes, in the set of variables used to fit a synthetic control. 

\section{An Auto-Regressive Model}
\label{section:autoregressive}

Although we have thus far used a linear factor structure in our simulations, it is important to notice that the validity of synthetic control estimators is not limited to the linear factor model in \eqref{equation:factor}. In particular, \cite{AbaDiaHai2010} study the bias behavior of synthetic control estimators under a generative model with a vector auto-regressive structure in covariates and outcomes. In this section, we illustrate the applicability of synthetic control estimators under an auto-regressive data generating process. For that purpose, we consider a simple version, without covariates, of the auto-regressive model in \cite{AbaDiaHai2010},
\[
Y_{jt}^N = \alpha_{jt} Y_{jt-1}^N + \epsilon_{jt},
\]
for $j=1, \ldots, J+1$ and $t=2,\ldots, T$, and $Y_{1t}^I=Y_{1t}^N$, for $t=T_0+1, \ldots, T$. This is an $AR(1)$ model with coefficients $\alpha_{jt}$ that are allowed to vary in time and across units.
For the simulation results, we adopt $J+1=50$ and, as before, $T_0=20$ and $T=30$. We model $Y^N_{j1}$---the initial value of the outcome---for units $j=1, \ldots J+1$, as a Gaussian variable with mean 100 and standard deviation equal to 20. The error terms, $\epsilon_{jt}$, are standard Gaussian and independent of each other, for $j=1, \ldots J+1$ and $t=1,\ldots, T$. For $\alpha_{jt}$, we adopt a grouped structure, with five groups of 10 units. Inside the groups, $\alpha_{j2}, \ldots, \alpha_{jT}$ are constant across units and drawn as independent Gaussian variables with mean equal to one and standard deviation equal to 0.1. The series $\alpha_{j2}, \ldots, \alpha_{jT}$ are independent across groups. 

\begin{figure}[ht!]
\centering
  	\includegraphics[width=0.5\linewidth]{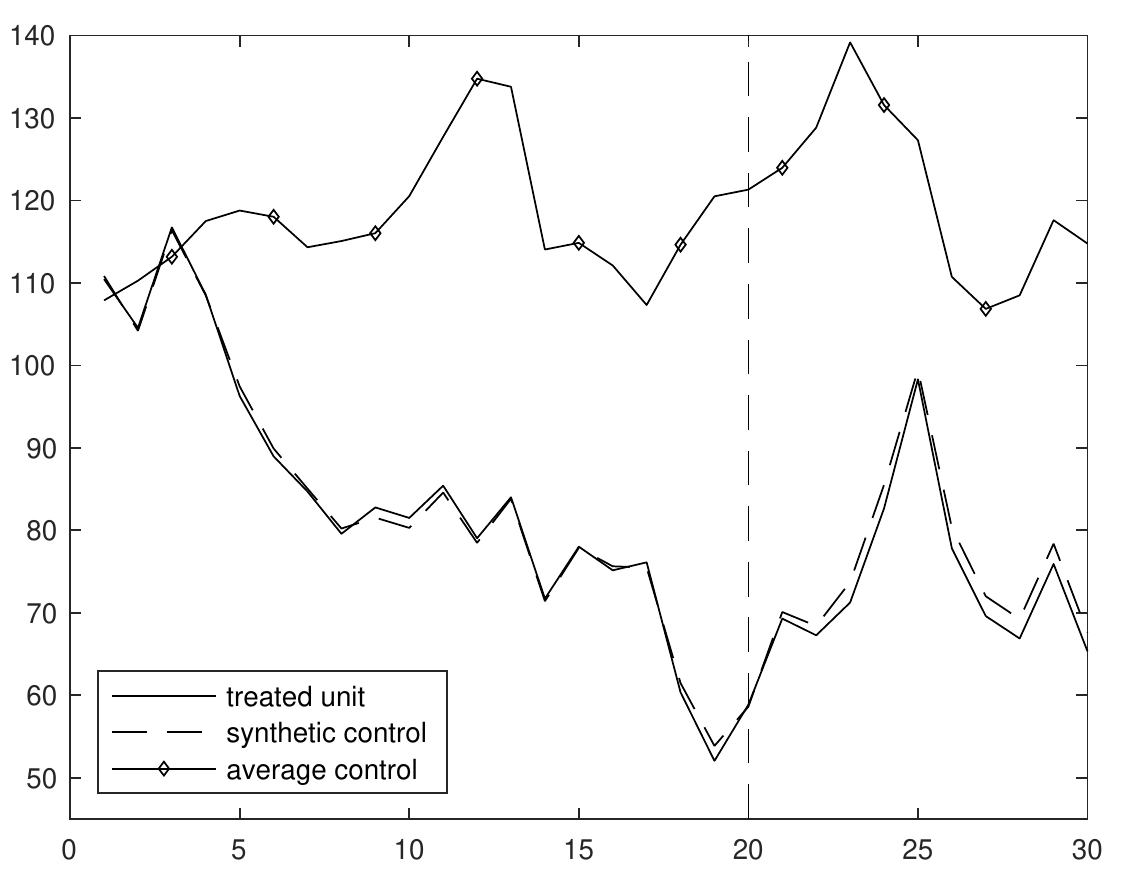}
  \caption{Synthetic control with an auto-regressive process.}
  \label{figure:ar}
  \floatfoot{{\em Note:} Synthetic control under an auto-regressive model for the outcome variable. $T_0 = 20$, $J+1=50$. See text of Section \ref{section:autoregressive} for a detailed description of the design of the simulation.}
\end{figure}

The result of a random simulation under this data generating process is reported in Figure \ref{figure:ar}. In this simulation, because of the assumed group-level heterogeneity in the trajectory of the outcome, the average of the outcome variable among all the untreated units is not able to reproduce the outcome for the treated unit, before or after the treatment is implemented. 
In contrast, the synthetic control estimates closely follow the trajectory of the outcome for the treated, demonstrating the applicability of synthetic control estimators outside the linear factor model considered in previous sections. 

\section{Conclusion}
\label{section:conclusions}

Synthetic controls are intuitive, transparent, and produce reliable estimates for a variety of data generating processes. However, like for any other statistical or econometric method to estimate treatment effects, there are settings where synthetic controls may fail. In this article, we have described the settings where the synthetic control method provides reliable estimates, and those where it does not. Of particular importance, we have discussed the nature of over-fitting in synthetic control estimators, when it arises, how to recognize it, and ways to alleviate or eliminate over-fitting biases. 

Moreover, we have described intuitive and effective diagnosis checks for the validity of synthetic control estimates that can be easily carried out in empirical research. They pertain to the quality of the fit and length of the pre-intervention period, the ability to fit the treated unit using a relatively small donor pool of similar units, the ability to validate out-of-sample, and that a treatment effect estimate emerges at the time of the actual treatment implementation, even when the treatment is artificially backdated in the data. 

Our recommendations are summarized in seven guiding principles for the research design of empirical studies that employ synthetic controls.

\clearpage
\bibliography{Bibliography}

\newpage
\section*{Appendix}

Table \ref{tab:sim} summarizes the results for all figures by providing the average pre-RMSE, post-RMSE and $W_2$ (the weight assigned to the second unit) over 10000 simulations. Figures \ref{figure:sim_noise_st}-\ref{figure:ar} show 95\% bands over 10000 simulations of the estimation error (prediction minus actual) for the same simulation designs as in Figures \ref{figure:noise_st}-\ref{figure:ar}, respectively.  

\setcounter{figure}{0}
\renewcommand{\thetable}{A.\arabic{table}}
\setcounter{figure}{3}
\renewcommand{\thefigure}{A.\arabic{figure}}

% Table generated by Excel2LaTeX from sheet 'table'
\begin{table}[htbp]
  \thispagestyle{empty}
  \centering
  \caption{Simulation results for all figures.}
    \begin{tabular}{lrrrrrrr}
    \multicolumn{1}{l}{Figure} & \multicolumn{1}{r}{$J$} & \multicolumn{1}{r}{$T_0$} & \multicolumn{1}{r}{$\sigma$} & \multicolumn{1}{r}{$\rho$} & \multicolumn{1}{r}{\text{post-RMSE}} & \multicolumn{1}{r}{\text{pre-RMSE}} & \multicolumn{1}{r}{$W^*_2$} \\
          &       &       &       &       &       &       &  \\
    Figure \ref{figure:noise} (a) & 19    & 20    & 0.25  & 0.5   & 0.365 & 0.315 & 0.890 \\
    Figure \ref{figure:noise} (b) & 19    & 20    & 0.5   & 0.5   & 0.723 & 0.588 & 0.730 \\
    Figure \ref{figure:noise} (c) & 19    & 20    & 1     & 0.5   & 1.362 & 1.032 & 0.429 \\
    Figure \ref{figure:noise} (d) & 19    & 20    & 2     & 0.5   & 2.421 & 1.782 & 0.171 \\
          &       &       &       &       &       &       &  \\
    Figure \ref{figure:noise_st} (a) & 19    & 20    & 0.25  & 1     & 0.387 & 0.341 & 0.983 \\
    Figure \ref{figure:noise_st} (b) & 19    & 20    & 0.5   & 1     & 0.795 & 0.675 & 0.953 \\
    Figure \ref{figure:noise_st} (c) & 19    & 20    & 1     & 1     & 1.642 & 1.315 & 0.882 \\
    Figure \ref{figure:noise_st} (d) & 19    & 20    & 2     & 1     & 3.272 & 2.513 & 0.763 \\
          &       &       &       &       &       &       &  \\
    Figure \ref{figure:noise_rho1} (a) & 19    & 20    & 0.25  & 1     & 0.439 & 0.320 & 0.920 \\
    Figure \ref{figure:noise_rho1} (b) & 19    & 20    & 0.5   & 1     & 0.921 & 0.609 & 0.799 \\
    Figure \ref{figure:noise_rho1} (c) & 19    & 20    & 1     & 1     & 1.799 & 1.108 & 0.564 \\
    Figure \ref{figure:noise_rho1} (d) & 19    & 20    & 2     & 1     & 3.117 & 1.965 & 0.316 \\
          &       &       &       &       &       &       &  \\
    Figure \ref{figure:overfitting} (a) & 19    & 4     & 0.5   & 1     & 2.595 & 0.282 & 0.397 \\
    Figure \ref{figure:overfitting} (b) & 199   & 4     & 0.5   & 1     & 3.305 & 0.057 & 0.080 \\
    Figure \ref{figure:overfitting} (c) & 19    & 20    & 0.5   & 1     & 0.921 & 0.609 & 0.799 \\
    Figure \ref{figure:overfitting} (d) & 199   & 20    & 0.5   & 1     & 1.308 & 0.484 & 0.529 \\
          &       &       &       &       &       &       &  \\
    Figure \ref{figure:flex} (a) & 19    & 15    & 0.25  & 0.5   & 0.381 & 0.304 & 0.874 \\
    Figure \ref{figure:flex} (b) & 19    & 15    & 0.25  & 0.5   & 0.405 & 0.289 & 0.860 \\
    Figure \ref{figure:flex} (c) & 19    & 15    & 0.25  & 0.5   & 1.019 & 0.000 &  \\
          &       &       &       &       &       &       &  \\
    Figure \ref{figure:validation} (a) & 19    & 20    & 0.25  & 0.5   & 0.415 & 0.356 & 0.841 \\
    Figure \ref{figure:validation} (b) & 19    & 20    & 0.5   & 0.5   & 0.829 & 0.692 & 0.631 \\
    Figure \ref{figure:validation} (c) & 19    & 20    & 1     & 0.5   & 1.515 & 1.234 & 0.320 \\
    Figure \ref{figure:validation} (d) & 19    & 20    & 2     & 0.5   & 2.579 & 2.114 & 0.131 \\
          &       &       &       &       &       &       &  \\
    Figure \ref{figure:validation_teffect} (a) & 19    & 20    & 0.25  & 0.5   & 1.694 & 0.356 & 0.841 \\
    Figure \ref{figure:validation_teffect} (b) & 19    & 20    & 0.25  & 0.5   & 1.682 & 0.315 & 0.890 \\
              &       &       &       &       &       &       &  \\
    Figure \ref{figure:trimming} (d) & 199   & 20    & 0.25  & 1     & 0.495 & 0.315 & 0.875 \\
    Figure \ref{figure:trimming} (c)& 199   & 20    & 0.25  & 1     & 0.524 & 0.305 & 0.853 \\
    Figure \ref{figure:trimming} (b) & 199   & 20    & 0.25  & 1     & 0.562 & 0.291 & 0.826 \\
     Figure \ref{figure:trimming} (a) & 199   & 20    & 0.25  & 1     & 0.637 & 0.274 & 0.786 \\
          &       &       &       &       &       &       &  \\
    Figure \ref{figure:factors} (a) & 1000  & 4     & 5     & 1     & 38.153 & 0.661 &  \\
    Figure \ref{figure:factors} (b) & 1000  & 4     & 5     & 1     & 34.967 & 1.080 &  \\
    Figure \ref{figure:factors} (c) & 1000  & 4     & 5     & 1     & 26.531 & 1.250 &  \\
    Figure \ref{figure:factors} (d) & 1000  & 4     & 5     & 1     & 20.531 & 1.438 &  \\
    Figure \ref{figure:factors} (e) & 1000  & 4     & 5     & 1     & 12.776 & 1.675 &  \\
    Figure \ref{figure:factors} (f) & 1000  & 4     & 5     & 1     & 7.616 & 1.811 &  \\
          &       &       &       &       &       &       &  \\
    Figure \ref{figure:ar} & 49  & 20     & 1     &      & 4.093 & 2.073 & \\ 
    \end{tabular}%
  \label{tab:sim}%
\end{table}%

\begin{figure}[ht!]
\centering
  \begin{subfigure}[b]{0.5\textwidth}
  \centering
  	\includegraphics[width=1\linewidth]{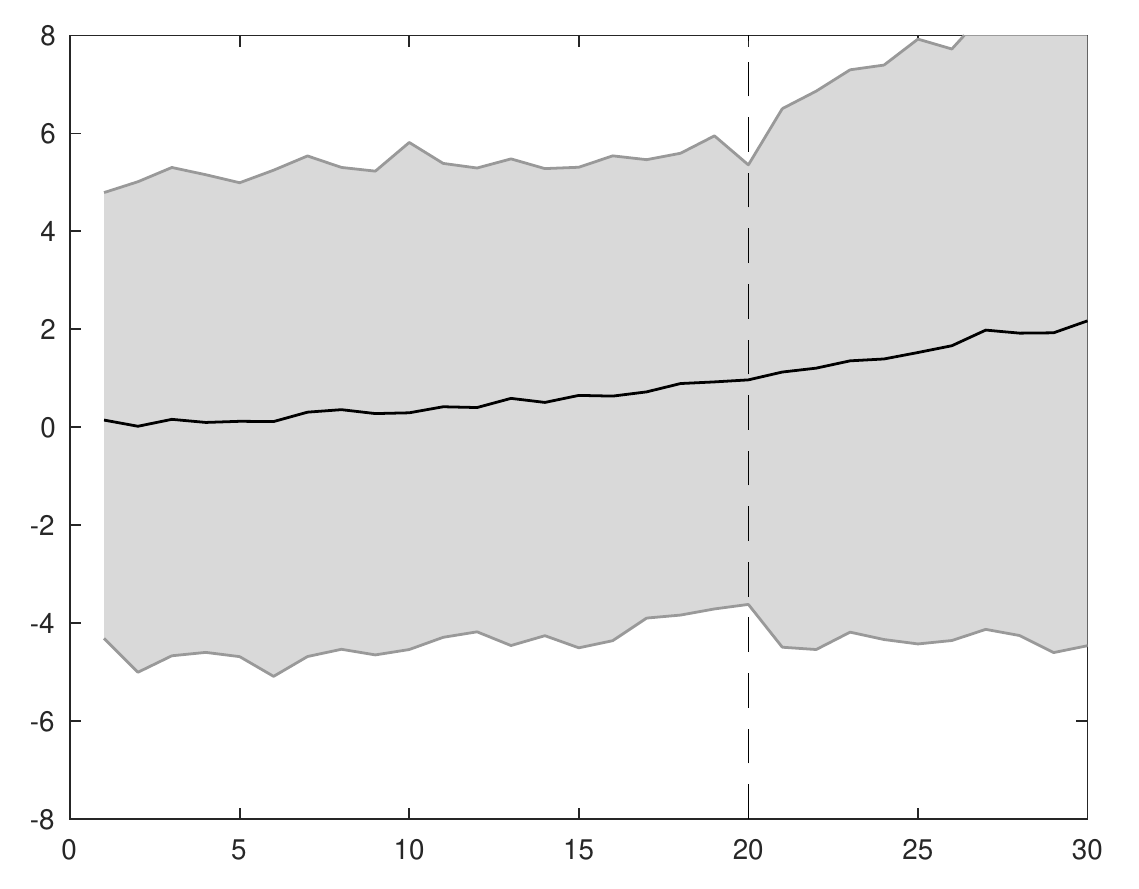}
  	\caption{$\sigma = 2$}
  \end{subfigure}%
  \begin{subfigure}[b]{0.5\textwidth}
  \centering
  	\includegraphics[width=1\linewidth]{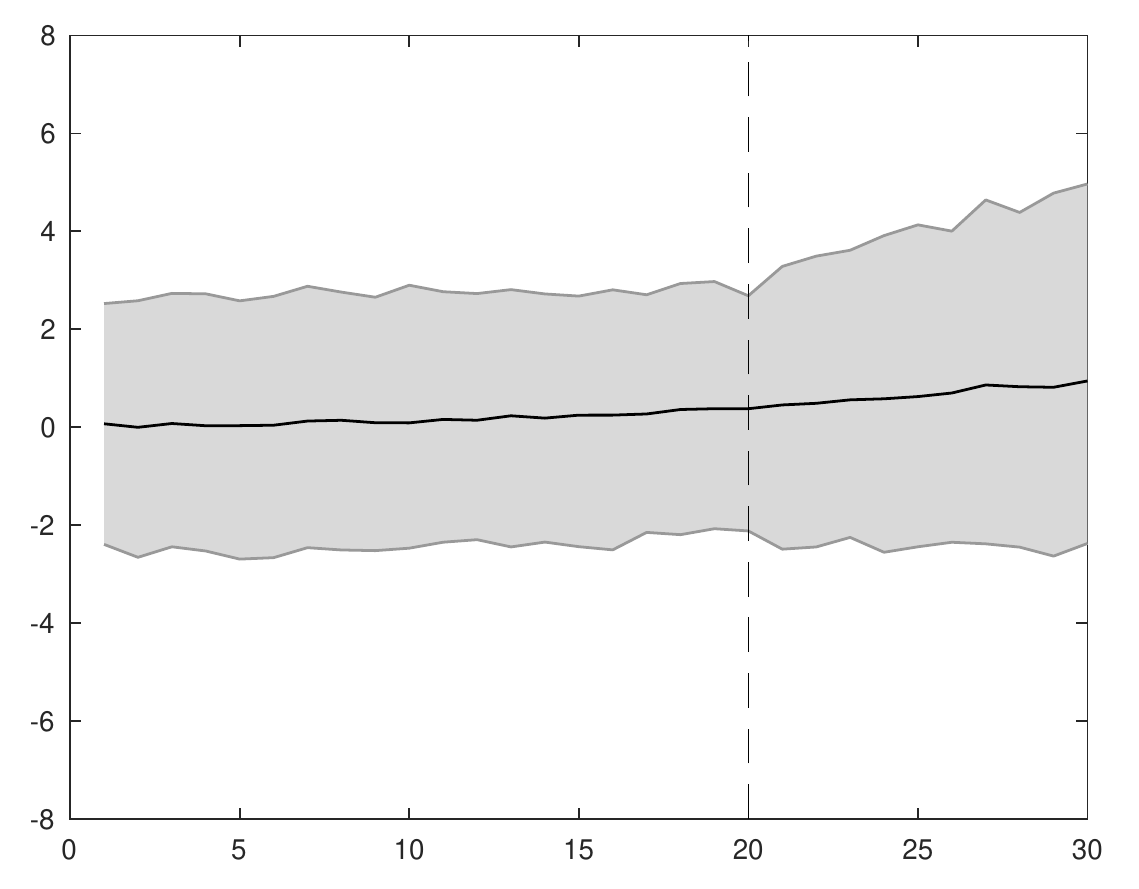}
  	\caption{$\sigma = 1$}
  \end{subfigure}%
  
   \begin{subfigure}[b]{0.5\textwidth}
  \centering
  	\includegraphics[width=1\linewidth]{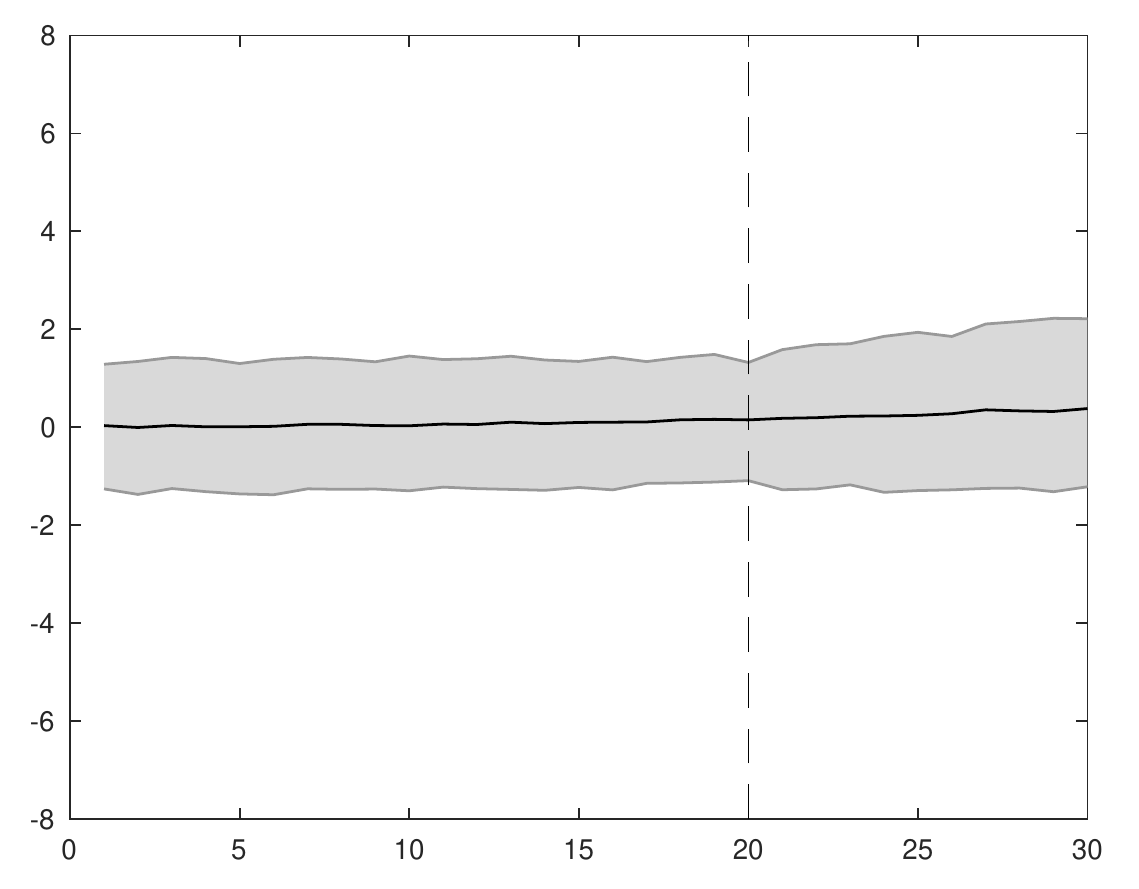}
  	\caption{$\sigma = 0.5$}
  \end{subfigure}%
  \begin{subfigure}[b]{0.5\textwidth}
  \centering
  	\includegraphics[width=1\linewidth]{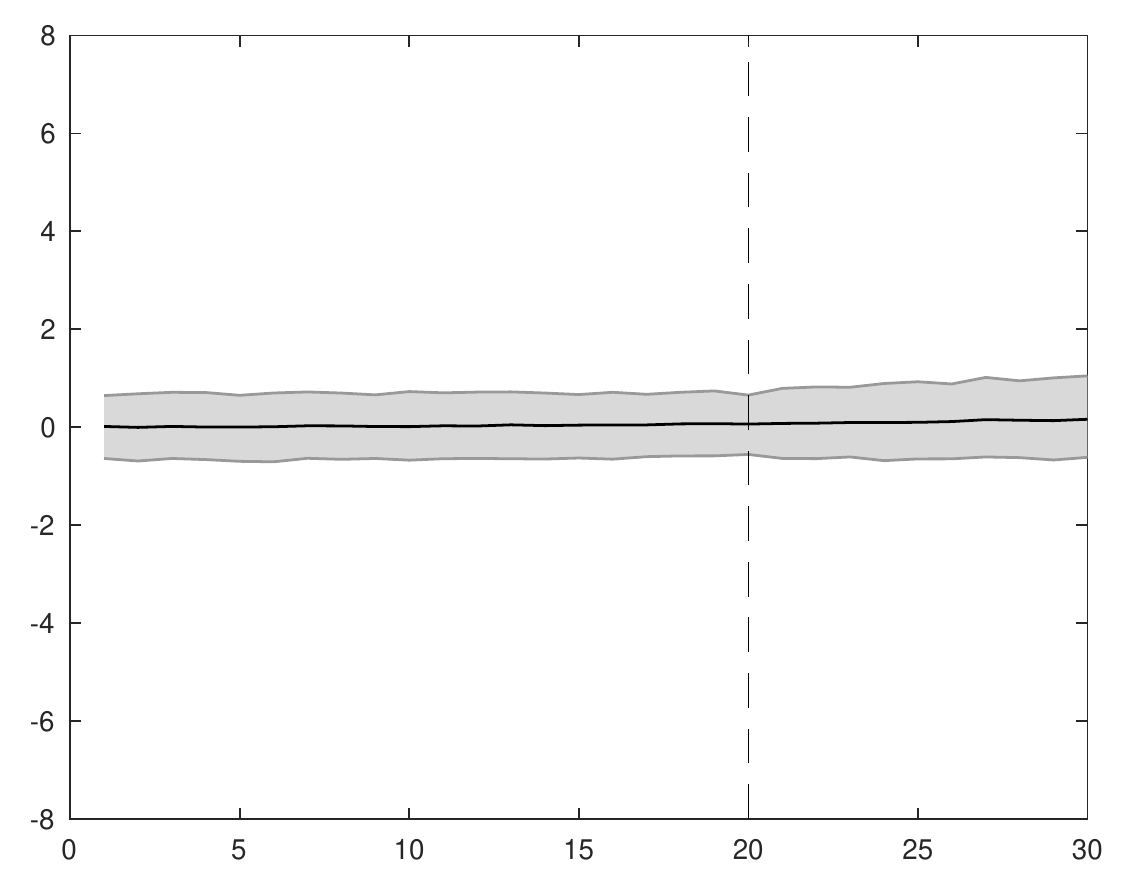}
  	\caption{$\sigma = 0.25$}
  \end{subfigure}%
  \caption{Synthetic control error simulations with a stochastic trend.}
  \label{figure:sim_noise_st}
  \floatfoot{{\em Note:} 95\% bands for the simulation design of Figure \ref{figure:noise_st}.}
\end{figure}
\clearpage
\begin{figure}[ht!]
\centering
  \begin{subfigure}[b]{0.5\textwidth}
  \centering
  	\includegraphics[width=1\linewidth]{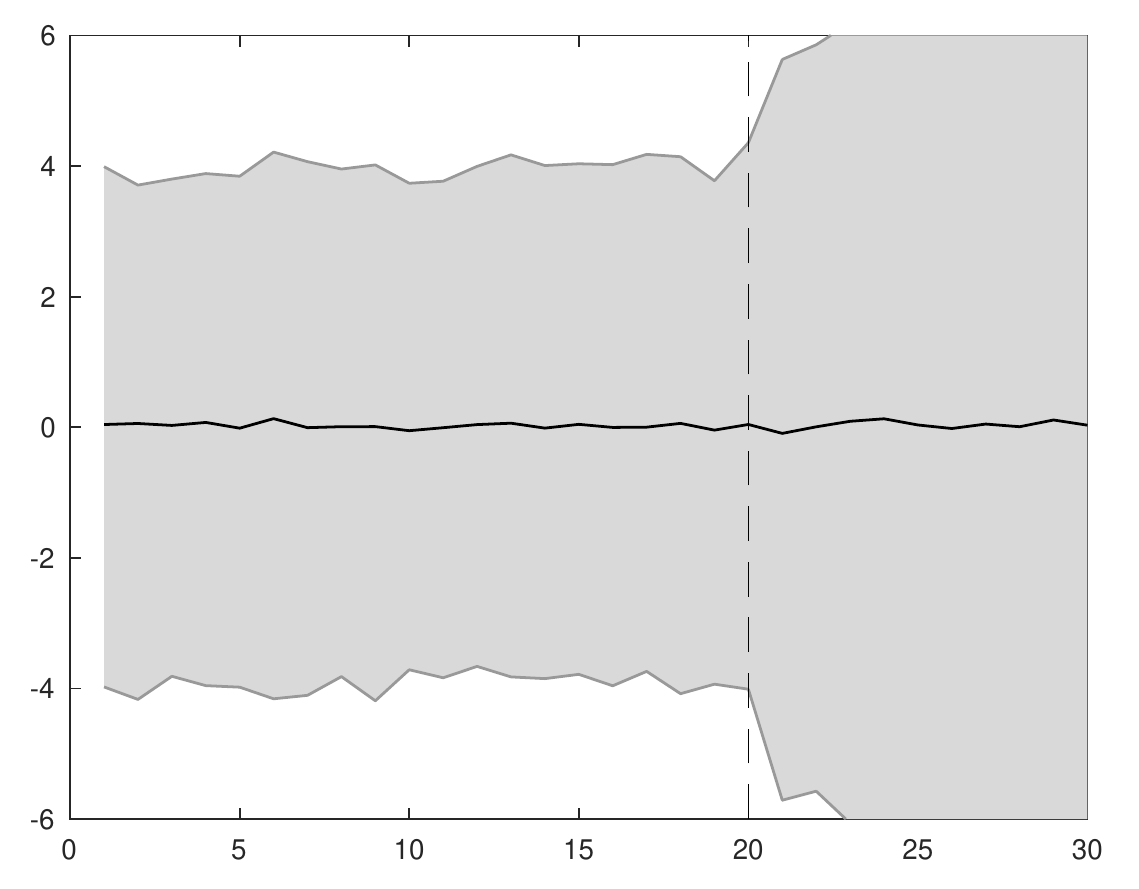}
  	\caption{$\sigma = 2$}
  \end{subfigure}%
  \begin{subfigure}[b]{0.5\textwidth}
  \centering
  	\includegraphics[width=1\linewidth]{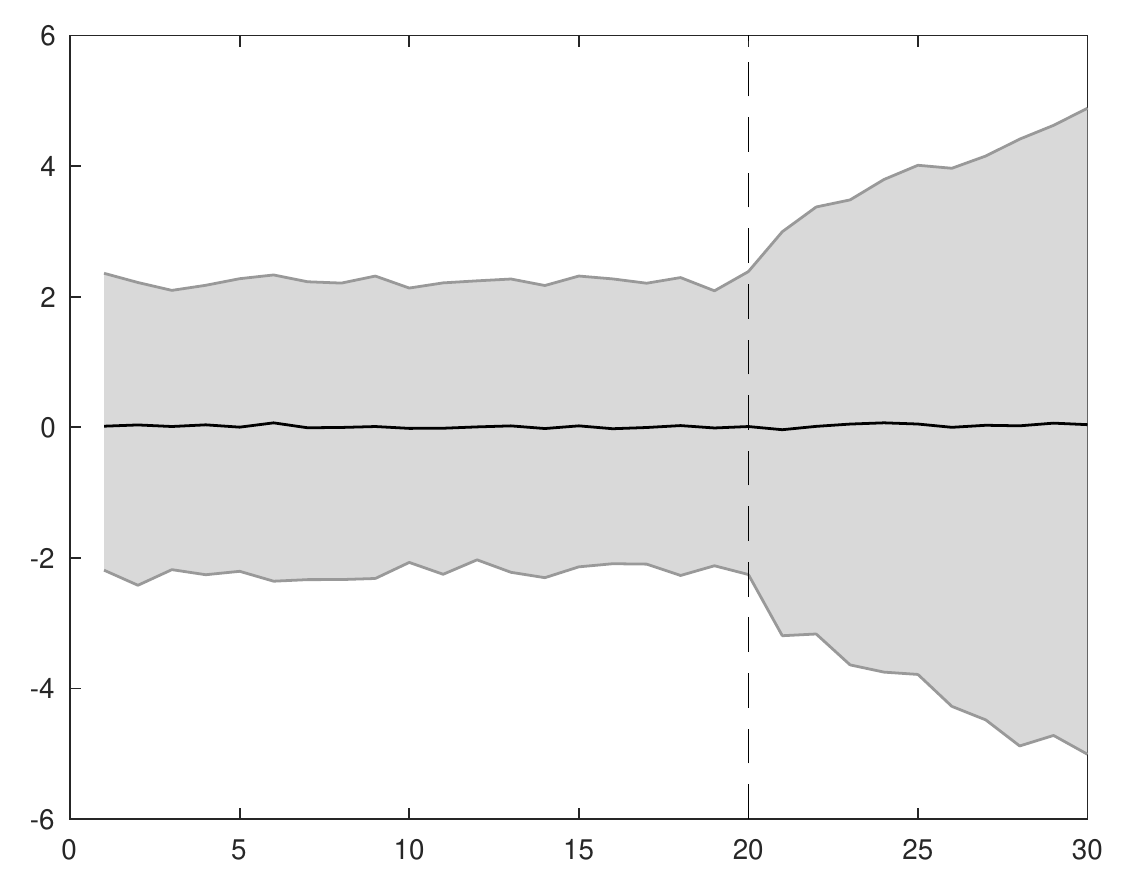}
  	\caption{$\sigma = 1$}
  \end{subfigure}%
  
   \begin{subfigure}[b]{0.5\textwidth}
  \centering
  	\includegraphics[width=1\linewidth]{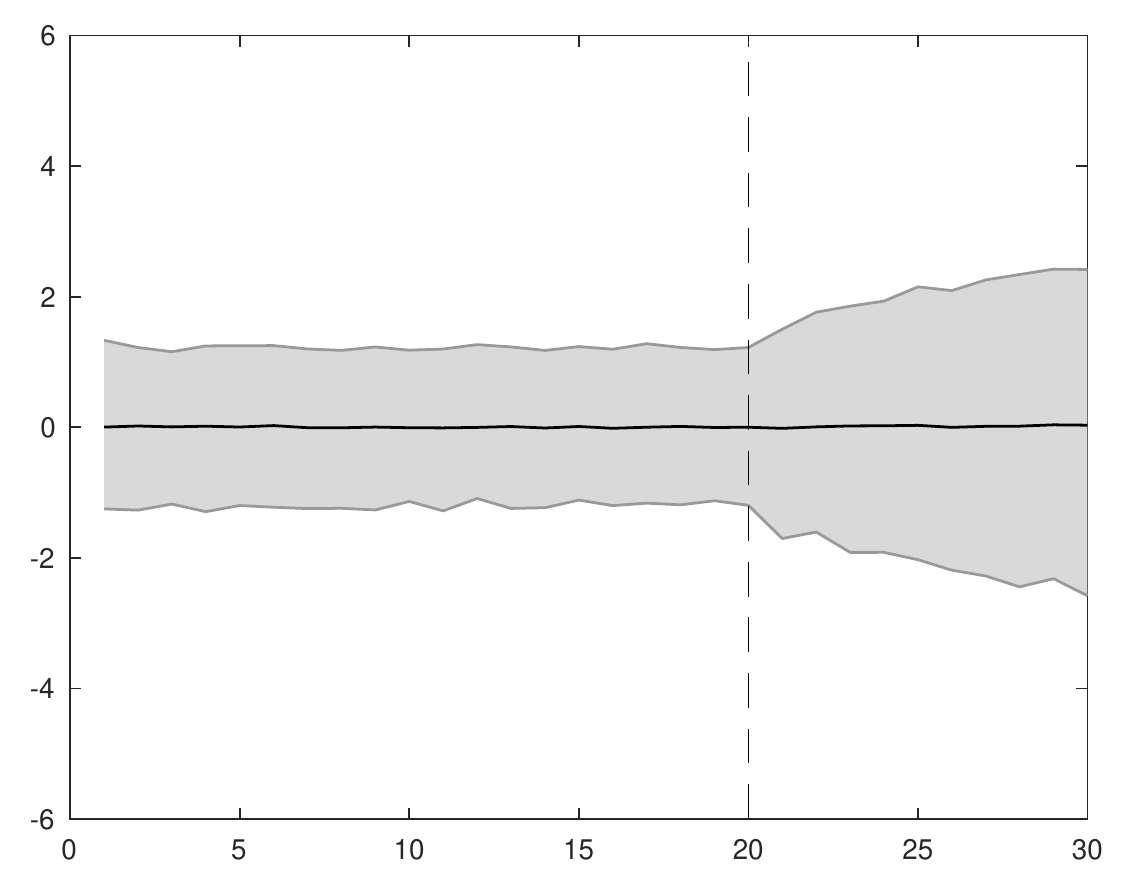}
  	\caption{$\sigma = 0.5$}
  \end{subfigure}%
  \begin{subfigure}[b]{0.5\textwidth}
  \centering
  	\includegraphics[width=1\linewidth]{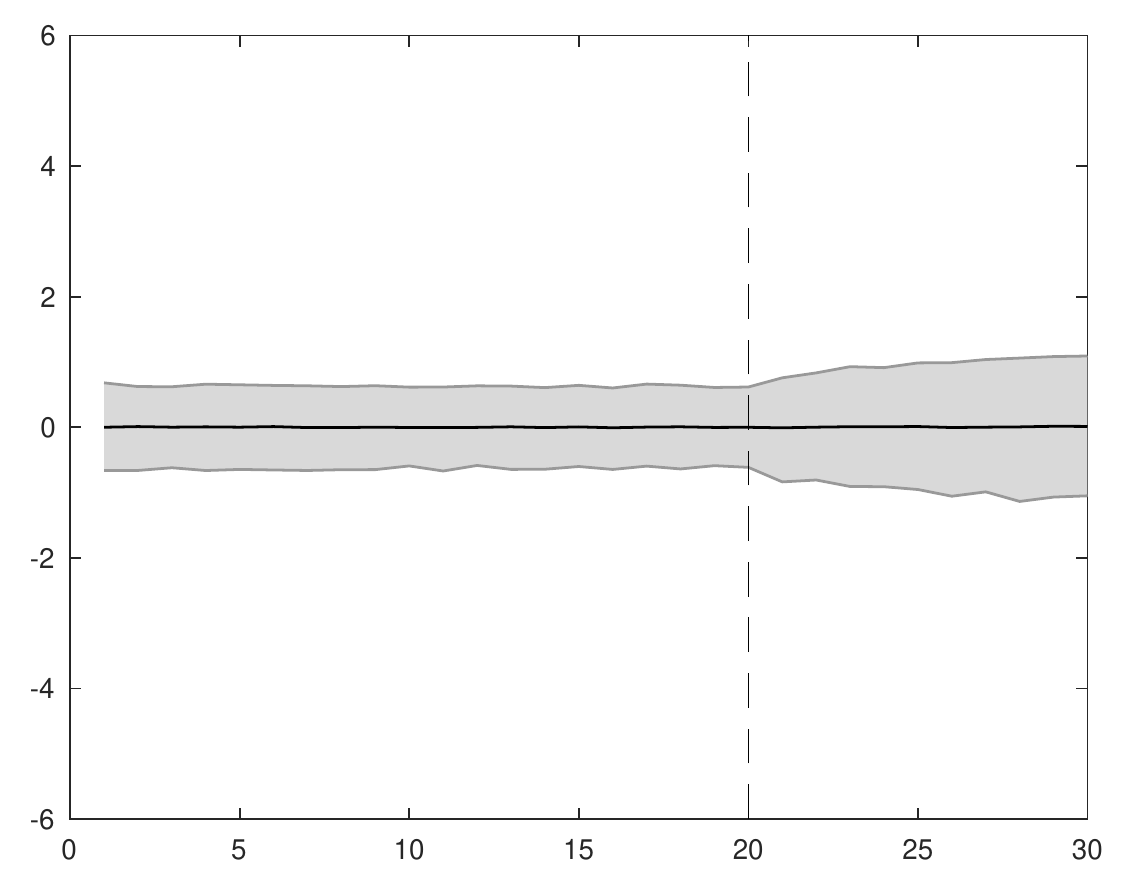}
  	\caption{$\sigma = 0.25$}
  \end{subfigure}%
  \caption{Synthetic control error simulations when $\rho = 1$.}
  \label{figure:sim_noise_rho1}
  \floatfoot{{\em Note:} 95\% bands for the simulation design of Figure \ref{figure:noise_rho1}.}
\end{figure}
\clearpage
\begin{figure}[ht!]
\centering
  \begin{subfigure}[b]{0.5\textwidth}
  \centering
  	\includegraphics[width=1\linewidth]{Tables_and_Figures/Brho100sigma50T020K10w.pdf}
  	\caption{$T_0 = 20$, $F = 10$ $(J = 19)$}
  \end{subfigure}%
  \begin{subfigure}[b]{0.5\textwidth}
  \centering
  	\includegraphics[width=1\linewidth]{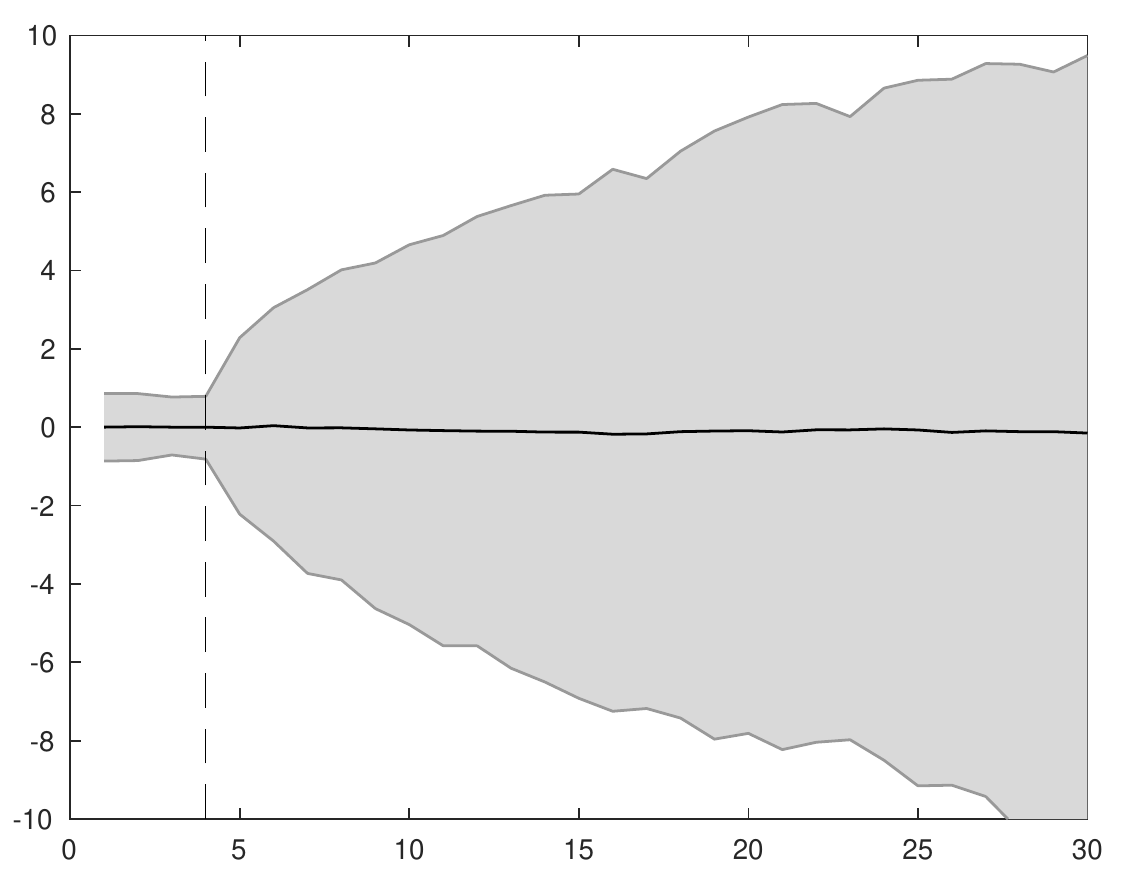}
  	\caption{$T_0 = 4$, $F = 10$ $(J = 19)$}
  \end{subfigure}%
  
   \begin{subfigure}[b]{0.5\textwidth}
  \centering
  	\includegraphics[width=1\linewidth]{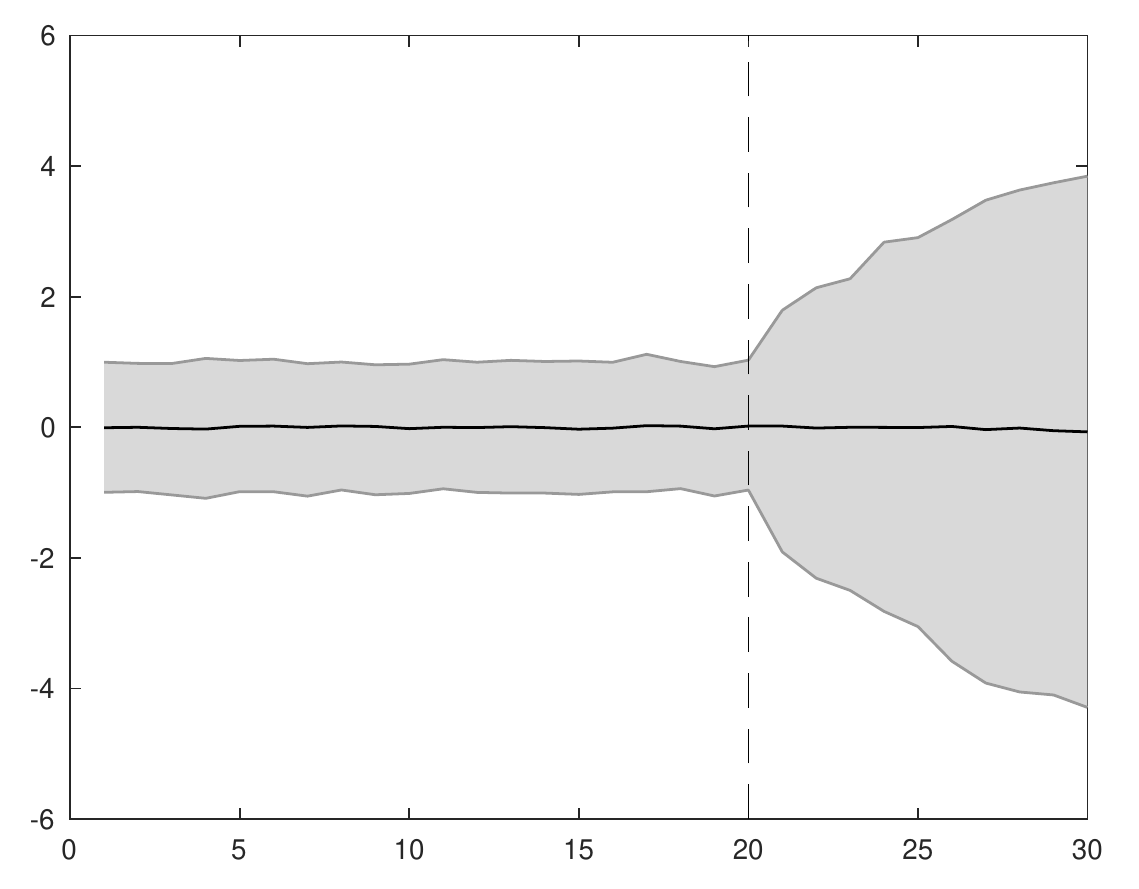}
  	\caption{$T_0 = 20$, $F = 100$ $(J = 199)$}
  \end{subfigure}%
  \begin{subfigure}[b]{0.5\textwidth}
  \centering
  	\includegraphics[width=1\linewidth]{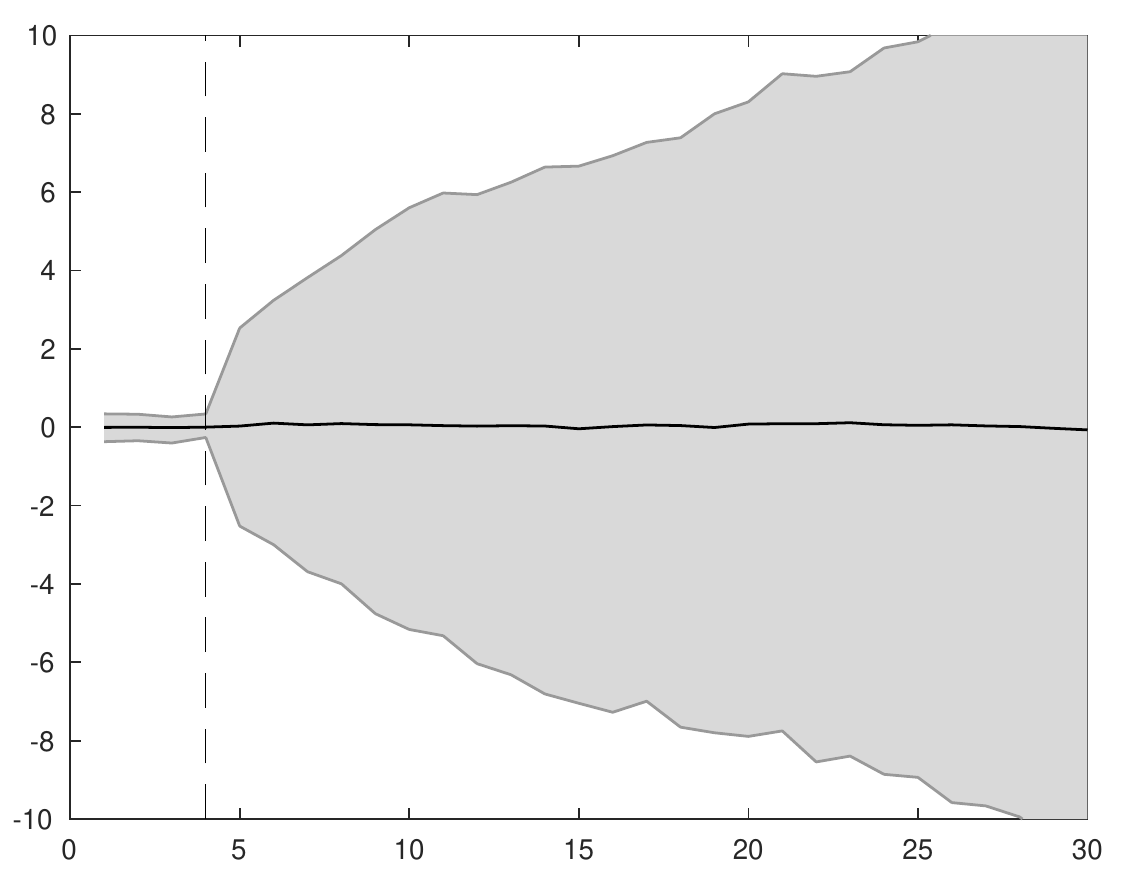}
  	\caption{$T_0 = 4$, $F = 100$ $(J = 199)$}
  \end{subfigure}%
  \caption{Synthetic control error simulations and over-fitting.}
\label{figure:sim_overfitting}
\floatfoot{{\em Note:} 95\% bands for the simulation design of Figure \ref{figure:overfitting}.}
\end{figure}

\clearpage
\begin{figure}[ht!]
\centering
  \begin{subfigure}[b]{0.5\textwidth}
  \centering
\includegraphics[width=1\linewidth]{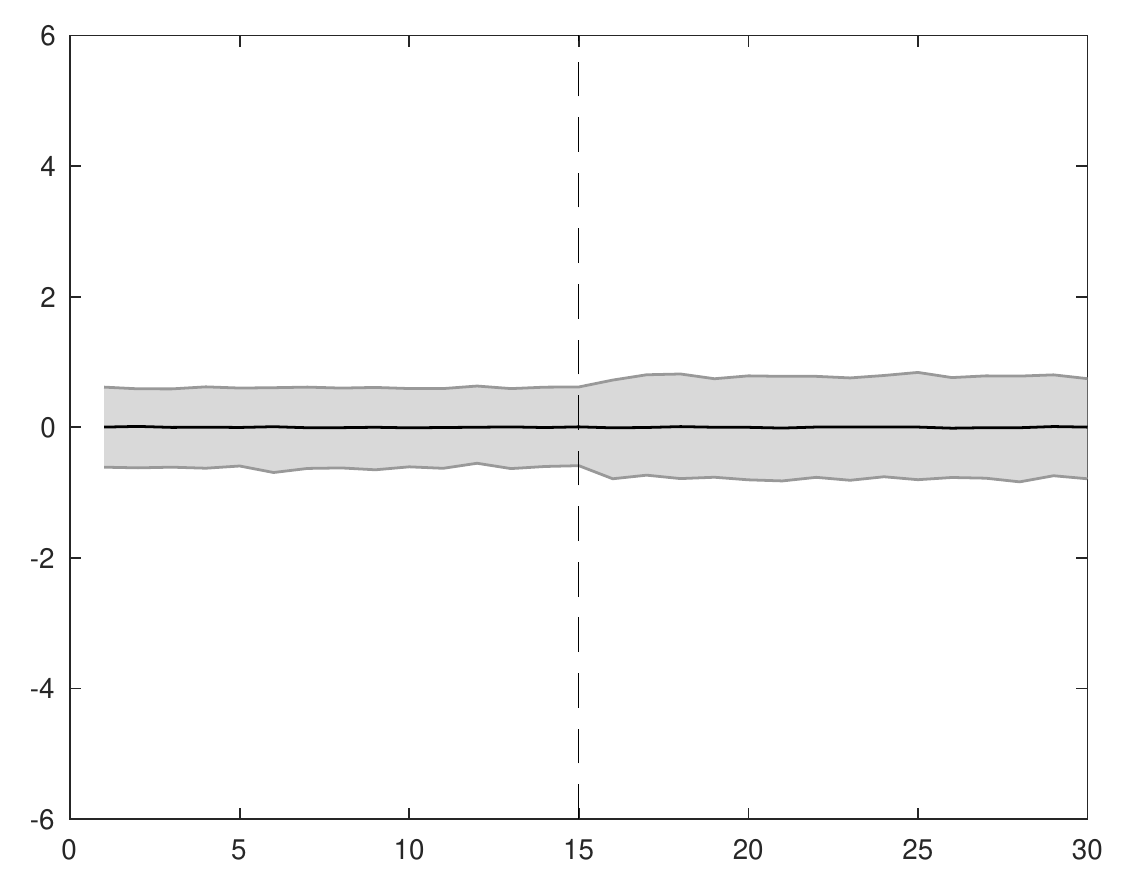}
  	\caption{Standard SC}
  \end{subfigure}%
  \begin{subfigure}[b]{0.5\textwidth}
  \centering
  	\includegraphics[width=1\linewidth]{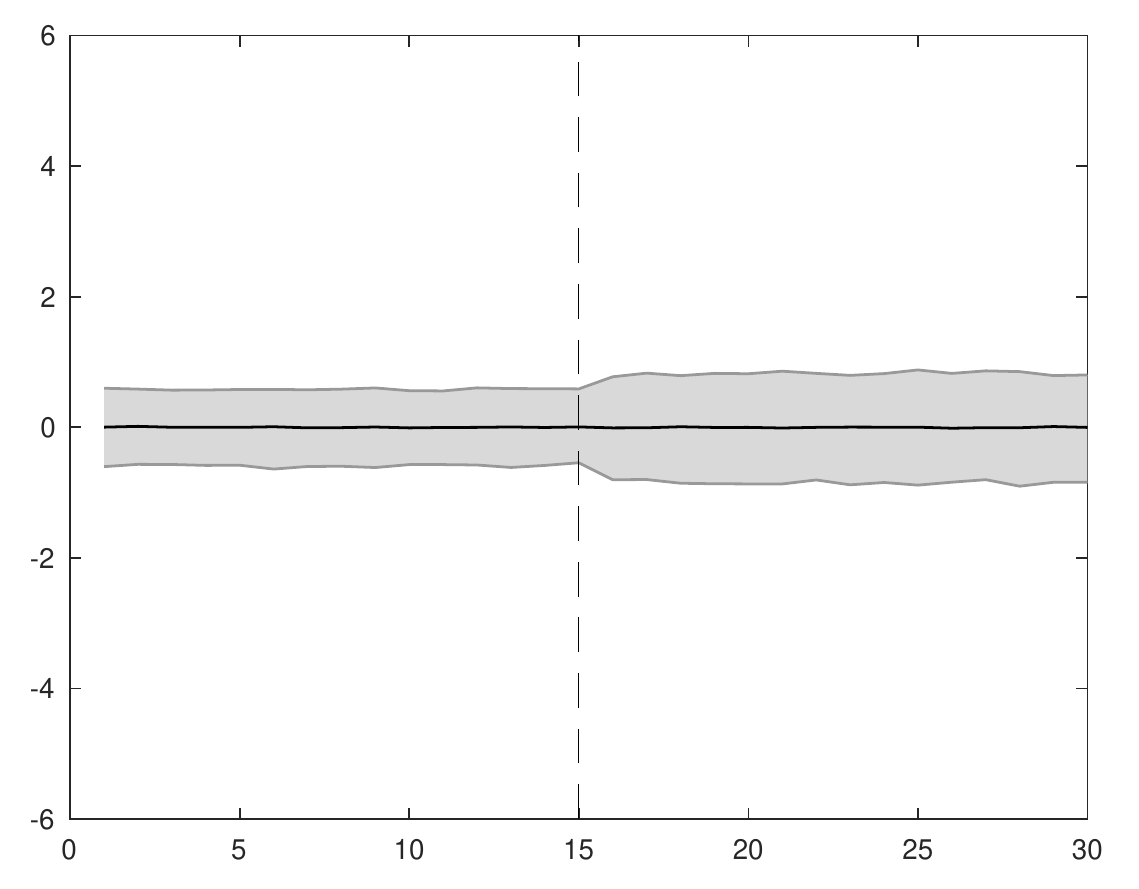}
  	\caption{SC with constant shift}
  \end{subfigure}%
  
   \begin{subfigure}[b]{0.5\textwidth}
  \centering
  	\includegraphics[width=1\linewidth]{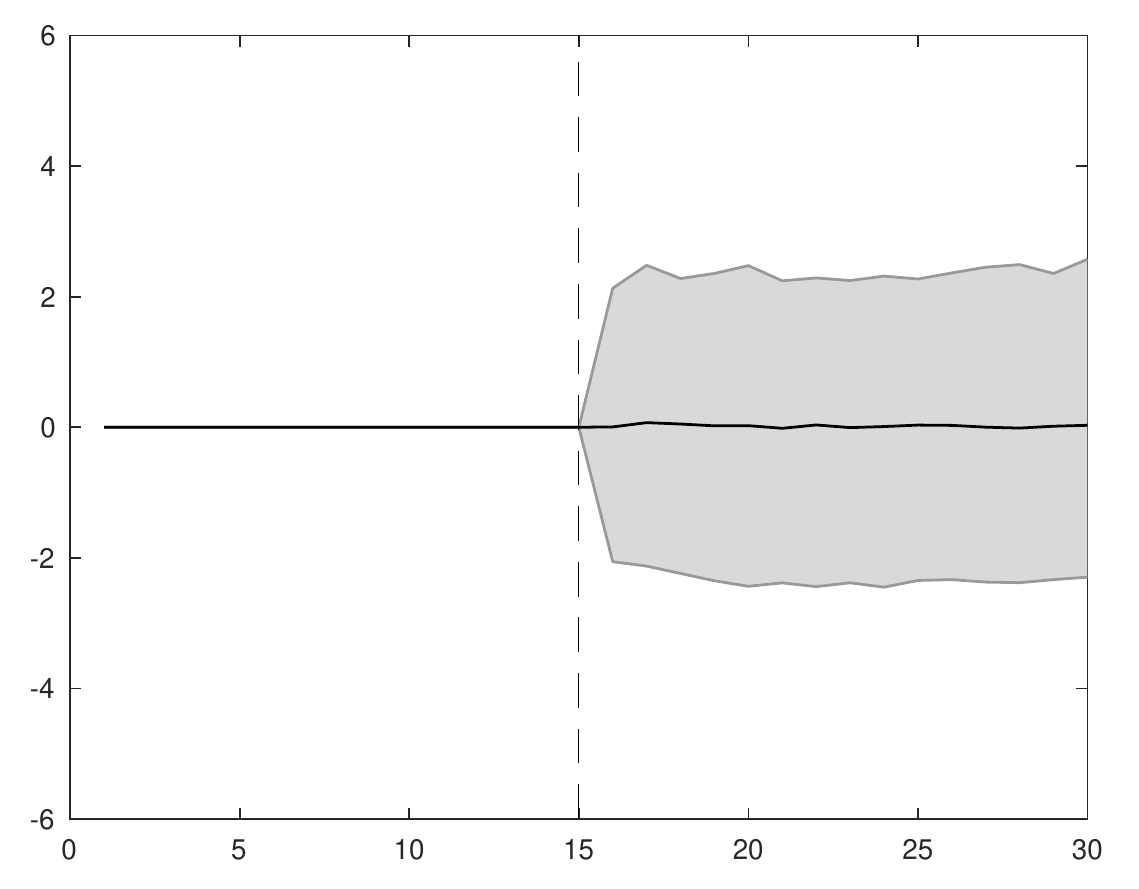}
  	\caption{SC via regression}
  \end{subfigure}%
  \caption{Synthetic control error simulations for different weight restrictions.}
  \label{figure:sim_flex}
  \floatfoot{{\em Note:} 95\% bands for the simulation design of Figure \ref{figure:flex}.}
\end{figure}
\clearpage
\begin{figure}[ht!]
\centering
  \begin{subfigure}[b]{0.5\textwidth}
  \centering
  	\includegraphics[width=1\linewidth]{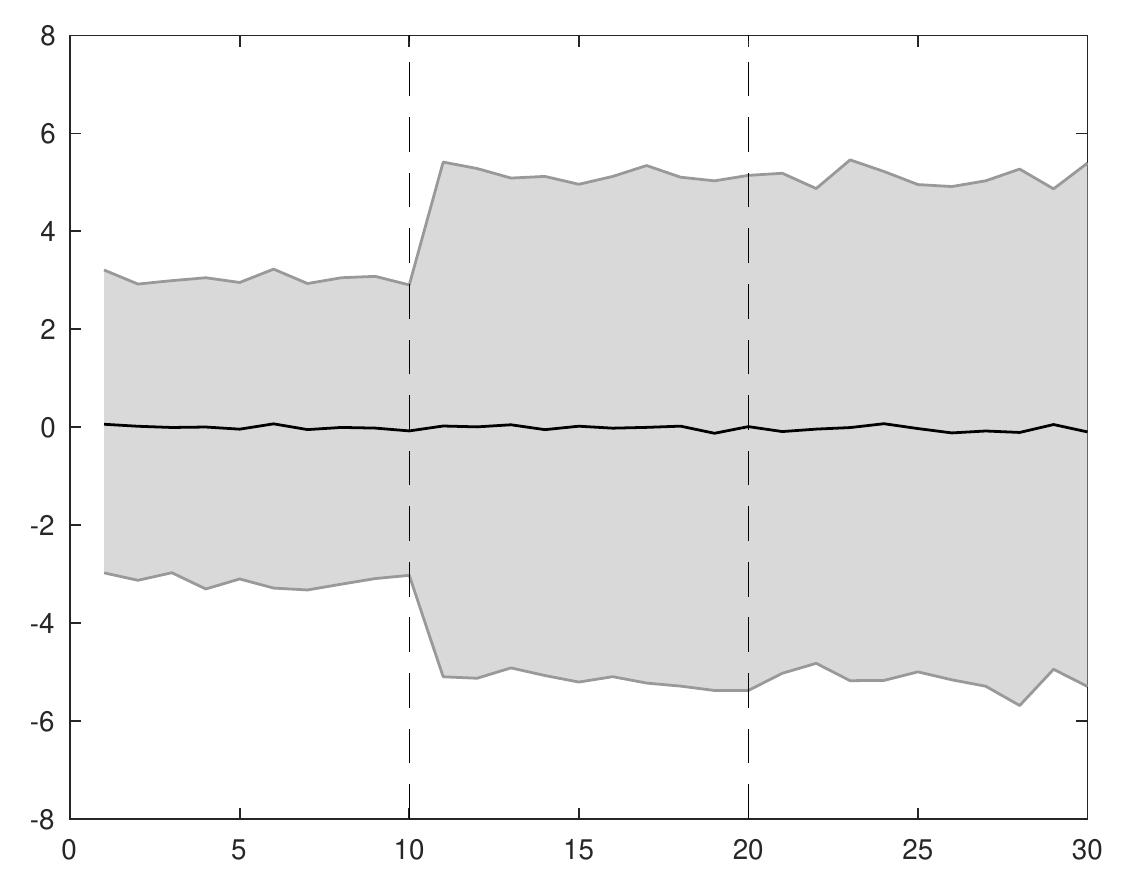}
  	\caption{$\sigma = 2$}
  \end{subfigure}%
  \begin{subfigure}[b]{0.5\textwidth}
  \centering
  	\includegraphics[width=1\linewidth]{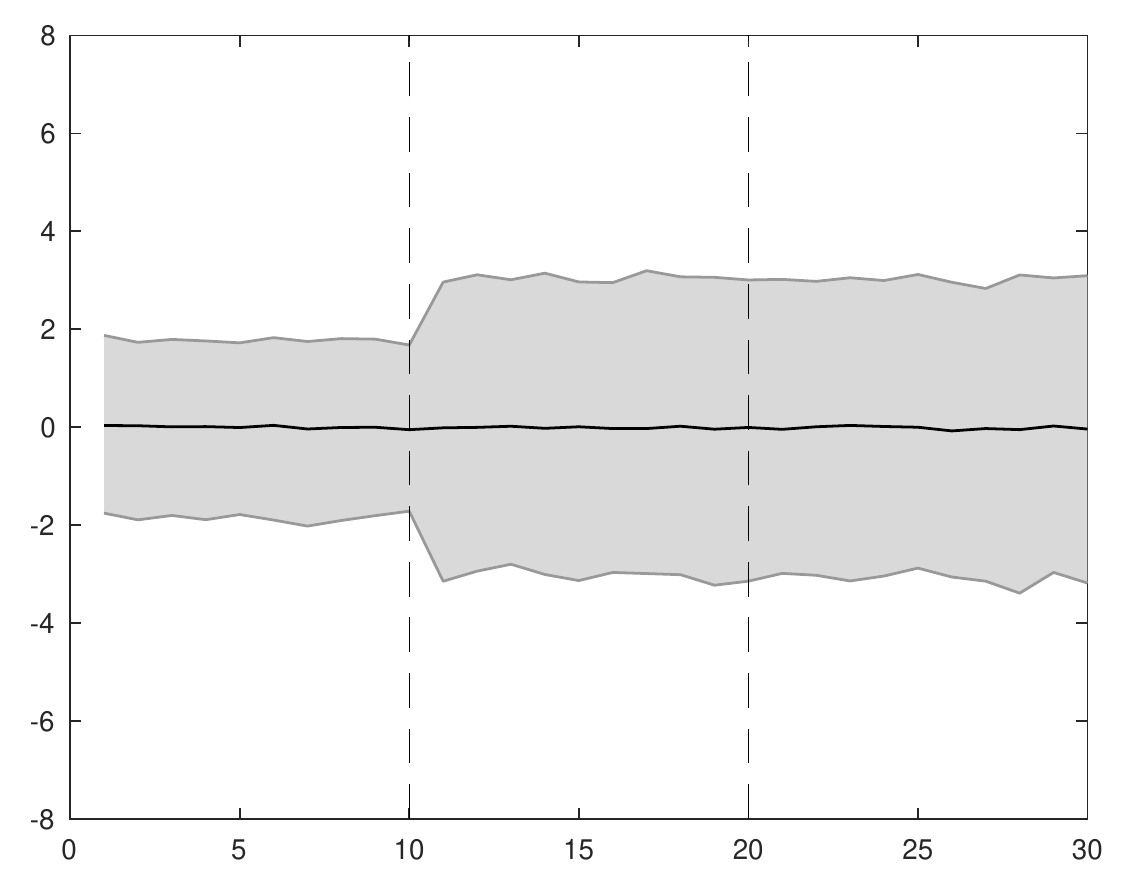}
  	\caption{$\sigma = 1$}
  \end{subfigure}%
  
   \begin{subfigure}[b]{0.5\textwidth}
  \centering
  	\includegraphics[width=1\linewidth]{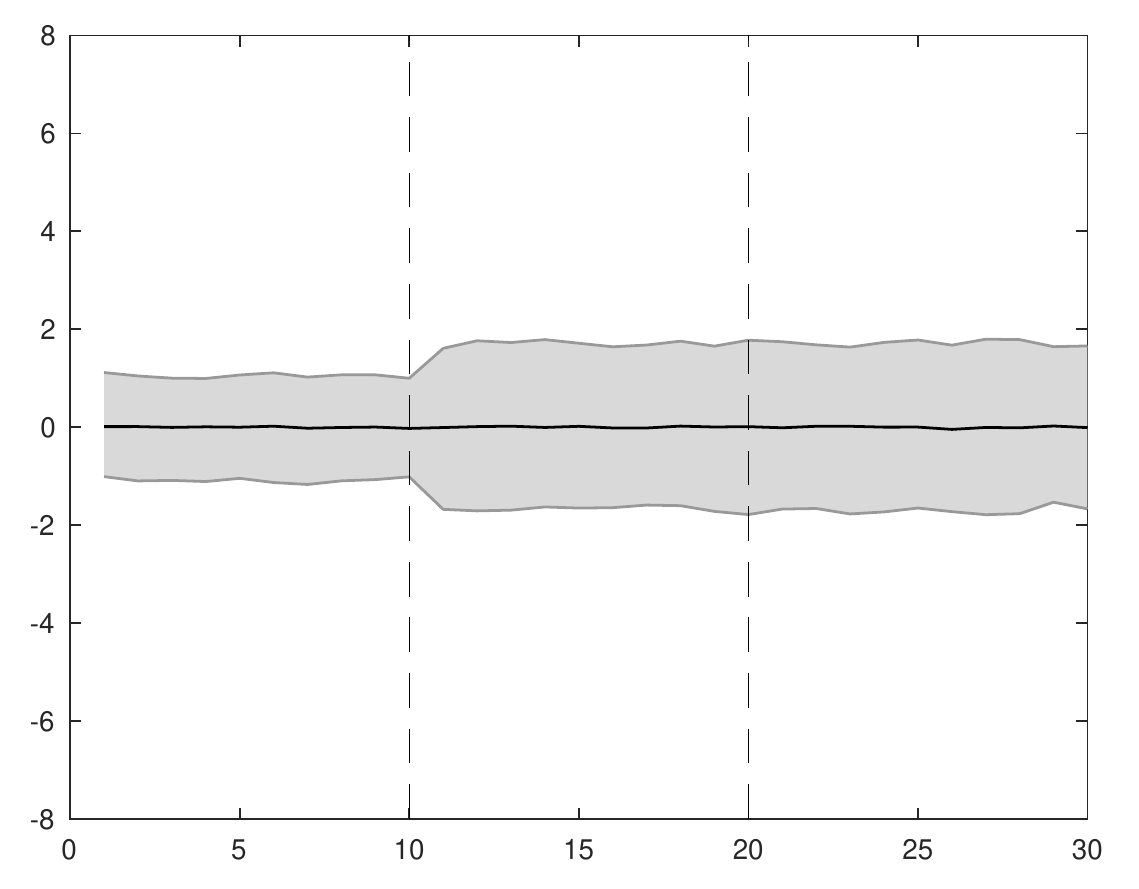}
  	\caption{$\sigma = 0.5$}
  \end{subfigure}%
  \begin{subfigure}[b]{0.5\textwidth}
  \centering
  	\includegraphics[width=1\linewidth]{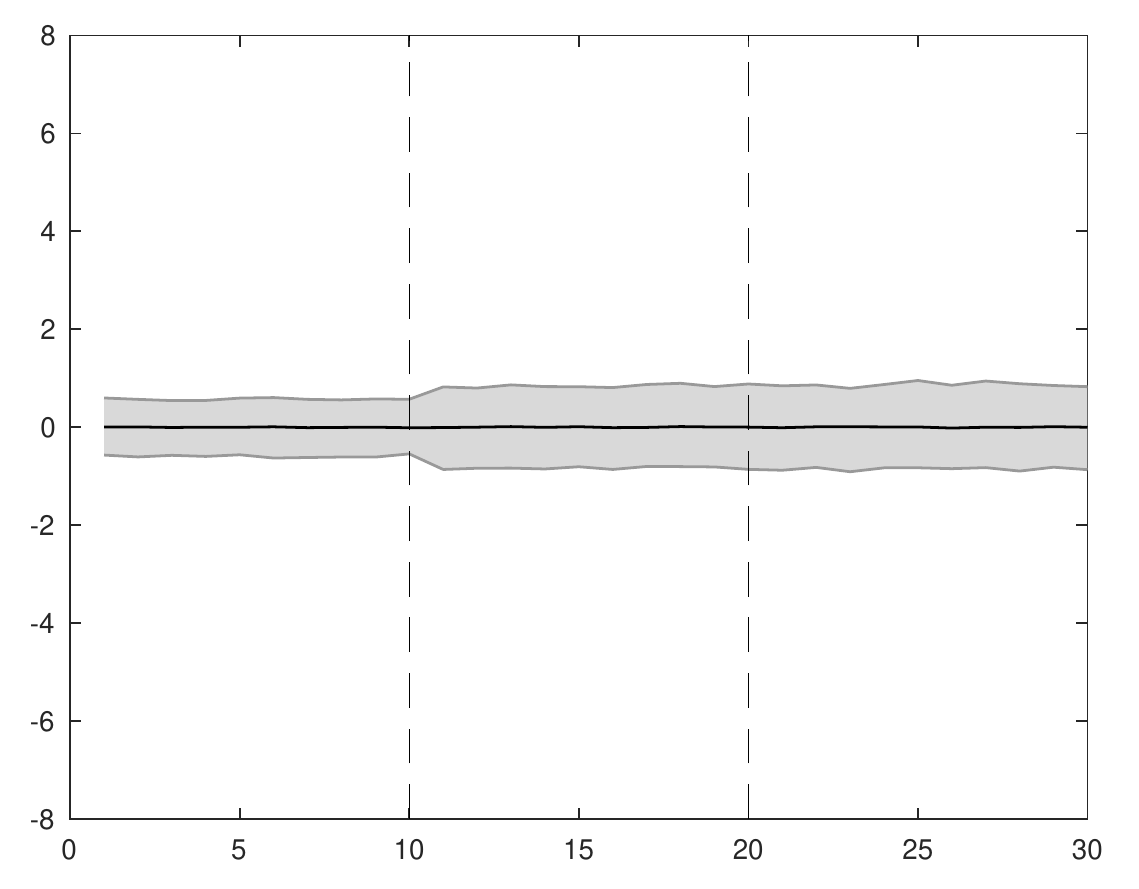}
  	\caption{$\sigma = 0.25$}
  \end{subfigure}%
  \caption{Synthetic control error simulations with a validation period.}
  \label{figure:sim_validation}
  \floatfoot{{\em Note:} 95\% bands for the simulation design of Figure \ref{figure:validation}.}
\end{figure}
\clearpage
\begin{figure}[ht!]
\centering
  \begin{subfigure}[b]{0.5\textwidth}
  \centering
  	\includegraphics[width=1\linewidth]{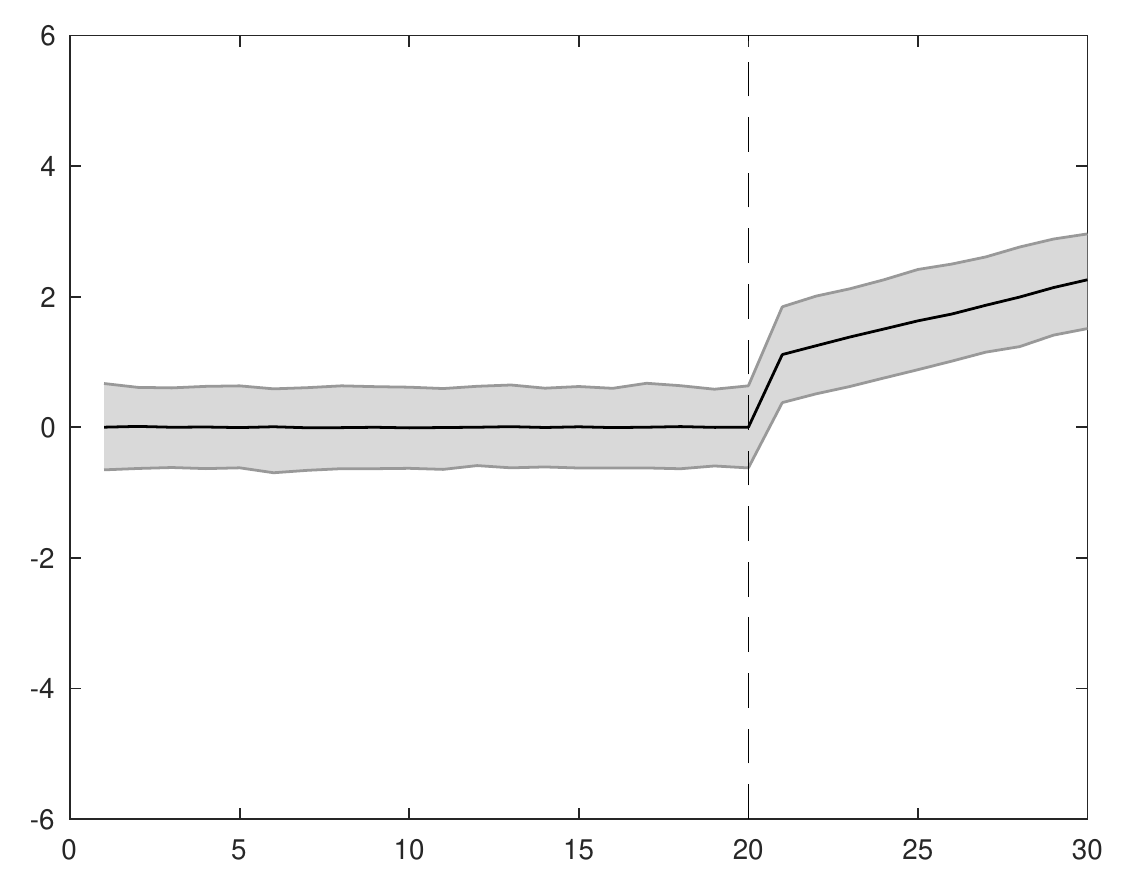}
  	\caption{Intervention at $T_0 =20$}
  \end{subfigure}%
  \begin{subfigure}[b]{0.5\textwidth}
  \centering
  	\includegraphics[width=1\linewidth]{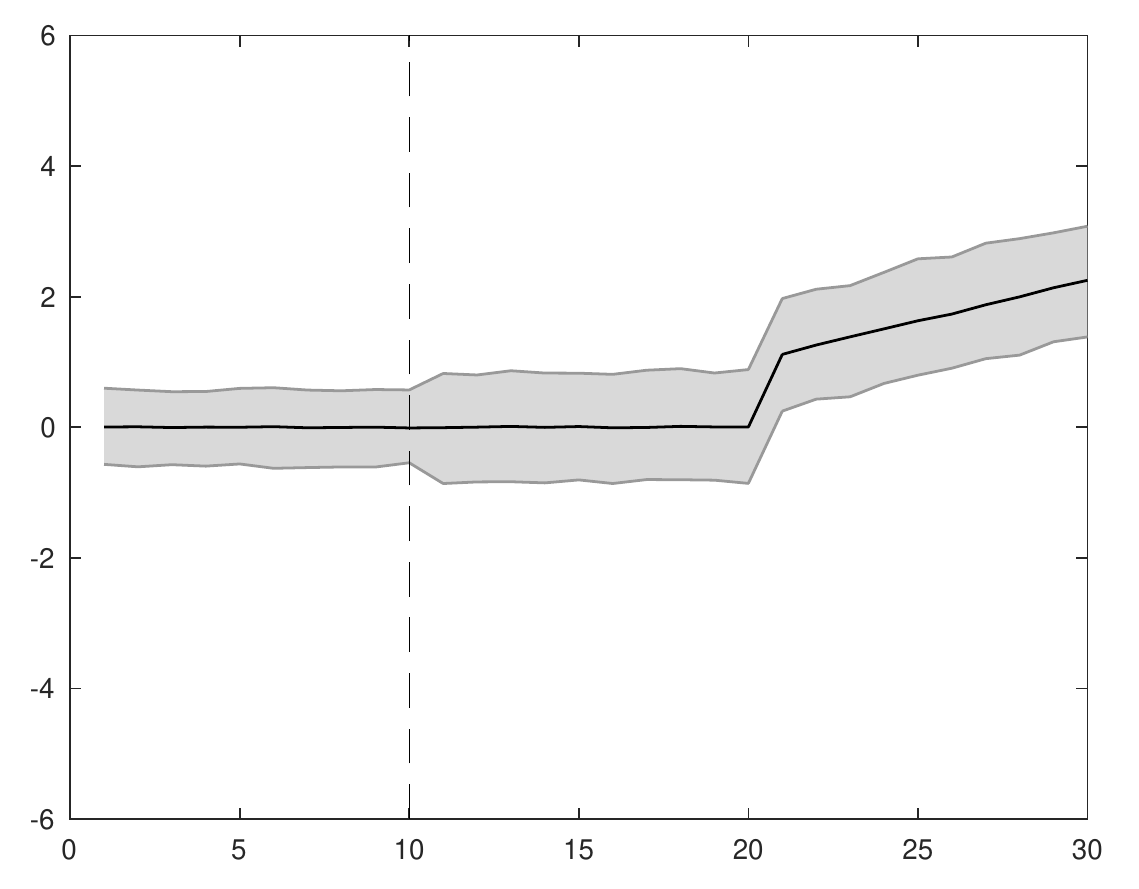}
  	\caption{Backdating to $T=10$}
  \end{subfigure}%
  \caption{Synthetic control error simulations with treatment effect.}
  \label{figure:sim_validation_teffect}
  \floatfoot{{\em Note:} 95\% bands for the simulation design of Figure \ref{figure:validation_teffect}.}
\end{figure}
\clearpage
\begin{figure}[ht!]
\centering
  \begin{subfigure}[b]{0.5\textwidth}
  \centering
  	\includegraphics[width=1\linewidth]{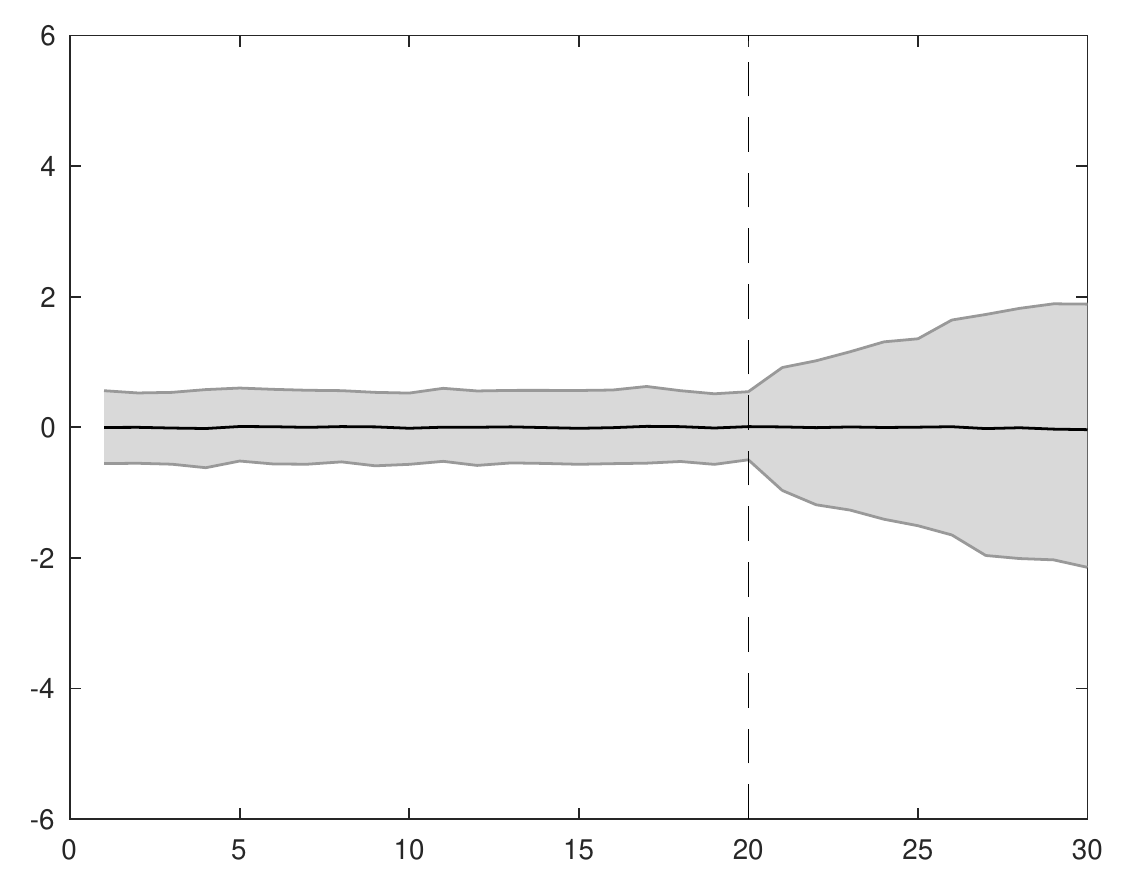}
  	\caption{Without trimming.}
  \end{subfigure}%
  \begin{subfigure}[b]{0.5\textwidth}
  \centering
  	\includegraphics[width=1\linewidth]{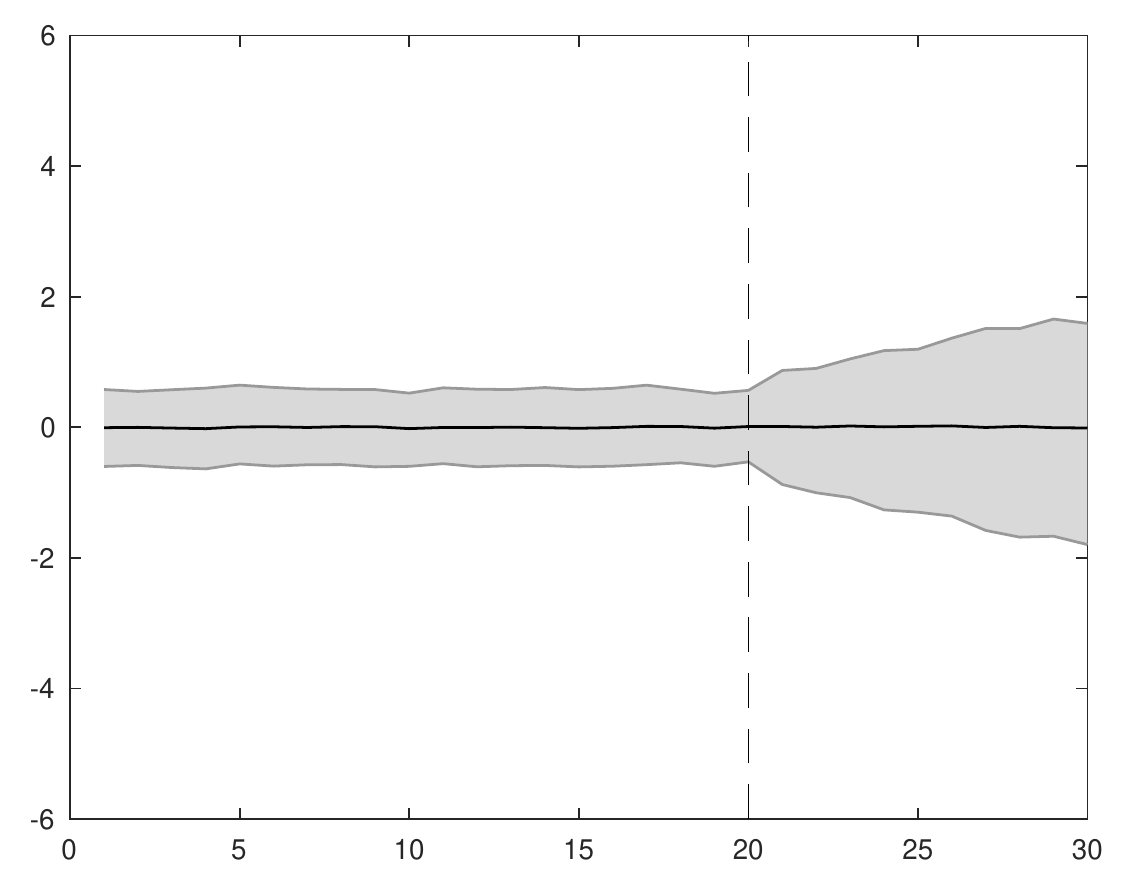}
  	\caption{75\% trimming.}
  \end{subfigure}
    \begin{subfigure}[b]{0.5\textwidth}
  \centering
  	\includegraphics[width=1\linewidth]{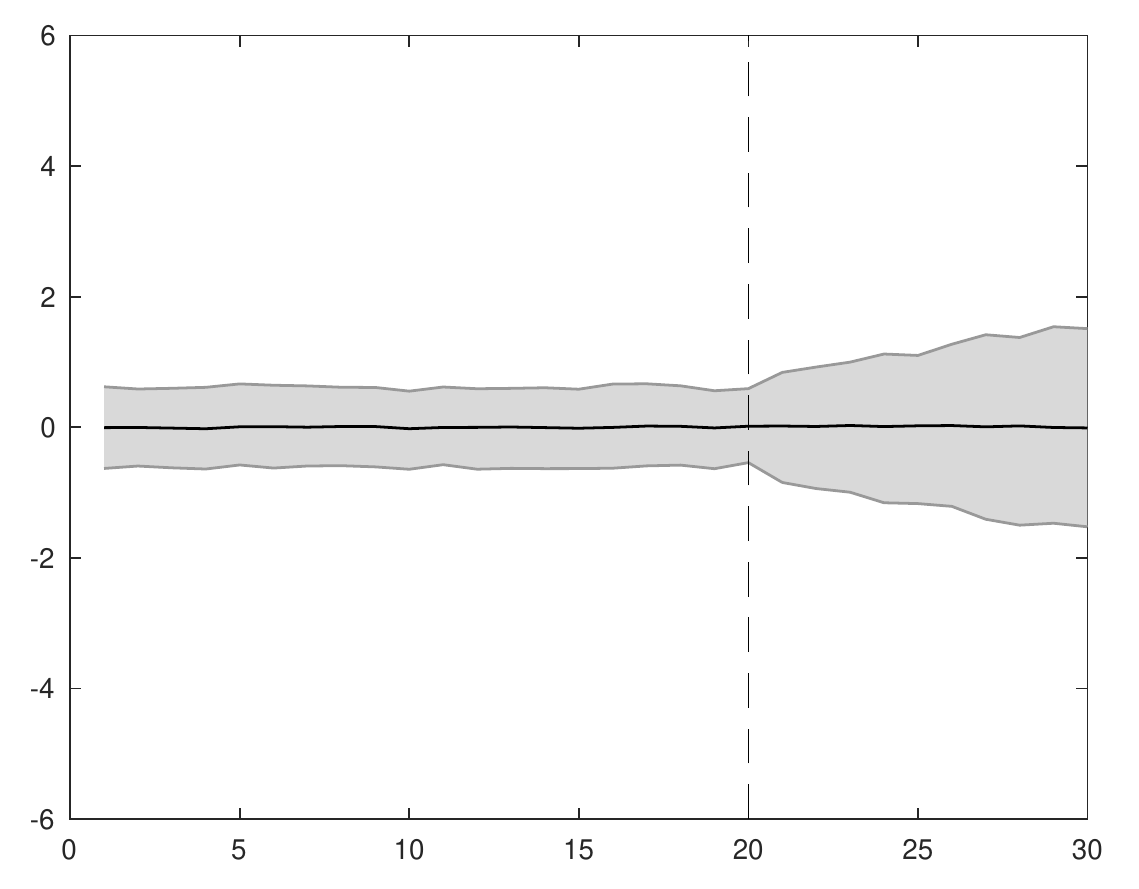}
  	\caption{90\% trimming.}
  \end{subfigure}%
  \begin{subfigure}[b]{0.5\textwidth}
  \centering
  	\includegraphics[width=1\linewidth]{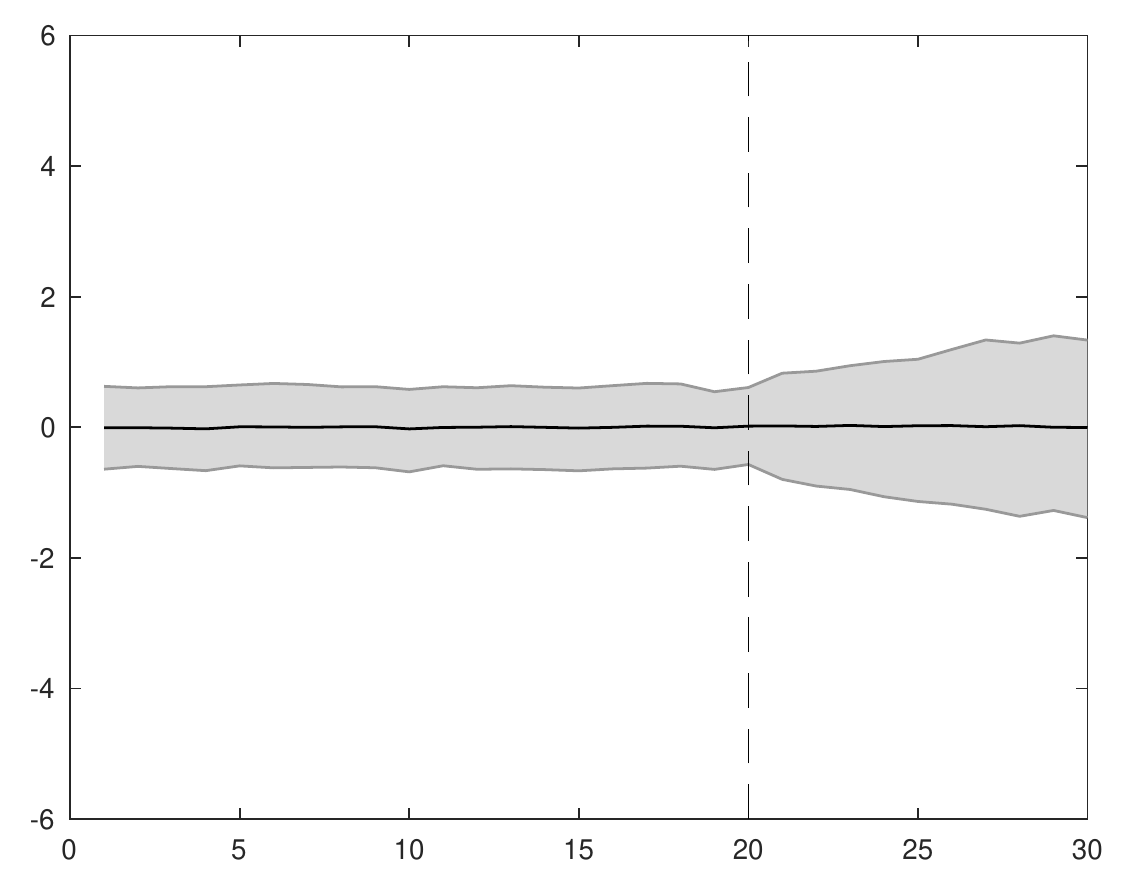}
  	\caption{95\% trimming.}
  \end{subfigure}
  \caption{Synthetic control error simulation with trimming.}
  \label{figure:sim_trimming}
  \floatfoot{{\em Note:} 95\% bands for the simulation design of Figure \ref{figure:trimming}}
\end{figure}
\clearpage
\begin{figure}[ht!]
\centering
  \begin{subfigure}[b]{0.45\textwidth}
  \centering
  	\includegraphics[width=1\linewidth]{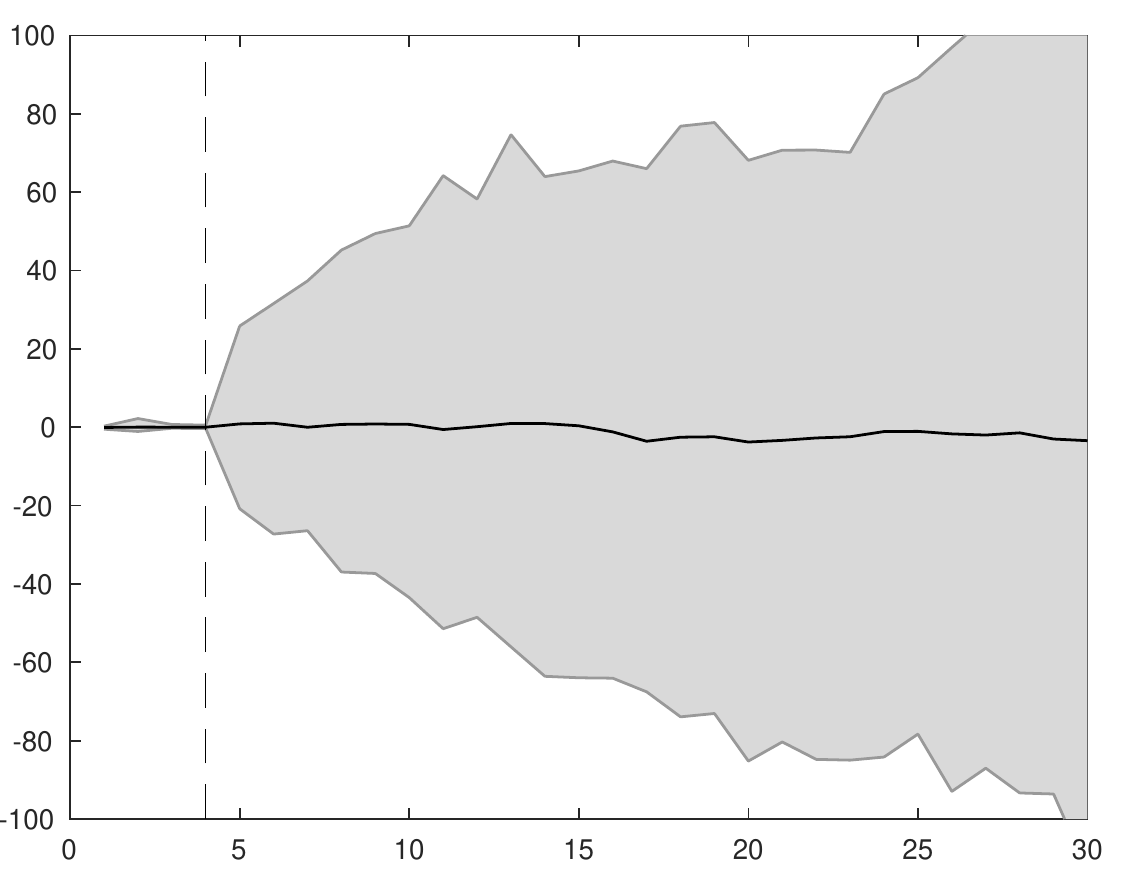}
  	\caption{$F = 5$}
  \end{subfigure}
  \begin{subfigure}[b]{0.45\textwidth}
  \centering
  	\includegraphics[width=1\linewidth]{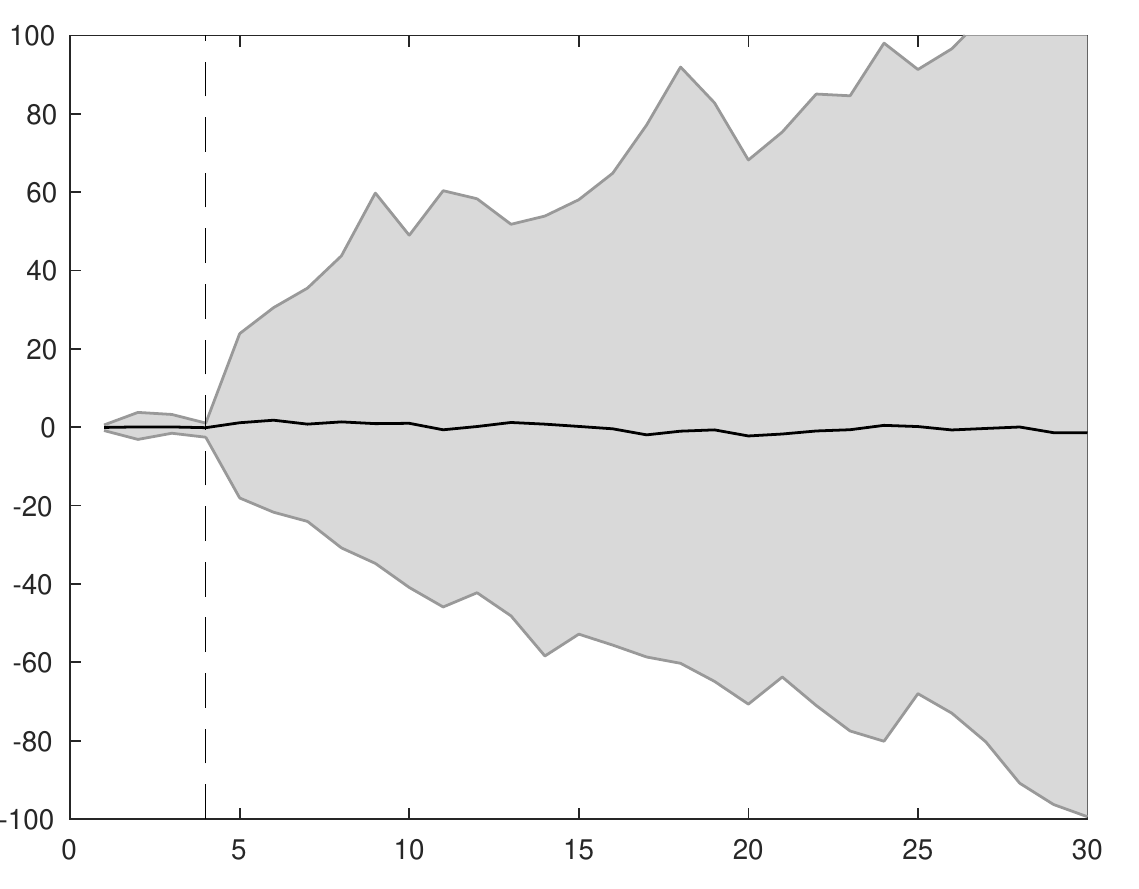}
  	\caption{$F = 4$}
  \end{subfigure}
  
   \begin{subfigure}[b]{0.45\textwidth}
  \centering
  	\includegraphics[width=1\linewidth]{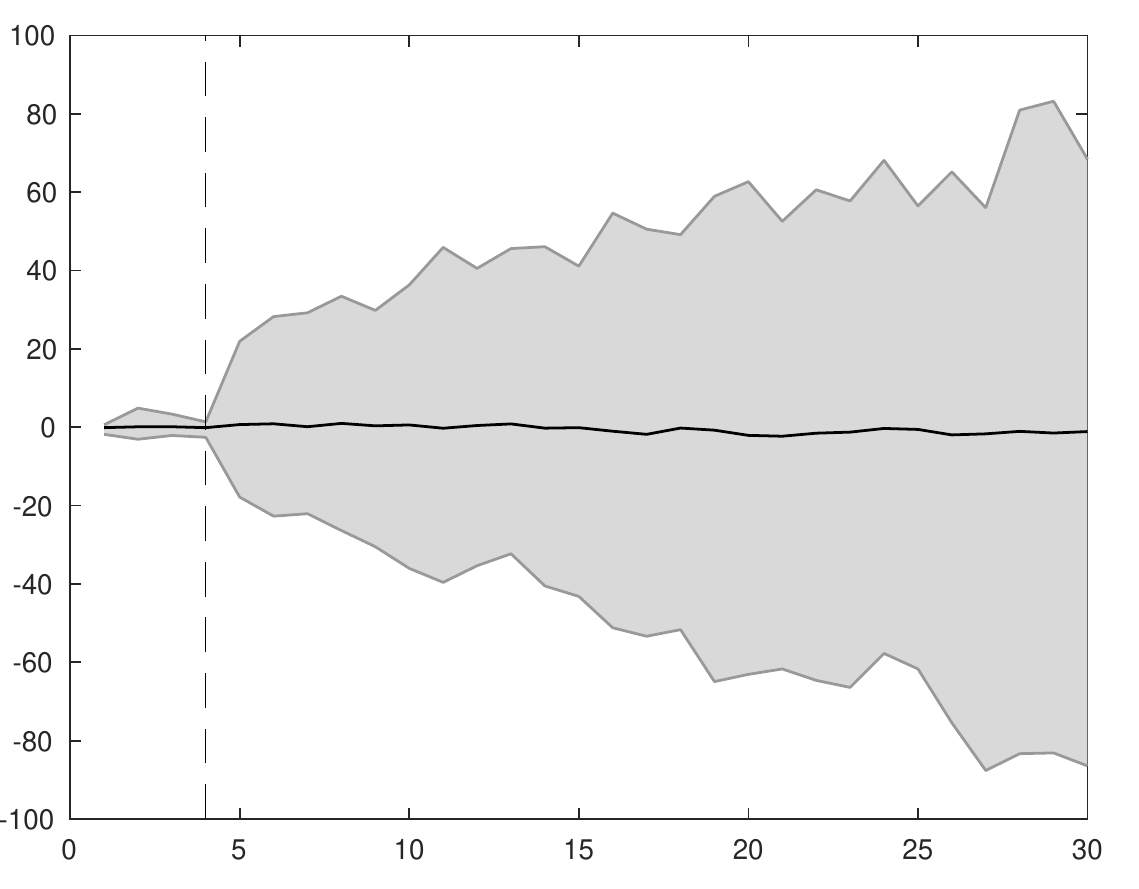}
  	\caption{$F = 3$}
  \end{subfigure}
  \begin{subfigure}[b]{0.45\textwidth}
  \centering
  	\includegraphics[width=1\linewidth]{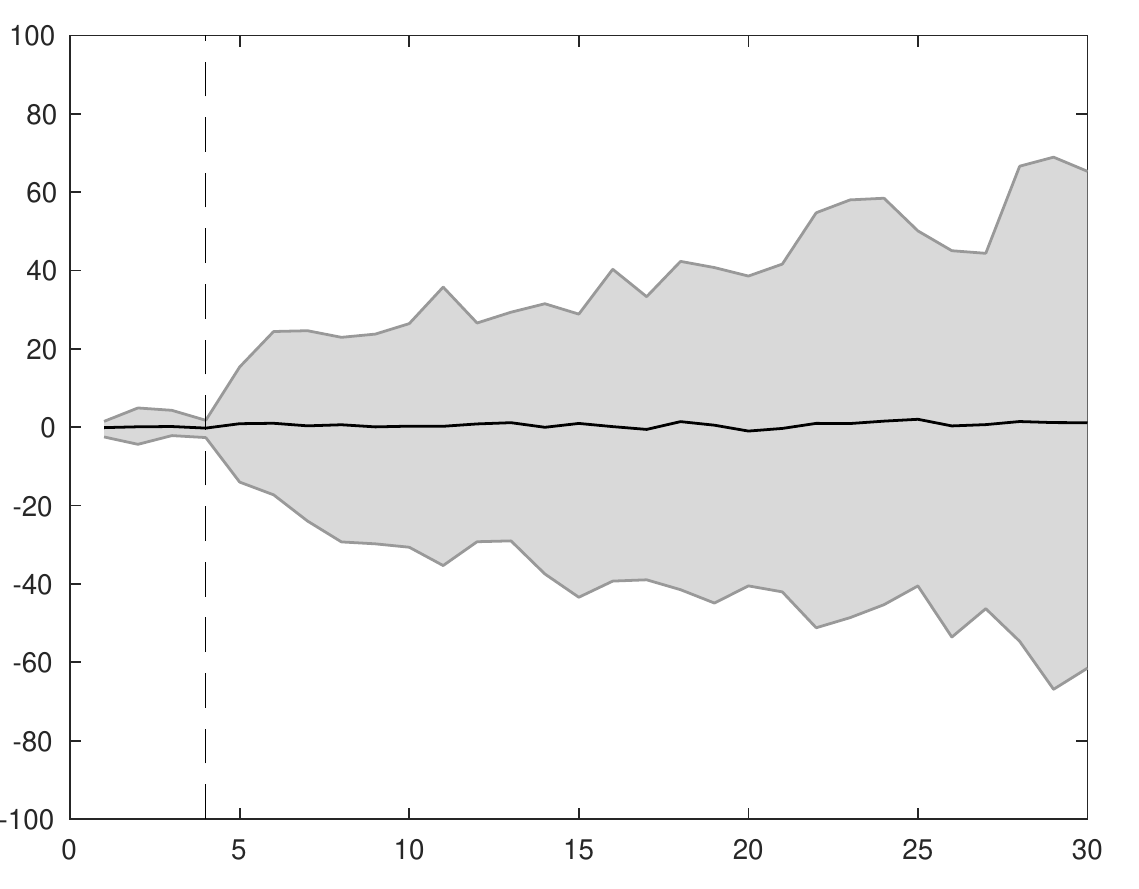}
  	\caption{$F = 2$}
  \end{subfigure}
  
     \begin{subfigure}[b]{0.45\textwidth}
  \centering
  	\includegraphics[width=1\linewidth]{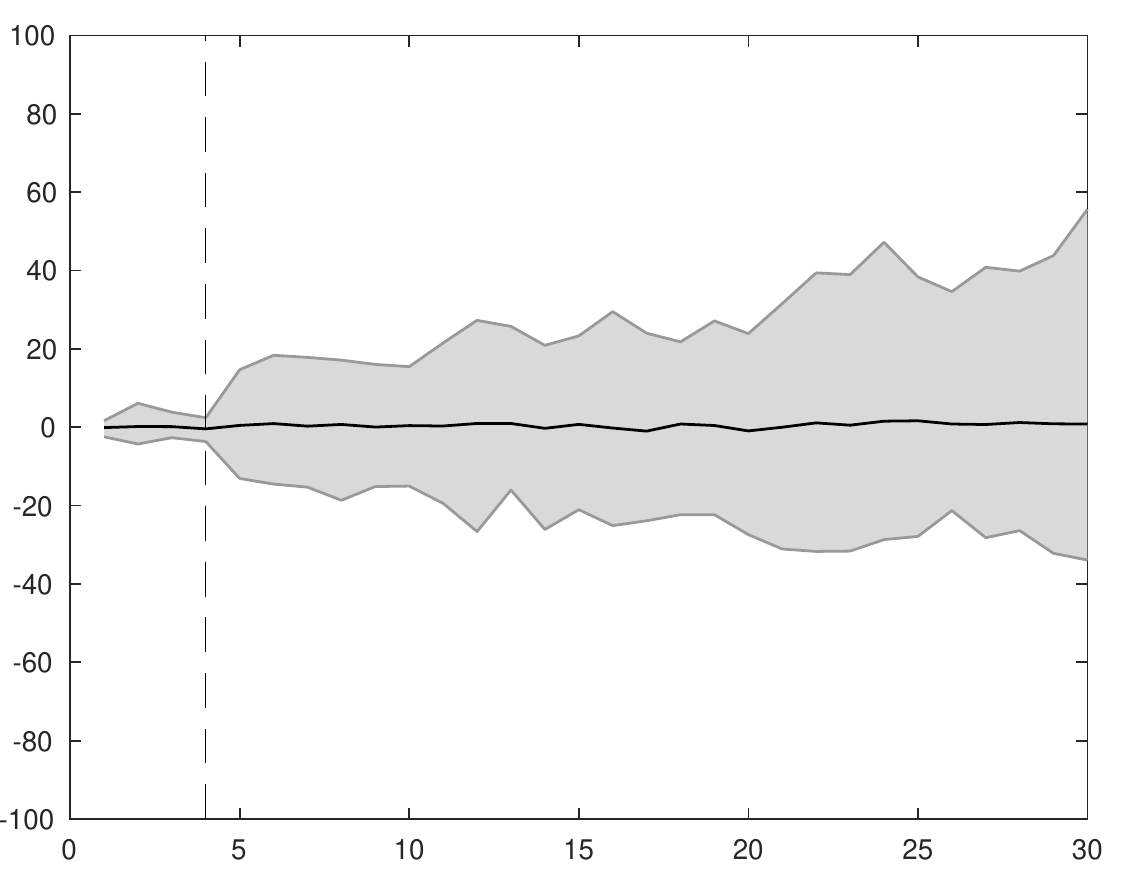}
  	\caption{$F = 1$}
  \end{subfigure}
  \begin{subfigure}[b]{0.45\textwidth}
  \centering
  	\includegraphics[width=1\linewidth]{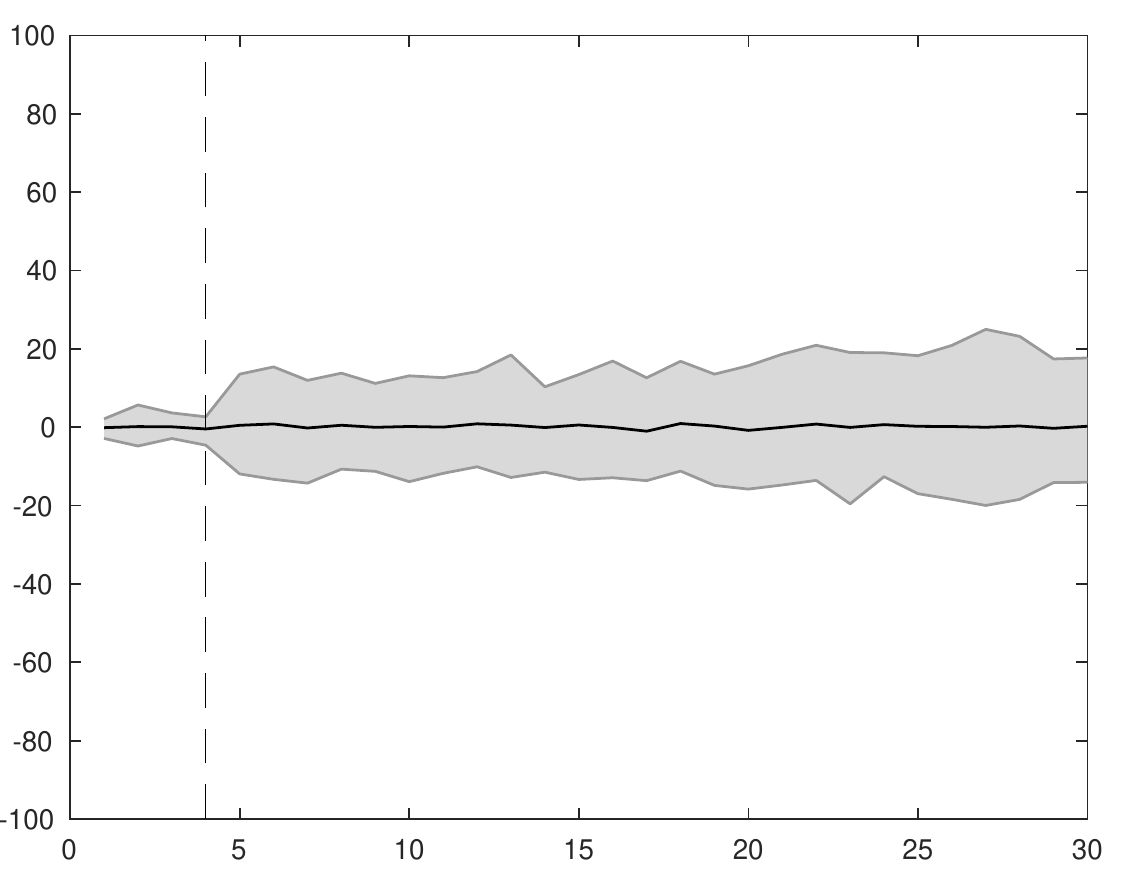}
  	\caption{$F = 0$}
  \end{subfigure}
  \caption{Synthetic control error simulations with observed covariates.}
  \label{figure:sim_factors}
  \floatfoot{{\em Note:} 95\% bands for the simulation design of Figure \ref{figure:factors}.}
\end{figure}
\clearpage
\begin{figure}[ht!]
\centering
  	\includegraphics[width=0.5\linewidth]{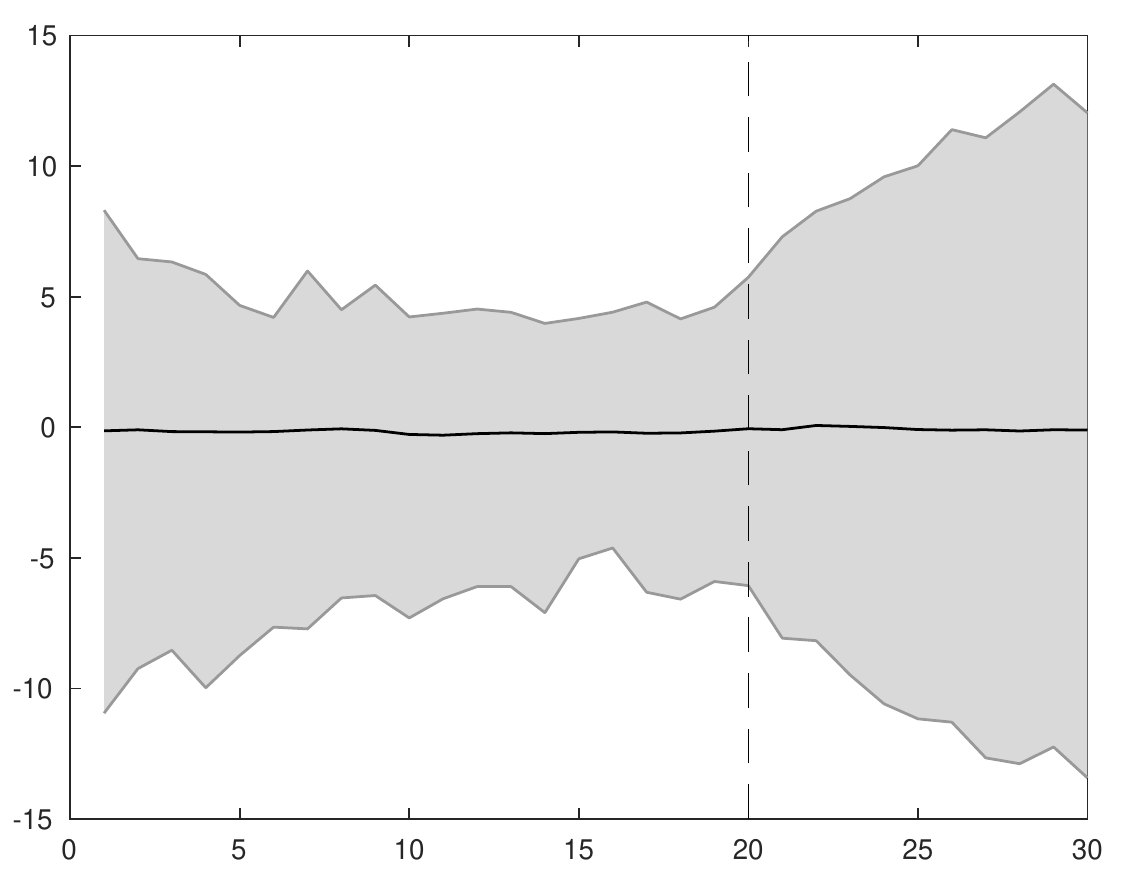}
  \caption{Synthetic control error simulations with an auto-regressive process.}
  \label{figure:sim_ar}
  \floatfoot{{\em Note:} 95\% bands for the simulation design of Figure \ref{figure:ar}.}
\end{figure}
\vfill

\end{document}